\documentclass[12pt,a4paper,reqno]{article}

\pdfoutput=1

\usepackage{amsmath}
\usepackage{amsfonts}
\usepackage{amssymb}

\usepackage[utf8]{inputenc}
\usepackage[T1]{fontenc}
\usepackage{lmodern}

\usepackage{comment}
\usepackage{enumerate}
\usepackage{geometry}
\usepackage{graphicx}
\usepackage[unicode,bookmarks]{hyperref}
\usepackage{mathbbol}
\usepackage[l2tabu, orthodox]{nag}
\usepackage[numbers,square]{natbib}
\usepackage[labelformat=empty]{subfig}
\usepackage[normalem]{ulem}
\usepackage{xcolor}

\newcommand{\mcal}[1]{\ensuremath{\mathcal{#1}}}
\newcommand{\abs}[1]{\ensuremath{\left \vert #1 \right \vert}}

\newcommand{\norm}[1]{\ensuremath{\left \lVert #1 \right \rVert}}
\newcommand{\pair}[2]{\ensuremath{\left < #1 \, , \, #2 \right >}}

\newcommand{\basis}[1]{\ensuremath{\mathrm{#1}}}
\newcommand{\coord}[1]{\ensuremath{#1}}

\newcommand{\bx}{\ensuremath{\basis{x}}}

\newcommand{\cx}{\ensuremath{\coord{x}}}
\newcommand{\cy}{\ensuremath{\coord{y}}}

\newcommand{\ct}{\ensuremath{\coord{t}}}

\newcommand{\pr}{\ensuremath{\mathrm{pr}}}

\newcommand{\rsp}{\ensuremath{\mathbb{R}}}
\newcommand{\eps}{\ensuremath{\epsilon}}
\newcommand{\phy}{\ensuremath{\varphi}}
\newcommand{\psie}{\ensuremath{\bar{\psi}}}

\newcommand{\unity}{\ensuremath{\mathbb{1}}}

\let\div\undefined
\DeclareMathOperator{\div}{div}

\usepackage{appendix}
\appendixtitletocoff

\title{\vspace{-2em}Variational Integrators for\\Reduced Magnetohydrodynamics}

\author{
\large{Michael Kraus}\\
\small{(michael.kraus@ipp.mpg.de)}
\vspace{.5em}\\
\normalsize{Max-Planck-Institut f\"ur Plasmaphysik}\\
\normalsize{Boltzmannstra\ss{}e 2, 85748 Garching, Deutschland}%
\vspace{.5em}\\
\normalsize{Technische Universit\"at M\"unchen, Zentrum Mathematik}\\
\normalsize{Boltzmannstra\ss{}e 3, 85748 Garching, Deutschland}%
\vspace{1em}\\
Emanuele Tassi\\
\small{(tassi@cpt.univ-mrs.fr)}
\vspace{.5em}\\
\normalsize{Aix-Marseille Universit\'e, Universit\'e de Toulon, CNRS, CPT, UMR 7332}\\
\normalsize{163 avenue de Luminy}\\
\normalsize{13288 case 907 cedex 9 Marseille, France}%
\vspace{1em}\\
Daniela Grasso\\
\small{(daniela.grasso@infm.polito.it)}
\vspace{.5em}\\
\normalsize{ISC-CNR and Politecnico di Torino, Dipartimento Energia}\\
\normalsize{C.so Duca degli Abruzzi 24, 10129 Torino, Italy}%
\vspace{1em}\\
}

\date{\today}

\begin{document}

\maketitle

\begin{abstract}
Reduced magnetohydrodynamics is a simplified set of magnetohydrodynamics equations with applications to both fusion and astrophysical plasmas, possessing a noncanonical Hamiltonian structure and consequently a number of conserved functionals.
We propose a new discretisation strategy for these equations based on a discrete variational principle applied to a formal Lagrangian. The resulting integrator preserves important quantities like the total energy, magnetic helicity and cross helicity exactly (up to machine precision). As the integrator is free of numerical resistivity, spurious reconnection along current sheets is absent in the ideal case. If effects of electron inertia are added, reconnection of magnetic field lines is allowed, although the resulting model still possesses a noncanonical Hamiltonian structure. After reviewing the conservation laws of the model equations, the adopted variational principle with the related conservation laws are described both at the continuous and discrete level. We verify the favourable properties of the variational integrator in particular with respect to the preservation of the invariants of the models under consideration and compare with results from the literature and those of a pseudo-spectral code.
\end{abstract}

\newpage

\tableofcontents

\vspace{1em}

\section{Introduction}

The large scale dynamics of many astrophysical and laboratory plasmas can be effectively modelled according to the magnetohydrodynamics (MHD) description, where the plasma is treated as a single fluid interacting with the magnetic field. Although MHD already provides  a highly simplified description of the plasma behaviour, in some cases, even more simplified and numerically tractable models are desirable.  In particular, for this purpose, the set of reduced magnetohydrodynamics (RMHD) equations was derived for low-$\beta$ plasmas~\cite{Str76}, where $\beta$ is the ratio between the plasma pressure and the magnetic pressure. According to such model, the plasma motion is approximately incompressible and nearly two-dimensional. This model aims at describing certain features, for instance, of tokamak plasmas, where the presence of an intense component of the magnetic field implies only weak variations along the toroidal direction.

The original RMHD model can be extended in order to account for effects of finite electron inertia, which can lead to the phenomenon of magnetic reconnection~\cite{Bis00}. In particular, electron inertia can provide reconnection events faster  than those induced by plasma resistivity, and consequently provide a closer agreement with what is observed during tokamak sawtooth collapses~\cite{Wes90}.

The equations of RMHD possess a noncanonical Hamiltonian structure~\cite{Mor84}, therefore they preserve the total energy of the system. Moreover, the system features two families of Casimir functional invariants, that is integrals of the dynamical variables, which are also preserved exactly.
The extended system, including effects of finite electron inertia, is still Hamiltonian and possesses similar families of Casimir invariants as the original RMHD system.
When constructing numerical methods, it is desirable to preserve these quantities in the discrete setting in order to obtain trustable and physically correct results in computer simulations.

Such numerical methods, which preserve certain structures of the continuous system, are referred to as structure-preserving or geometric numerical methods~\cite{Christiansen:2011, HairerLubichWanner:2006, BuddPiggot:2000}. 
For many years, numerical mathematics and computer science have been focused on the development of algorithms which are based on discretisations of the equations which aim at minimising the local error of the numerical solution, that is the difference between the numerical solution and the exact solution, during one solution step. Even though this error might be very small during one step, the error will sum up when computing many steps, so that the final result might be severely different from the exact solution. 
Geometric numerical integration methods, on the other hand, rather focus on the minimisation of global errors, which often results in better long-time fidelity of the numerical solution. In recent years, the study of such methods has become a flourishing field of research in numerical analysis and scientific computing.

Some of this effort geared towards the structure-preserving discretisation of ideal magnetohydrodynamics. \citeauthor{LiuWang:2001} \cite{LiuWang:2001} proposed to couple the MAC scheme \cite{Harlow:1965} for the Navier-Stokes equation with Yee's scheme \cite{Yee:1966} for the Maxwell equations. \citeauthor{Gawlik:2011} \cite{Gawlik:2011} followed a more geometric motivated approach based on discrete Euler-Poincar\'{e} reduction (see also~\cite{Pavlov:2011}), which yields a similar scheme as that of \citeauthor{LiuWang:2001}, but with different time discretisation.
A variational integrator in Lagrangian (material) variables, based on directly discretising Newcomb's Lagrangian \cite{Newcomb:1962}, has been derived by \citeauthor{Zhou:2014} \cite{Zhou:2014}. A variational integrator in Eulerian (spatial) variables, based on combining a similar method as the one described here with discrete differential forms, has been obtained by~\citet{KrausMaj:2016}.

In the reduced case, however, the situation is more complicated and we are not aware of any attempts to derive geometric numerical methods for RMHD so far.
The difficulty of this task is on the one hand that in contrast to ideal MHD there is no natural Lagrangian formulation for RMHD readily available, which would recommend itself for discretisation by variational methods~\cite{MarsdenPatrick:1998, Pavlov:2011, Gawlik:2011}.
On the other hand, although the RMHD equations have a Hamiltonian structure, this structure is not of canonical type, so that standard methods for Hamiltonian PDEs~\cite{Bridges:2001, Bridges:2006} cannot be applied.
Therefore, we propose here to derive variational integrators for a formal Lagrangian formulation of RMHD~\cite{KrausMaj:2015, Ibragimov:2006, AthertonHomsy:1975}. 
That is, we treat the RMHD equations as part of a larger system which has a variational formulation. The corresponding Lagrangian is simply constructed by multiplying each of the RMHD equations with an auxiliary variable and summing the resulting expressions. Applying Hamilton's principle of stationary action yields an extended system consisting of the RMHD equations plus a set of adjoint equations. These adjoint equations are not solved for in the code but are crucial in the analysis of the conservation laws with Noether's theorem.

Even though we apply a simple finite difference discretisation to the Lagrangian, we obtain a well working integrator, which preserves the total energy of the system, the $L^{2}$ norm of the magnetic potential as well as magnetic helicity and cross helicity exactly (up to machine accuracy).
One of the highlights of our approach is that conservation of these properties can be proved in a relatively straight forward manner, both on the continuous and the discrete level, by the application of Noether's theorem.
The flexibility of the variational integrator framework allows for an immediate generalisation towards more elaborate, higher-order discretisation approaches \cite{Chen:2008, Leok:2005, Leok:2012, Hall:2015, Kraus:2016:Evolution, Kraus:2016:Splines}.

\pagebreak

We proceed as follows.
In Section~\ref{sec:rmhd} we review the RMHD system with and without electron inertia effects and construct a formal Lagrangian from which this system can be obtained. This Lagrangian is then used to derive important conservation laws with Noether's theorem.
We show how to rewrite and symmetrise the formal Lagrangian in order to obtain a formulation that is better suited for discretisation.
In Section~\ref{sec:vi} we review the discrete action principle and detail the derivation of the discrete equations of motion from the discretised Lagrangian.
Using the previously obtained symmetry transformations and their infinitesimal generators, the discrete conservation laws follow immediately by applying the procedure outlined in Reference~\cite{KrausMaj:2015}.
In Section~\ref{sec:numerical_experiments} we perform several numerical experiments in order to demonstrate the numerical fidelity and the excellent conservation properties of the integrator and perform a comparison with numerical simulations carried out with a pseudo-spectral code.

\section{Reduced Magnetohydrodynamics}\label{sec:rmhd}

In this work, we consider a simplified version of the reduced MHD equations by~\citet{Str76}, which provide a toy-model for low beta tokamak dynamics. Our simplification concerns the 2D reduction, which is based on the property that variations along the toroidal magnetic field are much weaker than those in the poloidal plane. Also, we consider a slab geometry, which makes our model valid only locally. With these limitations the model acquires infinite Casimir invariants (among which the cross helicity) with respect to its three-dimensional counterpart. The two-dimensional approximation therefore appears appropriate for developing and testing discretisation methods which are particularly good at respecting conservation laws. The choice of a uniform density is made for simplicity. The proposed variational method could also be applied to more sophisticated models with a Lie-Poisson structure, including, for instance, gyrofluid models \cite{Wae12}, the four-field model for tokamak dynamics \cite{Haz87}, models for electrostatic turbulence \cite{WaelMorHor} and models for magnetic reconnection evolving temperature and heat flux fluctuations \cite{Gra15}.

The reduced ideal MHD equations in two dimensions, defined on a bounded domain $\Omega \subset \rsp^{2}$, read
\begin{subequations}\label{eq:rmhd_equations}
\begin{align}
\label{eq:rmhd_equations1}
\omega_{t} & + \{ \phi , \omega \} + \{ j, \psi \} = 0 , &
- \Delta \phi &= \omega , \\
\label{eq:rmhd_equations2}
\psi_{t} & + \{ \phi, \psi \} = 0 , &
-  \Delta \psi &= j ,
\end{align}
\end{subequations}
where $\phi$ is the stream function, $\omega$ the vorticity, $\psi$ the magnetic potential, and $j$ the current density. All four fields depend on $t$, $x$ and $y$.
The canonical Poisson bracket $\{ \cdot , \cdot \}$ is defined by
\begin{align}\label{eq:vorticity_brackets}
\{ \phi , \omega \} &= \phi_{\cx} \omega_{\cy} - \phi_{\cy} \omega_{\cx} ,
\end{align}
and subscripts $t$, $x$ and $y$ denote derivatives with respect to time and the spatial dimensions, respectively.
The first equation in (\ref{eq:rmhd_equations1}) is called the vorticity equation, and the first equation in (\ref{eq:rmhd_equations2}) the collisionless Ohm's law. Note that while the second equation in (\ref{eq:rmhd_equations1}) is a Poisson equation which has to be solved for $\phi$, the second equation in (\ref{eq:rmhd_equations2}) is merely an identity for $j$.

The system (\ref{eq:rmhd_equations}) possesses a noncanonical Hamiltonian structure \cite{morrison:1998} with Hamiltonian

\begin{align}\label{eq:rmhd_hamiltonian}
H
= \dfrac{1}{2} \int \limits_{\Omega} \left( \abs{ \nabla \psi }^{2} + \abs{ \nabla \phi }^{2} \right) dx \, dy
= \dfrac{1}{2} \int \limits_{\Omega} \left( \psi j + \phi \omega \right) dx \, dy ,
\end{align}
and Lie-Poisson bracket
\begin{align}
[F,G] = - \int \left( \omega \left\{ \frac{\delta F}{\delta \omega} ,\frac{\delta G}{\delta \omega}\right\} +\psi \left( \left\{ \frac{\delta F}{\delta \omega} ,\frac{\delta G}{\delta \psi}\right\} +\left\{ \frac{\delta F}{\delta \psi}, \frac{\delta G}{\delta \omega}\right\}  \right)\right) dx \, dy ,
\end{align}
subject to the elliptic constraints on the right-hand sides of~\eqref{eq:rmhd_equations}, where $F$ and $G$ are two functionals of $\omega, \psi, \phi, j$, so that the time evolution of any such functional is given by
\begin{align}
\dot{F} = [F, H] .
\end{align}
The first term on the right-hand side of~\eqref{eq:rmhd_hamiltonian} corresponds to magnetic energy and the second term to kinetic energy.
The system possesses two infinite families of Casimir invariants,
\begin{align}\label{eq:rmhd_casimirs}
C_{1} &= \int \limits_{\Omega} f (\psi) \, dx \, dy,  &
C_{2} &= \int \limits_{\Omega} \omega \, g (\psi) \, dx \, dy,
\end{align}
where $f$ and $g$ are arbitrary functions of $\psi$.
Notable examples  of $C_{1}$ are $C_{\mathrm{MH}}$, which is proportional to the magnetic helicity, and the $L^{2}$ norm of $\psi$, $C_{L^{2}}$, given by
\begin{align}\label{eq:rmhd_magnetic_helicity}
C_{\mathrm{MH}} = \int \limits_{\Omega} \psi \, dx \, dy ,
\end{align}
and
\begin{align}\label{eq:rmhd_l2norm}
C_{L^{2}} = \int \limits_{\Omega}\psi^{2} \, dx \, dy ,
\end{align}
respectively.
Note that for a magnetic field of the form
\begin{align*}
B = B_{0} \hat{z} + \nabla \psi \times \hat{z} ,
\end{align*}
with vector potential
\begin{align*}
A = \psi \hat{z} + \tfrac{1}{2} B_{0} x \hat{y} - \tfrac{1}{2} B_{0} y \hat{x} ,
\end{align*}
magnetic helicity is given by
\begin{align*}
\int \limits_{\Omega} A \cdot B \, dx \, dy 
= \int \limits_{\Omega} \left[ - \tfrac{1}{2} B_{0} y \partial_{y} \psi - \tfrac{1}{2} B_{0} x \partial_{x} \psi + B_{0} \psi \right] \, dx \, dy
= 2 B_{0} \int \limits_{\Omega} \psi \, dx \, dy ,
\end{align*}
and therefore proportional to $C_{\mathrm{MH}}$.
An important special case of $C_{2}$ is cross helicity, $C_{\mathrm{CH}}$, where $g (\psi) = \psi$,
\begin{align}\label{eq:rmhd_cross_helicity}
C_{\mathrm{CH}}
&= \int \limits_{\Omega} \nabla \phi \cdot \nabla \psi \, dx \, dy
 = \int \limits_{\Omega} \omega \psi \, dx \, dy .
\end{align}
In our discretisation, we aim at preserving the Hamiltonian (\ref{eq:rmhd_hamiltonian}) and the aforementioned Casimir invariants (\ref{eq:rmhd_magnetic_helicity})-(\ref{eq:rmhd_cross_helicity}).

\subsection{Formal Lagrangian}

The starting point for the derivation of variational integrators is a continuous action principle, more specifically a Lagrangian formulation of the RMHD system in Eulerian coordinates.
Unfortunately, such an Eulerian Lagrangian is not readily available, so that we have to resort to a formal Lagrangian formulation~\cite{KrausMaj:2015, Ibragimov:2006, AthertonHomsy:1975}.
The idea is to treat the RMHD system as part of a larger system, which can be derived from a Eulerian Lagrangian using Hamilton's principle of stationary action.
This Lagrangian is obtained by multiplying each equation of~(\ref{eq:rmhd_equations}) with an auxiliary variable, $\xi$, $\mu$, $\chi$ and $\zeta$, respectively.
The formal Lagrangian is given as the sum of the resulting expressions,
\begin{align}\label{eq:rmhd_formal_lagrangian}
\mcal{L} (\phy, \phy_{t}, \phy_{x}, \phy_{y})
\nonumber
&= \xi    \, \big( \omega_{t} + \{ \phi , \omega \} + \{ j, \psi \} \big)
 + \mu    \, \big( \omega + \Delta \phi \big) \\
&+ \chi   \, \big( \psi_{t} + \{ \phi, \psi \} \big)
 + \zeta  \, \big( j + \Delta \psi \big) .
\end{align}
For concise notation, we write $\phy$ to denote all variables,
\begin{align}
\phy = (\omega, \psi, \phi, j, \xi, \chi, \mu, \zeta)^{T} ,
\end{align}
and $\phy_{t}$, $\phy_{x}$ and $\phy_{y}$ to denote the corresponding derivatives with respect to $t$, $x$ and $y$.
Requiring stationarity of the action functional (Hamilton's principle), 
\begin{align}
\delta \mcal{A} [\phy] = \delta \int \limits_{\Omega} \mcal{L} (\phy, \phy_{t}, \phy_{x}, \phy_{y}) \, dt \, dx \, dy = 0 ,
\end{align}
for variations $\delta \phy$ of the variables $\phy$, which vanish at the boundaries but are otherwise arbitrary, we obtain the Euler-Lagrange field equations,
\begin{align}\label{eq:variational_continuous_euler_lagrange}
\dfrac{\partial \mcal{L}}{\partial \phy^{a}} \big( \phy, \phy_{t}, \phy_{x}, \phy_{y} \big) - \dfrac{\partial}{\partial \bx^{\mu}} \left( \dfrac{\partial \mcal{L}}{\partial \phy^{a}_{\mu}} \big( \phy, \phy_{t}, \phy_{x}, \phy_{y} \big) \right) &= 0 &
& \text{for all $a$} .
\end{align}
Here and in the following, $(\bx^{\mu})$ denotes all coordinates, both spatial and temporal, and summation over $\mu$ is implied following the Einstein summation convention.
This yields the RMHD equations~(\ref{eq:rmhd_equations}) and in addition the so called adjoint equations, which determine the evolution of the auxiliary variables,
\begin{subequations}\label{eq:rmhd_adjoint_equations}
\begin{align}
\xi_{t} &+ \{ \phi , \xi \} = \mu , &
\{ \xi , \omega \} + \{ \chi , \psi \} &= \Delta \mu , \\
\chi_{t} &+ \{ \phi , \chi \} + \{ j , \xi \} = \Delta \zeta , &
\{ \xi , \psi \} &= \zeta .
\end{align}
\end{subequations}
For the analysis of the conservation laws, it will be necessary to construct a solution of the adjoint variables in terms of the physical variables.
The extended system can be solved in terms of only the physical variables by using the embedding
\begin{align}\label{eq:rmhd_restriction}
\Phi : (\omega, \psi, \phi, j) \mapsto (\omega, \psi, \phi, j, \psi, \omega, 0, 0) ,
\end{align}
in order to restrict solutions of the extended system to solutions of the physical system (see Reference~\cite[Section~3.3]{KrausMaj:2015} for details).
Setting $\chi = \omega$ and $\xi = \psi$, we find that $\zeta = \{ \psi , \psi \} = 0$ as well as $\Delta \mu = \{ \psi , \omega \} + \{ \omega , \psi \} = 0$. As a particular solution of this equation, we can choose $\mu = 0$, which justifies identifying $\xi$ with $\psi$. Similarly, identifying $\chi$ with $\omega$ is only justified because as a result $\zeta$ vanishes identically.

\subsection{Conservation Laws}
\label{sec:rmhd_conservation_laws}

In the analysis of conservation laws, we use the Noether theorem~\cite{Noether:1918, KosmannSchwarzbach:2010}, which connects Lie point symmetries of the Lagrangian and conservation laws of the associated Euler-Lagrange equations.
We use the same formulation of the Noether theorem as in Reference~\citep{KrausMaj:2015} (see also Reference~\citep{MarsdenPatrick:1998}), which is summarised in the following.
As in Reference~\cite{KrausMaj:2015}, we restrict our attention to conservation laws which are generated by vertical transformations, that is transformations which only affect the field variables but leave the base space invariant\footnote{Interestingly, in the framework of formal Lagrangians, conservation of e.g. energy and momentum of the physical system does usually not follow from invariance of the Lagrangian with respect to temporal or spatial translations, but from invariance with respect to vertical transformations.
Therefore, these symmetries are in general not lost when introducing a discrete temporal and spatial grid as would be the case with natural Lagrangians.
On the other hand, temporal and spatial symmetries of formal Lagrangians are usually related to trivial conservation laws without physical relevance.}.

We directly prescribe the generating vector field $V$ of the transformation map $\sigma(\phy, \eps)$ and not the transformation itself.
The vector field is given in terms of its components $\eta^{a}$,
\begin{align}\label{eq:noether_continuous_vector_field}
V (\phy) &= \eta^{a} (\phy) \, \dfrac{\partial}{\partial \phy^{a}} &
& \text{with} &
\eta^{a} (\phy) &= \dfrac{d}{d\eps} \sigma^{a} (\phy, \eps) \bigg\vert_{\eps=0} ,
\end{align}
where $\sigma^{a}$ is the $a$-th component of the transformation map $\sigma$. For simplicity we assume that $V$ only depends on the fields $\phy$ but not explicitly on the base coordinates $(t,x,y)$.
As the Lagrangian is a function not only of the fields but also of their derivatives, we have to compute the action of the generating vector field on the derivatives induced by the transformation in the fields.
This is accounted for by the prolongation of $V$, which is given by
\begin{align}\label{eq:noether_continuous_vector_field_prolongation}
\pr V
= \eta^{a} \, \dfrac{\partial}{\partial \phy^{a}} + \phy^{b}_{\mu} \dfrac{\partial \eta^{a}}{\partial \phy^{b}} \dfrac{\partial}{\partial \phy^{a}_{\mu}}
= \eta^{a} \, \dfrac{\partial}{\partial \phy^{a}} + \eta^{a}_{\mu} \, \dfrac{\partial}{\partial \phy^{a}_{\mu}} .
\end{align}
Denoting the transformed fields by $\phy^{\eps} = \sigma(\phy, \eps)$, the condition for a transformation $\sigma (\phy, \eps)$ being a symmetry transformation of the Lagrangian $\mcal{L}$ reads
\begin{align}\label{eq:noether_continuous_symmetry_1}
\mcal{L} \big( \phy^{\eps}, \phy^{\eps}_{t}, \phy^{\eps}_{x}, \phy^{\eps}_{y} \big) &= \mcal{L} \big( \phy, \phy_{t}, \phy_{x}, \phy_{y} \big) .
\end{align}
Taking the $\eps$ derivative of (\ref{eq:noether_continuous_symmetry_1}), we obtain the infinitesimal invariance condition,
\begin{align}\label{eq:noether_continuous_symmetry_2}
\dfrac{d}{d\eps} \mcal{L} \big( \phy^{\eps}, \phy^{\eps}_{t}, \phy^{\eps}_{x}, \phy^{\eps}_{y} \big) \bigg\vert_{\eps=0}
= \pr V (\mcal{L})
= 0 ,
\end{align}
which is equivalent to (\ref{eq:noether_continuous_symmetry_1}).
The corresponding conservation law is given by 
\begin{align}\label{eq:noether_continuous_noether_theorem}
\div \mcal{J} (\phy, \phy_{t}, \phy_{x}, \phy_{y}) = 0 ,
\end{align}
where $\div$ denotes the spacetime divergence and $\mcal{J}$ denotes the so called Noether current~\cite{GotayMarsden:1998} with components
\begin{align}\label{eq:noether_continuous_current_definition}
\mcal{J}^{\mu} \big( \phy, \phy_{t}, \phy_{x}, \phy_{y} \big) = \dfrac{\partial \mcal{L}}{\partial \phy^{a}_{\mu}} \big( \phy, \phy_{t}, \phy_{x}, \phy_{y} \big) \cdot \eta^{a} (\phy) .
\end{align}
The fact that $\mcal{J}$ is divergence-free expresses the conservation law satisfied by solutions $\phy$ of the Euler-Lagrange field equations (\ref{eq:variational_continuous_euler_lagrange}).
Commonly, \eqref{eq:noether_continuous_noether_theorem} is integrated over the spatial domain $\Omega$, so that under the assumption of appropriate boundary conditions only the temporal component of the Noether current is retained,
\begin{align}\label{eq:noether_continuous_noether_charge}
  \dfrac{d}{dt} \int \limits_{\Omega} \div \mcal{J} (\phy, \phy_{t}, \phy_{x}, \phy_{y}) \, dx \, dy
= \dfrac{d}{dt} \int \limits_{\Omega} \dfrac{\partial \mcal{L}}{\partial \phy^{a}_{t}} \big( \phy, \phy_{t}, \phy_{x}, \phy_{y} \big) \cdot \eta^{a} (\phy) \, dx \, dy 
= 0 .
\end{align}
This relation states the conservation of the so called Noether charge\footnote{We remark that Noether charge and current do not correspond in general to electrical charge and current.}.

The restriction to vertical vector fields appears logical as in the discrete case we can only consider such vertical transformations due to the spatio-temporal grid being fixed.
However, as it turns out, the generating vector fields of the interesting conservation laws corresponding to (\ref{eq:rmhd_hamiltonian}) and (\ref{eq:rmhd_magnetic_helicity})-(\ref{eq:rmhd_cross_helicity}) feature horizontal components.
To circumvent this problem, we consider a simplified version of (\ref{eq:rmhd_formal_lagrangian}), where we assume that $\phi$ and $j$ are prescribed and constant in time. This amounts to a partial linearisation of the system. The reduced solution vector is $\bar{\phy} = (\omega, \psi, \xi, \chi)^{T}$ and the corresponding Lagrangian is given by
\begin{align}\label{eq:rmhd_formal_lagrangian_simplified}
\bar{\mcal{L}} (\bar{\phy}, \bar{\phy}_{t}, \bar{\phy}_{x}, \bar{\phy}_{y})
&= \xi    \, \big( \omega_{t} + \{ \phi , \omega \} + \{ j, \psi \} \big)
 + \chi   \, \big( \psi_{t} + \{ \phi, \psi \} \big) .
\end{align}
Using this simplification, the analysis of the conservation laws carries over to the discrete case in a straight forward way. As we will see, the conservation laws thus obtained will also be respected by the fully nonlinear continuous and discrete systems (see Reference~\cite[Section 4.3.2]{KrausMaj:2015} for a similar analysis for the vorticity equation).
The adjoint equations of~\eqref{eq:rmhd_formal_lagrangian_simplified} are
\begin{align}
\xi_{t} &+ \{ \phi , \xi \} = 0 , &
\chi_{t} &+ \{ \phi , \chi \} + \{ j , \xi \} = 0 ,
\end{align}
so that the system is immediately seen to be self-adjoint, using the embedding
\begin{align}\label{eq:rmhd_restriction_simplified}
\Phi : (\omega, \psi) \mapsto (\omega, \psi, \psi, \omega) .
\end{align}

For the invariants of interest, c.f. Equations~(\ref{eq:rmhd_hamiltonian}) and~(\ref{eq:rmhd_magnetic_helicity})-(\ref{eq:rmhd_cross_helicity}), the generating vector fields can easily be constructed by observation. That is knowledge of the exact form of the Noether charge~\eqref{eq:noether_continuous_noether_charge} together with the restriction map~\eqref{eq:rmhd_restriction_simplified}, immediately dictates the components $\eta^{a}$ of the generating vector field corresponding to a transformation of one of the physical variables $(\omega, \psi)$. The components of the adjoint variables can then be determined by the procedure outlined in References~\cite[Section~3.3]{KrausMaj:2015} and~\cite[Theorem~3.3]{Ibragimov:2007}.

The generating vector field for magnetic helicity reads
\begin{align}
\eta^{\omega} = 1 ,
\end{align}
with prolongation
\begin{align}
\pr V = \dfrac{\partial}{\partial \omega} , 
\end{align}
so that
\begin{align}
\pr V (\bar{\mcal{L}}) = 0 .
\end{align}
The corresponding Noether charge~\eqref{eq:noether_continuous_noether_charge} becomes
\begin{align}
\dfrac{d}{dt} \int \limits_{\Omega} \xi \, dx \, dy = 0 .
\end{align}
Restricted to solutions of the form $\bar{\phy} = (\omega, \psi, \psi, \omega)^{T}$, c.f. Equation~\eqref{eq:rmhd_restriction_simplified}, this corresponds to conservation of magnetic helicity~\eqref{eq:rmhd_magnetic_helicity},
\begin{align}
\dfrac{d}{dt} \int \limits_{\Omega} \psi \, dx \, dy = 0 .
\end{align}
The generating vector field for the $L^{2}$ norm of $\psi$ is
\begin{align}
\eta^{\omega} &= \psi , &
\eta^{\chi} &= - \xi ,
\end{align}
and its prolongation
\begin{align}
\pr V
&= \psi \, \dfrac{\partial}{\partial \omega}
 + \psi_{t} \, \dfrac{\partial}{\partial \omega_{t}}
 + \psi_{x} \, \dfrac{\partial}{\partial \omega_{x}}
 + \psi_{y} \, \dfrac{\partial}{\partial \omega_{y}}
 - \xi \, \dfrac{\partial}{\partial \chi}
 - \xi_{t} \, \dfrac{\partial}{\partial \chi_{t}}
 - \xi_{x} \, \dfrac{\partial}{\partial \chi_{x}}
 - \xi_{y} \, \dfrac{\partial}{\partial \chi_{y}} .
\end{align} 
The symmetry condition is verified to be satisfied,
\begin{align}
\pr V (\bar{\mcal{L}})
&= \xi \, \big( \psi_{t} + \{ \phi , \psi \} \big)
 - \xi \, \big( \psi_{t} + \{ \phi , \psi \} \big)
 = 0 ,
\end{align}
and conservation of the corresponding Noether charge~\eqref{eq:noether_continuous_noether_charge} becomes
\begin{align}
\dfrac{d}{dt} \int \limits_{\Omega} \xi \psi \, dx \, dy = 0 ,
\end{align}
which after restriction of the solution by~\eqref{eq:rmhd_restriction_simplified} becomes conservation of the $L^{2}$ norm of $\psi$ \eqref{eq:rmhd_l2norm}.
For cross helicity, we consider
\begin{align}
\eta^{\omega} &= \omega , &
\eta^{\xi}    &= - \xi ,
\end{align}
so that the prolongation of the generating vector field becomes
\begin{align}
\pr V
&= \omega     \, \dfrac{\partial}{\partial \omega}
 + \omega_{t} \, \dfrac{\partial}{\partial \omega_{t}}
 + \omega_{x} \, \dfrac{\partial}{\partial \omega_{x}}
 + \omega_{y} \, \dfrac{\partial}{\partial \omega_{y}}
 - \xi        \, \dfrac{\partial}{\partial \xi}
 - \xi_{t} \, \dfrac{\partial}{\partial \xi_{t}}
 - \xi_{x} \, \dfrac{\partial}{\partial \xi_{x}}
 - \xi_{y} \, \dfrac{\partial}{\partial \xi_{y}} .
\end{align}
The symmetry condition reads
\begin{align}
\pr V (\bar{\mcal{L}})
&= \xi    \, \big( \omega_{t} + \{ \phi , \omega \} + \{ j, \psi \} \big)
 - \xi    \, \big( \omega_{t} + \{ \phi , \omega \} + \{ j, \psi \} \big)
 = 0 ,
\end{align}
and conservation of the Noether charge~\eqref{eq:noether_continuous_noether_charge} becomes
\begin{align}
\dfrac{d}{dt} \int \limits_{\Omega} \xi \omega \, dx \, dy = 0 ,
\end{align}
which amounts to conservation of cross helicity~\eqref{eq:rmhd_cross_helicity} when restricting the solution with~\eqref{eq:rmhd_restriction_simplified}.
Energy conservation is obtained from the generating vector field
\begin{align}
\eta^{\omega} &= \tfrac{1}{2} j , &
\eta^{\psi} &= \tfrac{1}{2} \phi .
\end{align}
As $\phi$ and $j$ are treated as constant and not dynamical vector fields, the prolongation is trivial, that is
\begin{align}
\pr V &= \tfrac{1}{2} j \, \dfrac{\partial}{\partial \omega} + \tfrac{1}{2} \phi \, \dfrac{\partial}{\partial \psi} ,
\end{align}
and the symmetry condition is satisfied,
\begin{align}
\pr V (\bar{\mcal{L}})
&= 0 .
\end{align}
The corresponding conservation law reads
\begin{align}
\dfrac{d}{dt} \int \limits_{\Omega} \tfrac{1}{2} \big( \xi j + \chi \phi \big) \, dx \, dy = 0 ,
\end{align}
which upon restriction of the solution with~\eqref{eq:rmhd_restriction_simplified} states that the total energy of the system \eqref{eq:rmhd_hamiltonian} is conserved.

\subsection{Electron Inertia}\label{sec:rmhd_electron_inertia}

The RMHD model can be extended to account for collisionless reconnection by adding effects of electron inertia in the following way~\cite{Schep:1994},
\begin{subequations}\label{eq:rmhd_equations_ei}
\begin{align}
\label{eq:rmhd_equations_ei1}
\omega_{t} & + \{ \phi , \omega \} + \{ j, \psi \} = 0 , &
- \Delta \phi &= \omega , \\
\label{eq:rmhd_equations_ei2}
\psie_{t} & + \{ \phi, \psie \} = 0 , &
-  \Delta \psi &= j .
\end{align}
\end{subequations}
Here, $\psie = \psi + d_{e}^{2} j$ is the electron canonical momentum and $d_{e}$ denotes the electron skin depth, assumed to be constant.
This model is also Hamiltonian with
\begin{align}\label{eq:rmhd_ei_hamiltonian}
H
 = \dfrac{1}{2} \int \limits_{\Omega} \left( \abs{ \nabla \psi }^{2} + d_{e}^{2} ( \Delta \psi )^{2} + \abs{ \nabla \phi }^{2} \right) dx \, dy
 = \dfrac{1}{2} \int \limits_{\Omega} \left( \psie j + \phi \omega \right) dx \, dy ,
\end{align}
Lie-Poisson bracket
\begin{align}
[F,G] = - \int \left( \omega \left\{ \frac{\delta F}{\delta \omega}, \frac{\delta G}{\delta \omega}\right\} +\psie \left( \left\{ \frac{\delta F}{\delta \omega} ,\frac{\delta G}{\delta \psie}\right\} +\left\{ \frac{\delta F}{\delta \psie} ,\frac{\delta G}{\delta \omega}\right\}  \right)\right) dx \, dy ,
\end{align}
subject to the elliptic constraints on the right-hand sides of~\eqref{eq:rmhd_equations_ei}, and, just as the original RMHD model, possesses two families of Casimir invariants,
\begin{align}\label{eq:rmhd_ei_casimirs}
C_{1} &= \int \limits_{\Omega} f (\psie) \, dx \, dy,  &
C_{2} &= \int \limits_{\Omega} \omega \, g (\psie) \, dx \, dy .
\end{align}
Special cases of $C_{1}$ are the generalised magnetic helicity and the $L^{2}$ norm of $\psie$, given by
\begin{align}\label{eq:rmhd_ei_magnetic_helicity}
C_{\mathrm{MH}} = \int \limits_{\Omega} \psie \, dx \, dy ,
\end{align}
and
\begin{align}\label{eq:rmhd_ei_l2norm}
C_{L^{2}} = \int \limits_{\Omega} \psie^{2} \, dx \, dy ,
\end{align}
respectively, and a special case of $C_{2}$ is the generalised cross helicity, given by
\begin{align}\label{eq:rmhd_ei_cross_helicity}
C_{\mathrm{CH}} &= \int \limits_{\Omega} \nabla \phi \cdot \nabla \psie \, dx \, dy = \int \limits_{\Omega} \omega \psie \, dx \, dy .
\end{align}
For the model with electron inertia, we aim at preserving the Hamiltonian~(\ref{eq:rmhd_ei_hamiltonian}), the generalised magnetic helicity~(\ref{eq:rmhd_ei_magnetic_helicity}), the $L^{2}$ norm of $\psie$~(\ref{eq:rmhd_ei_l2norm}), and generalised cross helicity~(\ref{eq:rmhd_ei_cross_helicity}).
The verification of these conservation laws via Noether's theorem follows analogous to Section~\ref{sec:rmhd_conservation_laws}.

\subsection{Symmetrisation}\label{sec:rmhd_symmetrisation}

In order to avoid second order derivatives and therefore to simplify the discretisation of the formal Lagrangian~(\ref{eq:rmhd_formal_lagrangian}), we apply symmetrisations (corresponding to integration by parts in the action functional) in the terms that produce the Laplacians,
\begin{align}\label{eq:rmhd_formal_lagrangian_1}
\mcal{L}' (\phy, \phy_{t}, \phy_{x}, \phy_{y})
\nonumber
&= \xi \omega_{t}
 + \xi  \, \{ \phi , \omega \}
 + \xi  \, \{ j, \psi \} 
 + \chi \psi_{t}
 + \chi \, \{ \phi, \psi \} \\
&+ \mu  \omega - \nabla \mu  \cdot \nabla \phi 
 + \zeta  j      - \nabla \zeta \cdot \nabla \psi .
\end{align}
This provides an equivalent Lagrangian, which yields the same Euler-Lagrange equations, but depends only on first order derivatives of the fields.

For the derivation of one-step methods of arbitrary order, it will be important to symmetrise the time derivatives in a similar fashion as the Laplacian,
\begin{align}\label{eq:rmhd_formal_lagrangian_2}
\mcal{L}'' (\phy, \phy_{t}, \phy_{x}, \phy_{y})
\nonumber
&= \tfrac{1}{2} \phy_{t}^{T} \Lambda \, \phy
 + \xi  \, \{ \phi , \omega \}
 + \xi  \, \{ j    , \psi   \} 
 + \chi \, \{ \phi , \psi   \} \\
&+ \mu  \omega - \nabla \mu   \cdot \nabla \phi 
 + \zeta  j    - \nabla \zeta \cdot \nabla \psi ,
\end{align}
where $\Lambda$ is an anti-symmetric matrix, given by
\begin{align}
\Lambda = \begin{pmatrix}
\hphantom{-} 0 & \hphantom{-} 0 & \hphantom{-} 0 & \hphantom{-} 0 &           +  1 & \hphantom{-} 0 & \hphantom{-} 0 & \hphantom{-} 0 \\
\hphantom{-} 0 & \hphantom{-} 0 & \hphantom{-} 0 & \hphantom{-} 0 & \hphantom{-} 0 &           +  1 & \hphantom{-} 0 & \hphantom{-} 0 \\
\hphantom{-} 0 & \hphantom{-} 0 & \hphantom{-} 0 & \hphantom{-} 0 & \hphantom{-} 0 & \hphantom{-} 0 & \hphantom{-} 0 & \hphantom{-} 0 \\
\hphantom{-} 0 & \hphantom{-} 0 & \hphantom{-} 0 & \hphantom{-} 0 & \hphantom{-} 0 & \hphantom{-} 0 & \hphantom{-} 0 & \hphantom{-} 0 \\
          -  1 & \hphantom{-} 0 & \hphantom{-} 0 & \hphantom{-} 0 & \hphantom{-} 0 & \hphantom{-} 0 & \hphantom{-} 0 & \hphantom{-} 0 \\
\hphantom{-} 0 &           -  1 & \hphantom{-} 0 & \hphantom{-} 0 & \hphantom{-} 0 & \hphantom{-} 0 & \hphantom{-} 0 & \hphantom{-} 0 \\
\hphantom{-} 0 & \hphantom{-} 0 & \hphantom{-} 0 & \hphantom{-} 0 & \hphantom{-} 0 & \hphantom{-} 0 & \hphantom{-} 0 & \hphantom{-} 0 \\
\hphantom{-} 0 & \hphantom{-} 0 & \hphantom{-} 0 & \hphantom{-} 0 & \hphantom{-} 0 & \hphantom{-} 0 & \hphantom{-} 0 & \hphantom{-} 0 \\
\end{pmatrix} .
\end{align}

Further, care has to be taken when discretising the Poisson bracket~\eqref{eq:vorticity_brackets}.
In order to retain the antisymmetry property of the continuous bracket at the discrete level, another symmetrisation has to be introduced in the Lagrangian (c.f.~\citet{SalmonTalley:1989}).
Using integration by parts while assuming appropriate boundary conditions, it is seen that the cyclic permutations of the functions in the integrand are all identical, e.g.,
\begin{align}
  \int \xi    \, \{ \phi   , \omega \} \, d\cx \, d\cy
= \int \phi   \, \{ \omega , \xi    \} \, d\cx \, d\cy
= \int \omega \, \{ \xi    , \phi   \} \, d\cx \, d\cy .
\end{align}
Instead of selecting one of those equivalent forms, a convex combination can be considered, namely,
\begin{align}
\int \xi \, \{ \phi , \omega \} \, d\cx \, d\cy
&= \int \Big[ \alpha \, \xi    \, \{ \phi   , \omega \}
            + \beta  \, \phi   \, \{ \omega , \xi    \}
            + \gamma \, \omega \, \{ \xi    , \phi   \} \Big] \, d\cx \, d\cy ,
\end{align}
with $\alpha + \beta + \gamma = 1$.
As the symmetric case, $\alpha = \beta = \gamma = 1/3$, is the one that retains the properties of the bracket at the discrete level, we use the modified Lagrangian
\begin{align}\label{eq:rmhd_formal_lagrangian_3}
\mcal{L}''' (\phy, \phy_{t}, \phy_{x}, \phy_{y})
 = \tfrac{1}{2} \phy_{t}^{T} \Lambda \, \phy
\nonumber
&+ \dfrac{1}{3} \Big( \xi \, \{ \phi , \omega \} + \phi \, \{ \omega , \xi \} + \omega \, \{ \xi , \phi \} \Big) \\
\nonumber
&+ \dfrac{1}{3} \Big( \xi \, \{ j, \psi \} + j \, \{ \psi, \xi \} + \psi \, \{ \xi, j \} \Big) \\
\nonumber
&+ \dfrac{1}{3} \Big( \chi \, \{ \phi, \psi \} + \phi \, \{ \psi, \chi \} + \psi \, \{ \chi, \phi \} \Big) \\
&+ \mu    \omega - \nabla \mu   \cdot \nabla \phi
 + \zeta  j      - \nabla \zeta \cdot \nabla \psi 
\end{align}
as basis for the variational discretisation procedure described in Section~\ref{sec:vi}.
The conservation laws obtained in Section~\ref{sec:rmhd_conservation_laws} are still valid for the symmetrised Lagrangian. The only difference is that pure symmetries of the original Lagrangian will be divergence symmetries of the symmetrised Lagrangian (see~References~\cite{Olver:1993, Ibragimov:2007} or~\cite[Section 2.3]{KrausMaj:2015} for more details).

\section{Variational Integrators}
\label{sec:vi}

In the application of most numerical methods to physical systems, the starting point for discretisation are the dynamical equations of the state variables.
In the variational integrator methodology, instead, the starting point is the Lagrangian with the action integral.
It is not the dynamical equations that are discretised but the building blocks of the underlying field theory. That is, after discretising the Lagrangian and the action integral, a discrete variational principle is applied which leads to the discrete dynamical equations.

In general, the discretisation and the application of the action principle do not commute, that is the equations obtained from the discrete action principle are not the same as the ones obtained from directly discretising the dynamical equations.
The advantage of using the discrete action principle is that the resulting equations will automatically preserve momenta related to symmetries of the discrete Lagrangian as well as a discrete multisymplectic form.
Both properties are important for obtaining physically sound results as well as for good long-time fidelity of numerical simulations.

The general framework of variational integrators has been developed by \citet{MarsdenPatrick:1998}, while \citet{KrausMaj:2015} describe the particular application to formal Lagrangians like the one given in equation~(\ref{eq:rmhd_formal_lagrangian_3}).
For more details see also References~\cite{Veselov:1988, Veselov:1991, MoserVeselov:1991, WendlandtMarsden:1997, MarsdenWendlandt:1997, MarsdenWest:2001, Kraus:2016:Evolution, KrausMaj:2016, KrausMajSonnendruecker:2016}.

\subsection{Discrete Action Principle}

In this section, we outline the discretisation of the action functional $\mcal{A}$ on a simple cartesian grid with one temporal and two spatial dimensions, and apply a discrete version of Hamilton's principle of stationary action~\cite{MarsdenPatrick:1998, KrausMaj:2015}.
For simplicity, we consider a Veselov-type finite difference discretisation of the Lagrangian \cite{Veselov:1988, Veselov:1991, MoserVeselov:1991}, but the extension towards other, more elaborate, e.g. Galerkin-type approaches \cite{Leok:2005, Leok:2012, Hall:2015, Kraus:2016:Evolution}, finite elements \cite{Chen:2008} or splines \cite{Kraus:2016:Splines} is straight forward.

\begin{figure}[t]
	\centering
	\subfloat[Grid Points]{\label{fig:vi_gridbox_indices}
		\includegraphics[height=4cm]{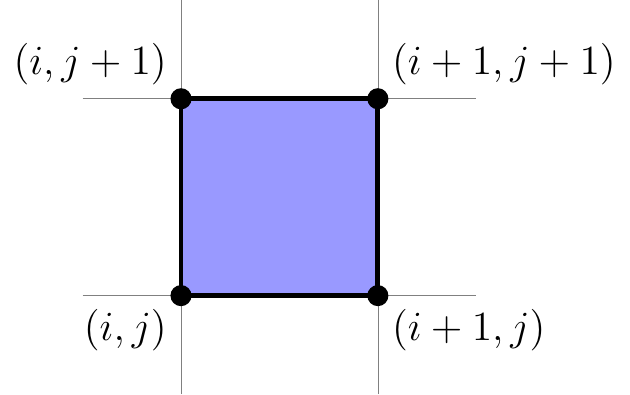}
	}
	\hspace{.5cm}
	\subfloat[Field Components]{\label{fig:vi_gridbox_fields}
		\includegraphics[height=4cm]{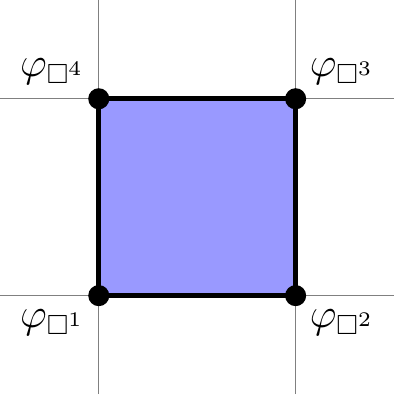}
	}
	\hspace{.5cm}
	\subfloat{
		\includegraphics[height=2cm]{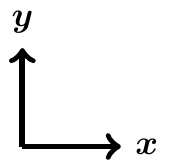}
	}
	
	\caption{Vertices of a primal grid cell in a two-dimensional rectangular grid and field components at those vertices.}
	\label{fig:vi_gridbox}
\end{figure}

We introduce a space-time split, that is first we consider the semi-discretisation in space and thereafter perform the discretisation in time.
The semi-discrete fields are denoted by
\begin{align}
\phy_{h} (t) = \left\{ \phy_{i,j} (t) \; \big\vert \; i = 1, ..., n_{x} , \; j = 1, ..., n_{y} \right\} ,
\end{align}
while the fully discrete fields are denoted by
\begin{align}
\phy_{d} = \left\{ \phy_{i,j}^{n} \; \big\vert \; i = 1, ..., n_{x} , \; j = 1, ..., n_{y} , \; n = 1, ..., n_{t} \right\} .
\end{align}
Here, $n_{x}$, $n_{y}$ and $n_{t}$ denote the number of grid points in space and time, respectively.
The semi-discrete Lagrangian $L_{h}$ approximates the integral of the continuous Lagrangian over one cell of the spatial grid, that is
\begin{multline}\label{eq:vi_semi_discrete_lagrangian}
\int \limits_{x_{i}}^{x_{i+1}} dx
\int \limits_{y_{j}}^{y_{j+1}} dy \, 
\mcal{L} \big( \phy, \phy_{t}, \phy_{x}, \phy_{y} \big)
\approx 
L_{h} \big( \phy_{i,j} (t), \phy_{i+1,j} (t), \phy_{i+1,j+1} (t), \phy_{i,j+1} (t), \\
\dot{\phy}_{i,j} (t), \dot{\phy}_{i+1,j} (t), \dot{\phy}_{i+1,j+1} (t), \dot{\phy}_{i,j+1} (t) \big) ,
\end{multline}
where the dot indicates the time derivative.
To make the manipulations more tractable, the semi-discrete Lagrangian is rewritten in a slightly more abstract way, namely in terms of cells rather than grid points. Let us consider a cell $\square$ determined by its vertices,
\begin{align}
\square = \big( (i, \, j  ), (i+1, \, j ), (i+1, \, j+1 ), (i, \, j+1 ) \big) .
\end{align}
The vertices $\square^{l}$ of a cell $\square$ with $1 \leq l \leq 4$ are counted counter-clockwise in the $x$-$y$-plane (c.f. Figure~\ref{fig:vi_gridbox}), namely,
\begin{align}\label{eq:vi_discrete_vertices}
\nonumber
\square^{1} &= (i,   j  ) , &
\square^{2} &= (i+1, j  ) , &
\square^{3} &= (i+1, j+1) , &
\square^{4} &= (i,   j+1) .
\end{align}
The field values on the vertices of that cell are denoted by
\begin{align}
( \phy_{\square^{1}}, \phy_{\square^{2}}, \phy_{\square^{3}}, \phy_{\square^{4}} ) = ( \phy_{i, j}, \phy_{i+1, j}, \phy_{i+1, j+1}, \phy_{i, j+1} ) .
\end{align}
The solution vector and the time derivatives on each grid cell are compactly denoted as
\begin{align}
\phy_{\square} (t) &= \big( \phy_{\square^{1}} (t), \phy_{\square^{2}} (t), \phy_{\square^{3}} (t), \phy_{\square^{4}} (t) \big)^{T} , \\
\dot{\phy}_{\square} (t) &= \big( \dot{\phy}_{\square^{1}} (t), \dot{\phy}_{\square^{2}} (t), \dot{\phy}_{\square^{3}} (t), \dot{\phy}_{\square^{4}} (t) \big)^{T} ,
\end{align}
so that the semi-discrete Lagrangian \eqref{eq:vi_semi_discrete_lagrangian} can be written as
\begin{align}
L_{h} (\phy_{\square}, \dot{\phy}_{\square} ) &\equiv
L_{h} (\phy_{\square^{1}},
       \phy_{\square^{2}}, 
       \phy_{\square^{3}}, 
       \phy_{\square^{4}},
       \dot{\phy}_{\square^{1}},
       \dot{\phy}_{\square^{2}}, 
       \dot{\phy}_{\square^{3}}, 
       \dot{\phy}_{\square^{4}}) .
\end{align}

In order to approximate the formal Lagrangian $\mcal{L}''' (\phy, \phy_{\ct}, \phy_{\cx}, \phy_{\cy})$ from~(\ref{eq:rmhd_formal_lagrangian_3}), we adopt a similar strategy as in Reference~\cite{KrausMaj:2015}, using simple finite differences to approximate the derivatives and the trapezoidal rule for the two spatial integrals (see also Reference~\cite{KrausMajSonnendruecker:2016}). On each cell $\square$, the Lagrangian $\mcal{L}'''$ is approximated by
\begingroup
\allowdisplaybreaks
\begin{align}\label{eq:vorticity_discrete_lagrangian}
\nonumber
L_{h} (\phy_{\square}, \dot{\phy}_{\square} ) 
\nonumber
&= \dfrac{h_{x} h_{y}}{4} \mcal{L}''' \Big(
  \phy_{\square^{1}} , \dot{\phy}_{\square^{1}} ,
  \dfrac{\phy_{\square^{2}} - \phy_{\square^{1}}}{h_{x}} , 
  \dfrac{\phy_{\square^{4}} - \phy_{\square^{1}}}{h_{y}}
\Big) \\
\nonumber
&+ \dfrac{h_{x} h_{y}}{4} \mcal{L}''' \Big(
  \phy_{\square^{2}} , \dot{\phy}_{\square^{2}} ,
  \dfrac{\phy_{\square^{2}} - \phy_{\square^{1}}}{h_{x}} ,
  \dfrac{\phy_{\square^{3}} - \phy_{\square^{2}}}{h_{y}}
\Big) \\
\nonumber
&+ \dfrac{h_{x} h_{y}}{4} \mcal{L}''' \Big(
  \phy_{\square^{3}} , \dot{\phy}_{\square^{3}} ,
  \dfrac{\phy_{\square^{3}} - \phy_{\square^{4}}}{h_{x}} ,
  \dfrac{\phy_{\square^{3}} - \phy_{\square^{2}}}{h_{y}}
\Big) \\
&+ \dfrac{h_{x} h_{y}}{4} \mcal{L}''' \Big(
  \phy_{\square^{4}} , \dot{\phy}_{\square^{4}} ,
  \dfrac{\phy_{\square^{3}} - \phy_{\square^{4}}}{h_{x}} ,
  \dfrac{\phy_{\square^{4}} - \phy_{\square^{1}}}{h_{y}}
\Big) ,
\end{align}
\endgroup
where $h_{x}$ and $h_{y}$ denote the grid step size.
In order to establish a flexible framework for the temporal discretisation, we employ a Hamilton-Pontryagin principle \cite{Pontryagin:1962, Sussmann:1998, Yoshimura:2006b, BouRabee:2008},
\begin{align}\label{eq:vi_semi_discrete_action}
\mcal{A}_{h} [\phy_{h}] = \int \limits_{0}^{T} \big[ \sum \limits_{\square} L_{h} ( \phy_{\square}, \upsilon_{\square} ) - \pair{ \lambda_{\square} }{ \dot{\phy}_{\square} - \upsilon_{\square} } \big] \, dt ,
\end{align}
where we introduced Lagrange multipliers $\lambda$ and the pairing $\pair{\cdot}{\cdot}$ defined by
\begin{align}
\pair{ \lambda_{\square} }{ \phy_{\square} } = \dfrac{1}{4} \sum \limits_{l=1}^{4} \lambda_{\square^{l}} \cdot \phy_{\square^{l}} .
\end{align}
The discretisation in time is performed in a similar way as described in Reference~\cite{Kraus:2016:Evolution} for the special case of one internal stage,
leading to the fully discrete action
\begin{multline}\label{eq:vi_discrete_action}
\mcal{A}_{d} [\phy_{d}]
= \sum \limits_{n}
  \sum \limits_{\square} \Big[ h_{t} \, L_{h} ( \phy_{\square}^{n+1/2} , \upsilon_{\square}^{n+1/2} )
- h_{t} \pair{ \lambda_{\square}^{n+1/2} }{ \phy_{\square}^{n+1/2} - \phy_{\square}^{n} - \tfrac{1}{2} h_{t} \upsilon_{\square}^{n+1/2} } \\
+ \pair{ \pi_{\square}^{n+1} }{ \phy_{\square}^{n+1} - \phy_{\square}^{n} - h_{t} \upsilon_{\square}^{n+1/2} } \Big] ,
\end{multline}
with $h_{t}$ being the time step.
The expression in the second line constitutes the continuity constraint connecting $\phy_{\square}^{n}$ with $\phy_{\square}^{n+1}$.
Requiring stationarity of the discrete action~\eqref{eq:vi_discrete_action}, 
\begin{align}
\delta \mcal{A}_{d}
 = \sum \limits_{n=0}^{n_{t}-1} \Big[
\nonumber
& h_{t} \sum \limits_{\square} \sum \limits_{l=1}^{4} \bigg(
   \dfrac{\partial L_{h}}{\partial \phy_{\square^{l}}} ( \phy_{\square}^{n+1/2} , \upsilon_{\square}^{n+1/2} ) \cdot \delta \phy_{\square^{l}}^{n+1/2}
 + \dfrac{\partial L_{h}}{\partial \upsilon_{\square^{l}}} ( \phy_{\square}^{n+1/2} , \upsilon_{\square}^{n+1/2} ) \cdot \delta \upsilon_{\square^{l}}^{n+1/2}
\bigg) \\
\nonumber
&- h_{t} \pair{ \delta \lambda_{\square}^{n+1/2} }{ \phy_{\square}^{n+1/2} - \phy_{\square}^{n} - \tfrac{1}{2} h_{t} \, \upsilon_{\square}^{n+1/2} } \\
\nonumber
&- h_{t} \pair{ \lambda_{\square}^{n+1/2} }{ \delta \phy_{\square}^{n+1/2} - \delta \phy_{\square}^{n} - \tfrac{1}{2} h_{t} \, \delta \upsilon_{\square}^{n+1/2} } \\
\nonumber
&+ \pair{ \delta \pi_{\square}^{n+1} }{ \phy_{\square}^{n+1} - \phy_{\square}^{n} - h_{t} \, \upsilon_{\square}^{n+1/2} } \\
&+ \pair{ \pi_{\square}^{n+1} }{ \delta \phy_{\square}^{n+1} - \delta \phy_{\square}^{n} - h_{t} \, \delta \upsilon_{\square}^{n+1/2} }
\Big] = 0 ,
\end{align}
leads to the discrete Euler-Lagrange field equations, 
\begin{subequations}
\begin{align}
\label{eq:vi_discrete_variations_1}
\lambda_{i,j}^{n+1/2} &= \sum \limits_{\square} \sum \limits_{\substack{l=1\\ \square^{l} = (i,j)}}^{4} \dfrac{\partial L_{h}}{\partial \phy_{\square^{l}}} ( \phy_{\square}^{n+1/2} , \upsilon_{\square}^{n+1/2} ) , \\
\label{eq:vi_discrete_variations_2}
\pi_{i,j}^{n+1\hphantom{/2}} &= \sum \limits_{\square} \sum \limits_{\substack{l=1\\ \square^{l} = (i,j)}}^{4} \dfrac{\partial L_{h}}{\partial \upsilon_{\square^{l}}} ( \phy_{\square}^{n+1/2} , \upsilon_{\square}^{n+1/2} ) + \tfrac{1}{2} h_{t} \, \lambda_{i,j}^{n+1/2} , \\
\label{eq:vi_discrete_variations_3}
\pi_{i,j}^{n+1\hphantom{/2}} &= \pi_{i,j}^{n} + h_{t} \, \lambda_{i,j}^{n+1/2} , \vphantom{\dfrac{1}{2}} \\
\label{eq:vi_discrete_variations_4}
\phy_{i,j}^{n+1/2} &= \phy_{i,j}^{n} + \tfrac{1}{2} h_{t} \, \upsilon_{i,j}^{n+1/2} , \vphantom{\dfrac{1}{2}} \\
\label{eq:vi_discrete_variations_5}
\phy_{i,j}^{n+1\hphantom{/2}} &= \phy_{i,j}^{n} + h_{t} \, \upsilon_{i,j}^{n+1/2} . \vphantom{\dfrac{1}{2}}
\end{align}
\end{subequations}
Upon defining
\begin{align}\label{eq:vi_discrete_variations_6}
\pi_{i,j}^{n+1/2}
&= \sum \limits_{\square} \sum \limits_{\substack{l=1\\ \square^{l} = (i,j)}}^{4} \dfrac{\partial L_{h}}{\partial \upsilon_{\square^{l}}} ( \phy_{\square}^{n+1/2} , \upsilon_{\square}^{n+1/2} ) 
= \tfrac{1}{2} \Lambda^{T} \phy_{i,j}^{n+1/2} ,
\end{align}
and using~\eqref{eq:vi_discrete_variations_3}, equation~\eqref{eq:vi_discrete_variations_2} can be rewritten as
\begin{align}\label{eq:vi_discrete_variations_7}
\pi_{i,j}^{n+1/2}
&= \pi_{i,j}^{n} + \tfrac{1}{2} h_{t} \, \lambda_{i,j}^{n+1/2} .
\end{align}
In the Hamilton-Pontryagin action~\eqref{eq:vi_discrete_action}, we introduced an additional set of variables $\pi$, which in practice are not needed to be solved for.
If $\pi$ is initialised by $\pi_{i,j}^{0} = \tfrac{1}{2} \Lambda^{T} \phy_{i,j}^{0}$, the variational integrator preserves the functional relation between $\pi$ and $\phy$, so that
\begin{align}
\pi_{i,j}^{n} = \tfrac{1}{2} \Lambda^{T} \phy_{i,j}^{n} 
\hspace{1em}
\text{for all $n$} .
\end{align}
To see this, we equate relations \eqref{eq:vi_discrete_variations_6} and \eqref{eq:vi_discrete_variations_7} for $\pi_{i,j}^{n+1/2}$,
\begin{align}
\tfrac{1}{2} \Lambda^{T} \phy_{i,j}^{n+1/2} = \pi_{i,j}^{n} + \tfrac{1}{2} h_{t} \lambda_{i,j}^{n+1/2} .
\end{align}
We insert the definition of $\phy_{i,j}^{n+1/2}$ from~\eqref{eq:vi_discrete_variations_4} and replace $\pi_{i,j}^{n}$ with the initial condition $\pi_{i,j}^{n} = \tfrac{1}{2} \Lambda^{T} \phy_{i,j}^{n}$, hence
\begin{align}
\tfrac{1}{2} \Lambda^{T} \big( \phy_{i,j}^{n} + \tfrac{1}{2} h_{t} \upsilon_{i,j}^{n+1/2} \big) = \tfrac{1}{2} \Lambda^{T} \phy_{i,j}^{n} + \tfrac{1}{2} h_{t} \lambda_{i,j}^{n+1/2} .
\end{align}
This shows that
\begin{align}
\lambda_{i,j}^{n+1/2} = \tfrac{1}{2} \Lambda^{T} \upsilon_{i,j}^{n+1/2} ,
\end{align}
so that upon insertion into the definition of $\pi_{i,j}^{n+1}$ from \eqref{eq:vi_discrete_variations_3}, we find 
\begin{align}
\pi_{i,j}^{n+1} &= \tfrac{1}{2} \Lambda^{T} \big( \phy_{i,j}^{n} + h_{t} \upsilon_{i,j}^{n+1/2} \big) = \tfrac{1}{2} \Lambda^{T} \phy_{i,j}^{n+1} .
\end{align}
That is if $\pi_{i,j}^{n} = \tfrac{1}{2} \Lambda^{T} \phy_{i,j}^{n}$ then $\pi_{i,j}^{n+1} = \tfrac{1}{2} \Lambda^{T} \phy_{i,j}^{n+1}$.
This allows us to eliminate $\pi$ from the scheme.
Combining equations \eqref{eq:vi_discrete_variations_1}, \eqref{eq:vi_discrete_variations_6} and \eqref{eq:vi_discrete_variations_7},
\begin{align}\label{eq:vi_discrete_euler_lagrange_equations}
\Lambda^{T} \left( \dfrac{\phy_{i,j}^{n+1/2} - \phy_{i,j}^{n}}{h_{t}} \right)
&= \sum \limits_{\square} \sum \limits_{\substack{l=1\\ \square^{l} = (i,j)}}^{4} \dfrac{\partial L_{h}}{\partial \phy_{\square^{l}}} ( \phy_{\square}^{n+1/2} , \upsilon_{\square}^{n+1/2} ) ,
\end{align}
provides integrators for the vorticity equation and Ohm's law,
\begin{align}\label{eq:rmhd_integrator_vorticity}
\dfrac{\omega_{i,j}^{n+1/2} - \omega_{i,j}^{n}}{h_{t}}
&+ \tfrac{1}{2} A_{i,j} (\phi^{n+1/2}, \omega^{n+1/2})
 + \tfrac{1}{2} A_{i,j} (j^{n+1/2}, \psi^{n+1/2}) = 0 ,
\\
\label{eq:rmhd_integrator_ohms_law}
\dfrac{\psi_{i,j}^{n+1/2} - \psi_{i,j}^{n}}{h_{t}} 
&+ \tfrac{1}{2} A_{i,j} (\phi^{n+1/2}, \psi^{n+1/2}) = 0 ,
\end{align}
where $A$ denotes Arakawa's discretisation of the Poisson bracket~\cite{Arakawa:1966}, given by
\begin{align}\label{eq:arakawa_bracket_1}
A_{i,j} (\phi, \omega) = \tfrac{1}{3} \big[ A_{i,j}^{++} (\phi, \omega) + A_{i,j}^{+ \times} (\phi, \omega) + A_{i,j}^{\times +} (\phi, \omega) \big] ,
\end{align}
with
\begingroup
\allowdisplaybreaks
\begin{subequations}\label{eq:arakawa_bracket_2}
\begin{align}
A_{i,j}^{++} (\phi, \omega)
\nonumber
= \dfrac{1}{4 h_{x} h_{y}} \Big[
     & \big( \phi_{i+1,j} - \phi_{i-1,j} \big) \big( \omega_{i,j+1} - \omega_{i,j-1} \big) \\
- \, & \big( \phi_{i,j+1} - \phi_{i,j-1} \big) \big( \omega_{i+1,j} - \omega_{i-1,j} \big) \Big] , \\
A_{i,j}^{+ \times} (\phi, \omega)
\nonumber
= \dfrac{1}{4 h_{x} h_{y}} \Big[
     & \phi_{i+1,j} \big( \omega_{i+1,j-1} - \omega_{i+1,j+1} \big) - \phi_{i-1,j} \big( \omega_{i-1,j-1} - \omega_{i-1,j+1} \big) \\
- \, & \phi_{i,j+1} \big( \omega_{i-1,j+1} - \omega_{i+1,j+1} \big) + \phi_{i,j-1} \big( \omega_{i-1,j-1} - \omega_{i+1,j-1} \big) \Big] , \\
A_{i,j}^{\times +} (\phi, \omega)
\nonumber
= \dfrac{1}{4 h_{x} h_{y}} \Big[
     & \phi_{i+1,j+1} \big( \omega_{i+1,j} - \omega_{i,j+1} \big) - \phi_{i-1,j-1} \big( \omega_{i,j-1} - \omega_{i-1,j} \big) \\
- \, & \phi_{i-1,j+1} \big( \omega_{i-1,j} - \omega_{i,j+1} \big) + \phi_{i+1,j-1} \big( \omega_{i,j-1} - \omega_{i+1,j} \big) \Big] ,
\end{align}
\end{subequations}
\endgroup
as well as an integrator for the Poisson equation for $\phi$ and a discrete rule for computing $j$,
\begin{align}
\label{eq:rmhd_integrator_poisson1}
 - \Delta_{x} \phi_{i,j}^{n+1/2}
 - \Delta_{y} \phi_{i,j}^{n+1/2}
&= \omega_{i,j}^{n+1/2} ,
\\
\label{eq:rmhd_integrator_poisson2}
 - \Delta_{x} \psi_{i,j}^{n+1/2}
 - \Delta_{y} \psi_{i,j}^{n+1/2}
&= j_{i,j}^{n+1/2} .
\end{align}
Here, $\Delta_{x}$ and $\Delta_{y}$ denote the standard centred finite difference second-order derivative with respect to $x$ and $y$, i.e.,
\begin{align}\label{eq:rmhd_integrator_poisson_definition}
\Delta_{x} \phi_{i,j} &= \dfrac{\phi_{i-1,j} - 2 \phi_{i,j} + \phi_{i+1,j}}{h_{x}^{2}} , &
\Delta_{y} \phi_{i,j} &= \dfrac{\phi_{i,j-1} - 2 \phi_{i,j} + \phi_{i,j+1}}{h_{y}^{2}} .
\end{align}
In addition, we obtain discrete versions of the adjoint equations \eqref{eq:rmhd_adjoint_equations}.
Using \eqref{eq:vi_discrete_variations_4}, the update rule \eqref{eq:vi_discrete_variations_5} can be rewritten as
\begin{align}
\tfrac{1}{2} \big[ \phy_{i,j}^{n+1} + \phy_{i,j}^{n} \big] = \phy_{i,j}^{n+1/2} .
\end{align}
As all discrete operators $A_{i,j}$, $\Delta_{x}$ and $\Delta_{y}$ are linear in their arguments, we can use this relation to eliminate $\phy_{i,j}^{n+1/2}$ and rewrite the integrator solely in terms of $\phy_{i,j}^{n}$ and $\phy_{i,j}^{n+1}$, namely
\begin{subequations}\label{eq:rmhd_integrator_simplified}
\begin{align}
\label{eq:rmhd_integrator_vorticity_simplified}
0 = \dfrac{\omega_{i,j}^{n+1} - \omega_{i,j}^{n}}{h_{t}}
\nonumber
&+ \dfrac{1}{4} \big[ A_{i,j} (\phi^{n+1}, \omega^{n+1}) + A_{i,j} (\phi^{n}, \omega^{n+1}) + A_{i,j} (\phi^{n+1}, \omega^{n}) + A_{i,j} (\phi^{n}, \omega^{n}) \big]
\\
&+ \dfrac{1}{4} \big[ A_{i,j} (j^{n+1}, \psi^{n+1}) + A_{i,j} (j^{n}, \psi^{n+1}) + A_{i,j} (j^{n+1}, \psi^{n}) + A_{i,j} (j^{n}, \psi^{n}) \big] ,
\\
\label{eq:rmhd_integrator_ohms_law_simplified}
0 = \dfrac{\psi_{i,j}^{n+1} - \psi_{i,j}^{n}}{h_{t}}
&+ \dfrac{1}{4} \big[ A_{i,j} (\phi^{n+1}, \psi^{n+1}) + A_{i,j} (\phi^{n}, \psi^{n+1}) + A_{i,j} (\phi^{n+1}, \psi^{n}) + A_{i,j} (\phi^{n}, \psi^{n}) \big] ,
\\
\label{eq:rmhd_integrator_poisson1_simplified}
\omega_{i,j}^{n+1}
&= - \Delta_{x} \phi_{i,j}^{n+1}
   - \Delta_{y} \phi_{i,j}^{n+1} ,
\vphantom{\dfrac{1}{2}}
\\
\label{eq:rmhd_integrator_poisson2_simplified}
j_{i,j}^{n+1} 
&= - \Delta_{x} \psi_{i,j}^{n+1}
   - \Delta_{y} \psi_{i,j}^{n+1} ,
\vphantom{\dfrac{1}{2}}
\end{align}
\end{subequations}
where the last two equations follow under the assumption that $\phi_{i,j}^{0}$ and $j_{i,j}^{0}$ are initialised using~\eqref{eq:rmhd_integrator_poisson1_simplified} and~\eqref{eq:rmhd_integrator_poisson2_simplified}.
In summary, we obtain a scheme of second-order, which consists of the Crank-Nicolson method for the time derivatives, Arakawa's method for the Poisson bracket and the standard centred finite difference approximation for the Laplacian.

\subsection{Discrete Conservation Laws}

The derivation of the discrete conservation laws follows analogously to the case of the vorticity equation described by \citet{KrausMaj:2015}.
Here, we only provide the discrete expressions of~(\ref{eq:rmhd_hamiltonian}), (\ref{eq:rmhd_magnetic_helicity}), (\ref{eq:rmhd_l2norm}) and~(\ref{eq:rmhd_cross_helicity}), that is
\begin{enumerate}[(a)]
\begin{subequations}\label{eq:rmhd_discrete_conservation_laws}
\item total energy
\begin{align}
E = \dfrac{h_x h_y}{2} \sum \limits_{i,j} \big( \phi_{i,j} \, \omega_{i,j} + \psi_{i,j} \, j_{i,j} \big) = \mathrm{const.} ,
\end{align}
\item magnetic helicity
\begin{align}
C_{\mathrm{MH}} = h_x h_y \sum \limits_{i,j} \psi_{i,j}  = \mathrm{const.} ,
\end{align}
\item $L^{2}$ norm of $\psi$
\begin{align}
C_{L^{2}} = h_x h_y \sum \limits_{i,j} \psi_{i,j}^{2} = \mathrm{const.} ,
\end{align}
\item and cross helicity
\begin{align}
C_{\mathrm{CH}} = h_x h_y \sum \limits_{i,j} \omega_{i,j} \psi_{i,j} = \mathrm{const.} .
\end{align}
\end{subequations}
\end{enumerate}
The corresponding discrete expressions of~(\ref{eq:rmhd_ei_hamiltonian}) and~(\ref{eq:rmhd_ei_magnetic_helicity})-(\ref{eq:rmhd_ei_cross_helicity}) are defined analogously.
In practice, the tolerance of the nonlinear solver plays a crucial role in conserving these quantities exactly (i.e., up to machine precision). We observe that if the tolerance is set to a value above a certain threshold, the errors in the conservation laws exhibit a drift on top of the oscillating behaviour usually observed (see e.g. Figures~\ref{fig:orszag_tang_vortex_timetraces}, \ref{fig:mhd_current_sheet_timetraces}, and \ref{fig:mhd_reconnection_timetraces}).

Variational integrators also preserve a discrete symplectic structure arising from the boundary terms in Hamilton's action principle~\cite{MarsdenPatrick:1998}. Considering that, it seems striking that the RMHD integrator preserves energy exactly. According to the famous theorem by Ge and Marsden~\cite{ZhongMarsden:1988} there are no symplectic integrators that preserve energy exactly.
This is an intricate delicacy of variational integrators for formal Lagrangians. To understand this issue, it is important to realise that the energy, which is preserved, is the energy of the physical system, whereas the symplectic structure, which is preserved, is the symplectic structure of the extended system. Thinking in terms of energy as the conservation law that originates from invariance of the Lagrangian with respect to temporal translations, we find that the corresponding ``energy'' of the formal Lagrangian is not preserved in the discrete case. Therefore there is no contradiction to the Ge-Marsden theorem.

This observation has further implications. In practice it also implies that we cannot make an immediate statement about the symplecticity of the resulting algorithm for the physical system. We only know that the discrete extended system has a symplectic structure which is indeed preserved. On the attempt of reducing this symplectic structure to the physical system by restricting solutions of the extended system to solutions of the physical system, c.f. Equation~\eqref{eq:rmhd_restriction_simplified}, we usually find that the restricted symplectic structure vanishes identically.

\section{Numerical Experiments}\label{sec:numerical_experiments}

In this section we verify the properties of the variational integrator \eqref{eq:rmhd_integrator_simplified} by considering several standard test cases from the literature.
In the first example, we simulate the evolution of the Orszag-Tang vortex, which is a challenging problem due to the appearance of strong nonlinear flows.
In the second example, we consider a current sheet model often used in reconnection studies in order to verify that the integrator is free of numerical resistivity, so that spurious reconnection along current sheets is absent. 
In the last example, we show that, when adding effects of electron inertia, collisionless magnetic reconnection is observed.
We find that important features of the reconnection process are in very good qualitative and quantitative agreement with a pseudo-spectral code previously used in reconnection studies.

A reference implementation of the variational integrator is provided based on Python~\cite{ScopatzHuff:2015, Langtangen:2014}, Cython~\cite{Behnel:2010}, PETSc~\cite{petsc-web-page, petsc-user-ref} and petsc4py~\cite{Dalcin:2011}. Visualisation is done using NumPy~\cite{vanDerWalt:2011}, SciPy~\cite{scipy-web-page}, matplotlib~\cite{Hunter:2007} and the IPython notebook~\cite{Perez:2007}.
The nonlinear system is solved with Newton's method, where in each iteration the linear system is solved via GMRES.
An important problem for solving the linear system is efficient preconditioning. In this work, we use a simplified version of the physics-based preconditioner described by~\citet{Chacon:2002}. For details on the solver and the preconditioner the reader is referred to Appendix~\ref{sec:app_solver}.
The absolute tolerance of the nonlinear solver is set to  $5 \times 10^{-16}$ and the relative tolerance to $10^{-10}$, which is usually reached after $2-5$ iterations.

For comparison we use an initial value code, based on a Fast Fourier Transform scheme for the spatial operators and a third order Adams-Bashforth scheme for advancing in time. This scheme has already been adopted by some of the authors in many investigations of collisionless reconnection~\cite{Grasso:2009, GrassoTassi:2010, TassiGrasso:2010, TassiMorrison:2010}.

\pagebreak

\subsection{Orszag-Tang Vortex}

The first example describes the evolution of current sheets in an Orszag-Tang vortex according to the RMHD equations~\eqref{eq:rmhd_equations} with initial conditions from \citet{CordobaMarliani:2000}, given by
\begin{align*}
\phi &= 2 \cos (x) - 2 \sin (y) , &
\psi &= 2 \cos (x) - \cos (2y) .
\end{align*}
The spatial domain is $(x,y) \in [0,2 \pi) \times [0,2 \pi)$ with periodic boundaries. We consider a spatial resolution of $n_{x} \times n_{y} = 64 \times 64$ grid points and the time step $h_{t} = 0.01$.

\begin{figure}[bth]
\centering
\subfloat[$\norm{j}_{2}$]{
\includegraphics[width=.4\textwidth]{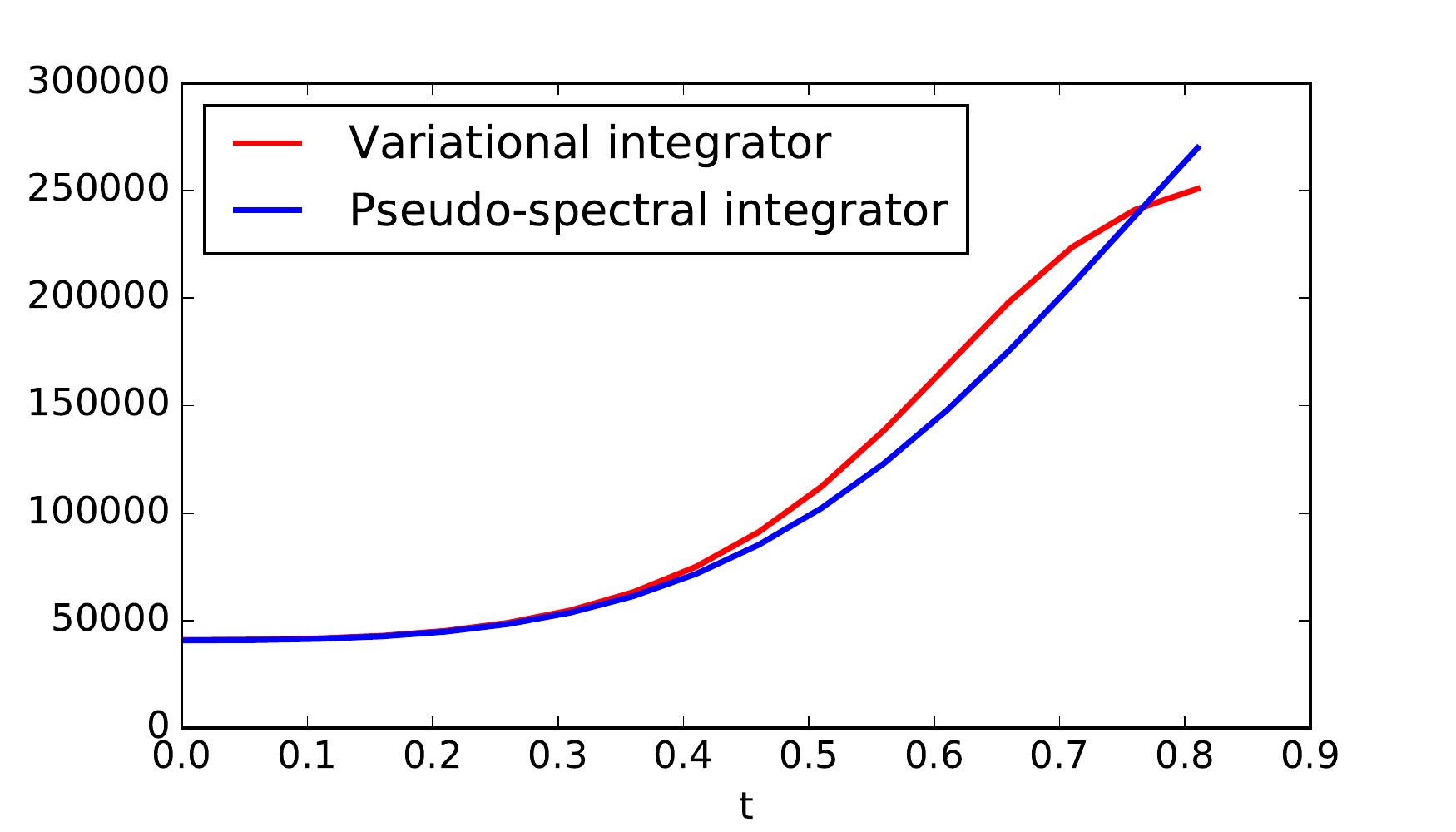}
}
\subfloat[$E_{\text{kinetic}}$]{
\includegraphics[width=.4\textwidth]{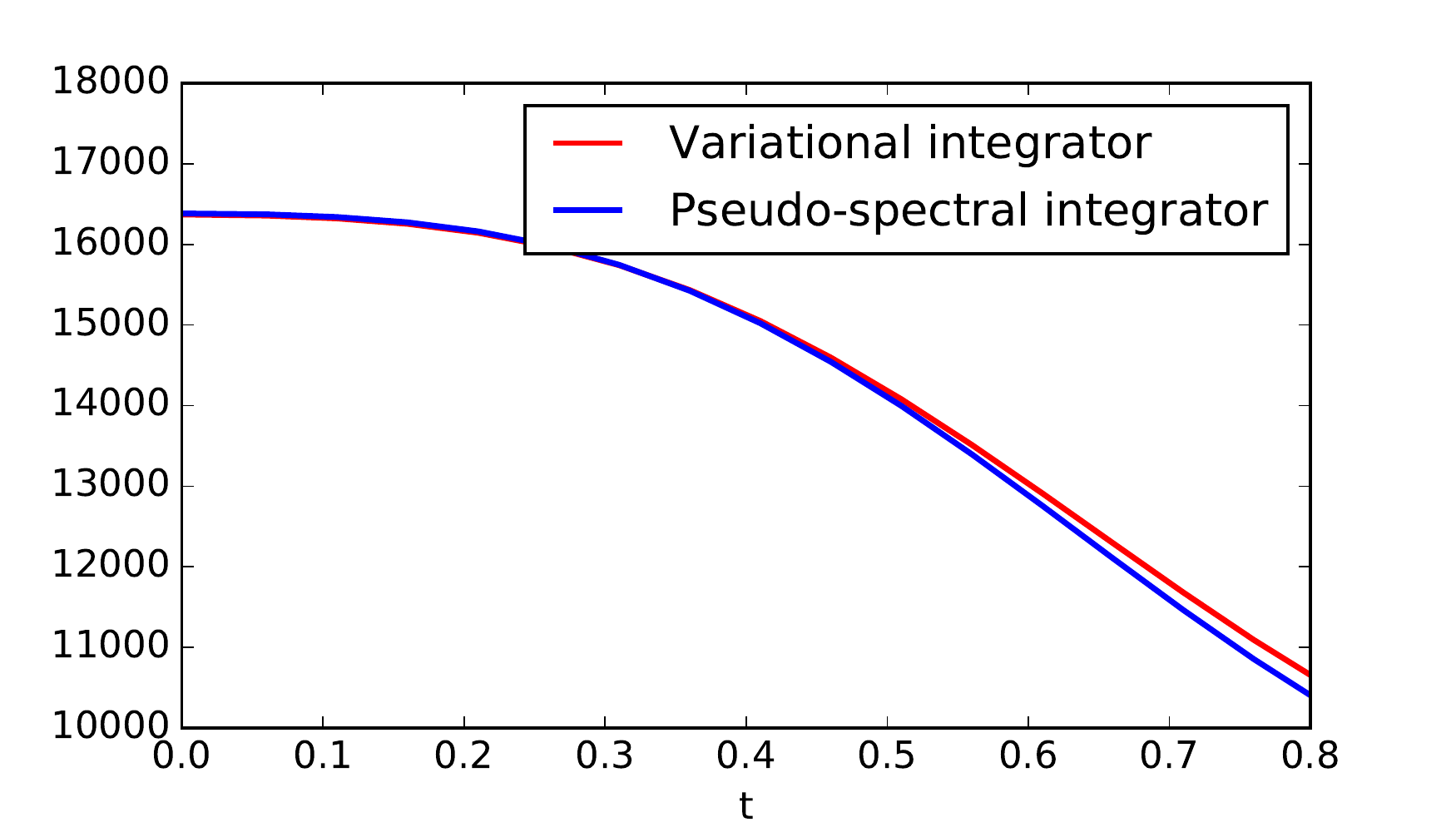}
}

\subfloat[$\norm{\omega}_{2}$]{
\includegraphics[width=.4\textwidth]{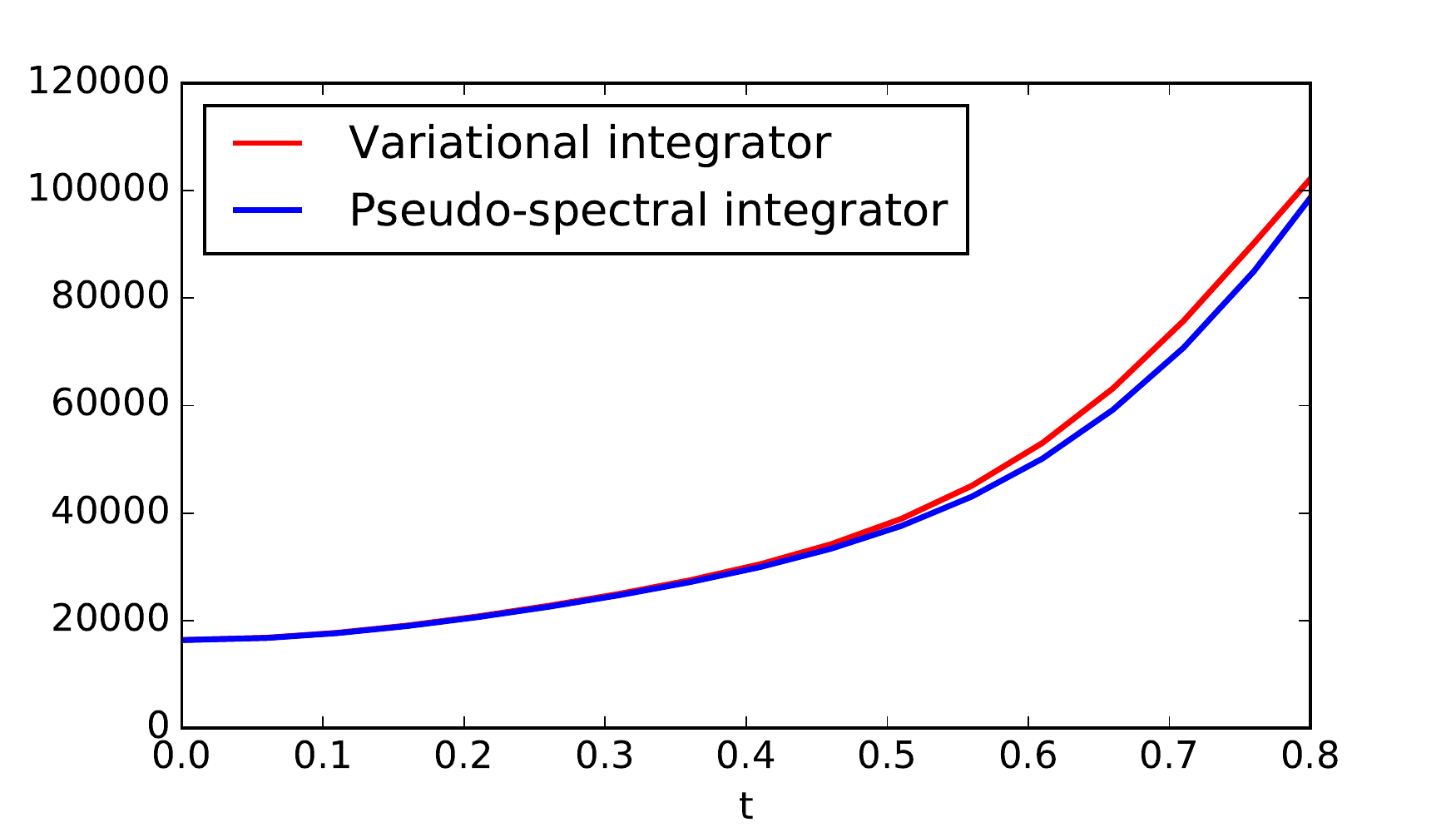}
}
\subfloat[$E_{\text{magnetic}}$]{
\includegraphics[width=.4\textwidth]{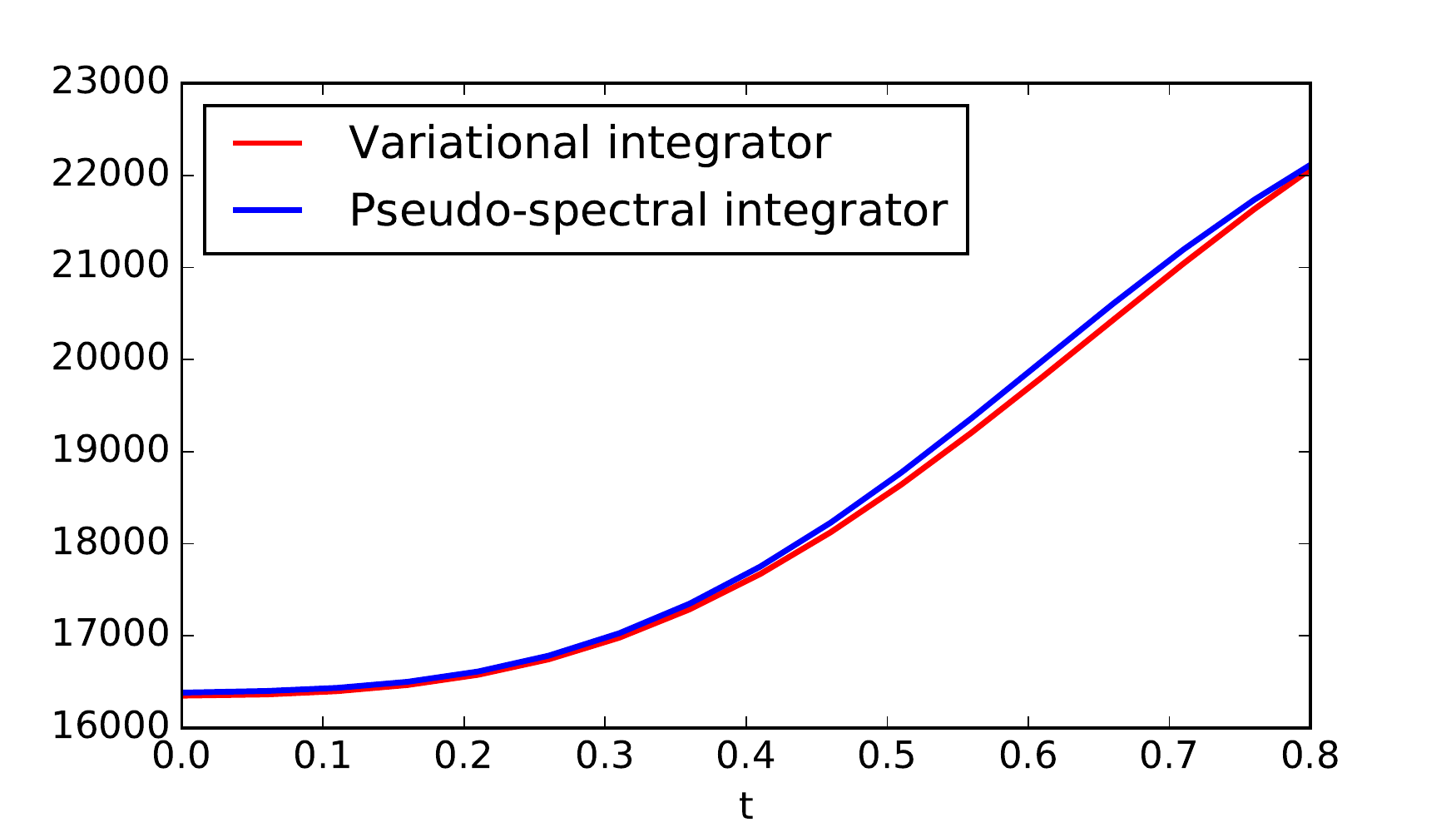}
}

\caption{Orszag-Tang vortex. Time evolution of the $L_{2}$ norms of $j$ and $\omega$ as well as the kinetic and magnetic energy.}
\label{fig:orszag_tang_vortex_norms}
\end{figure}

\begin{figure}[p]
\centering
\vspace{-3em}

\subfloat{
\includegraphics[width=.42\textwidth]{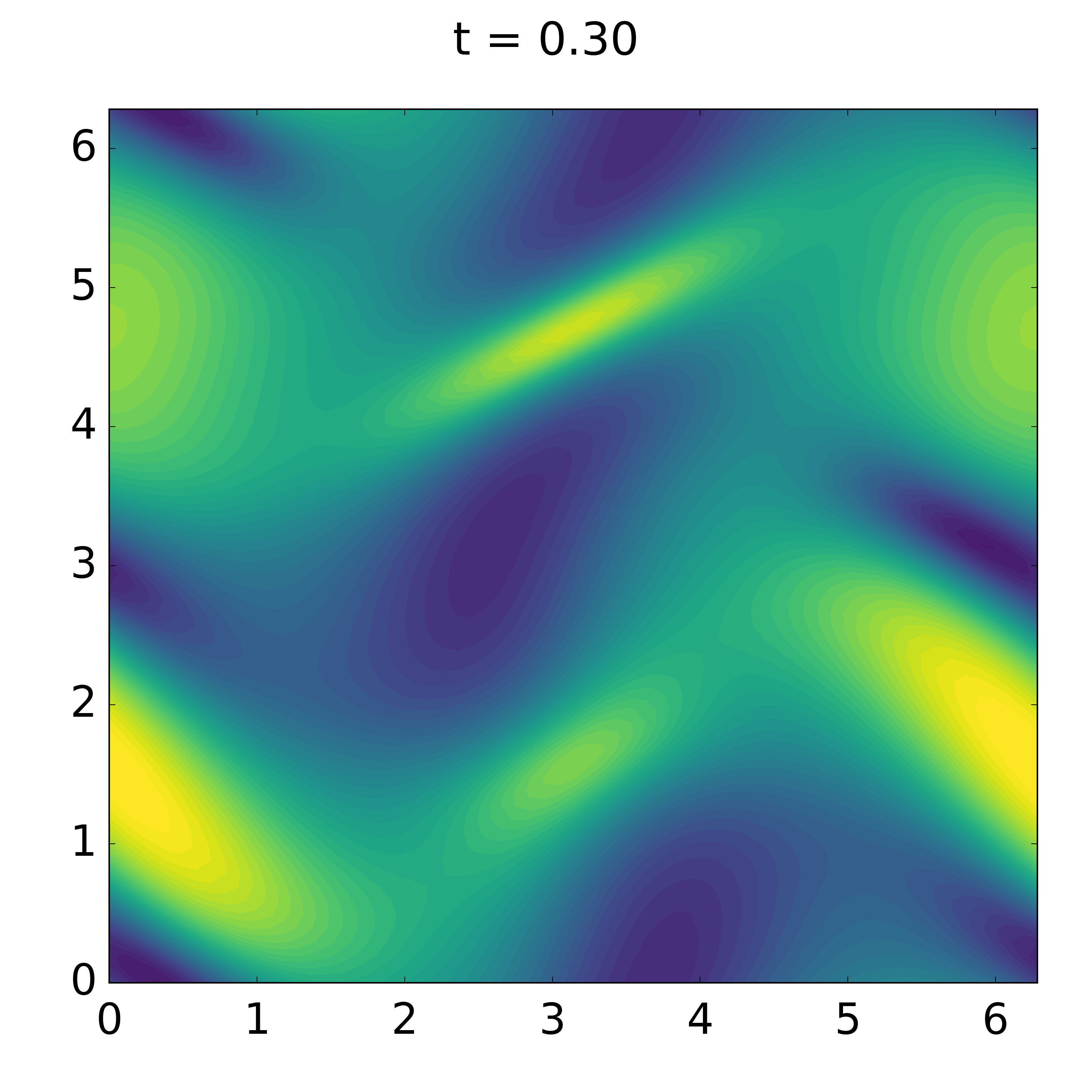}
}
\subfloat{
\includegraphics[width=.42\textwidth]{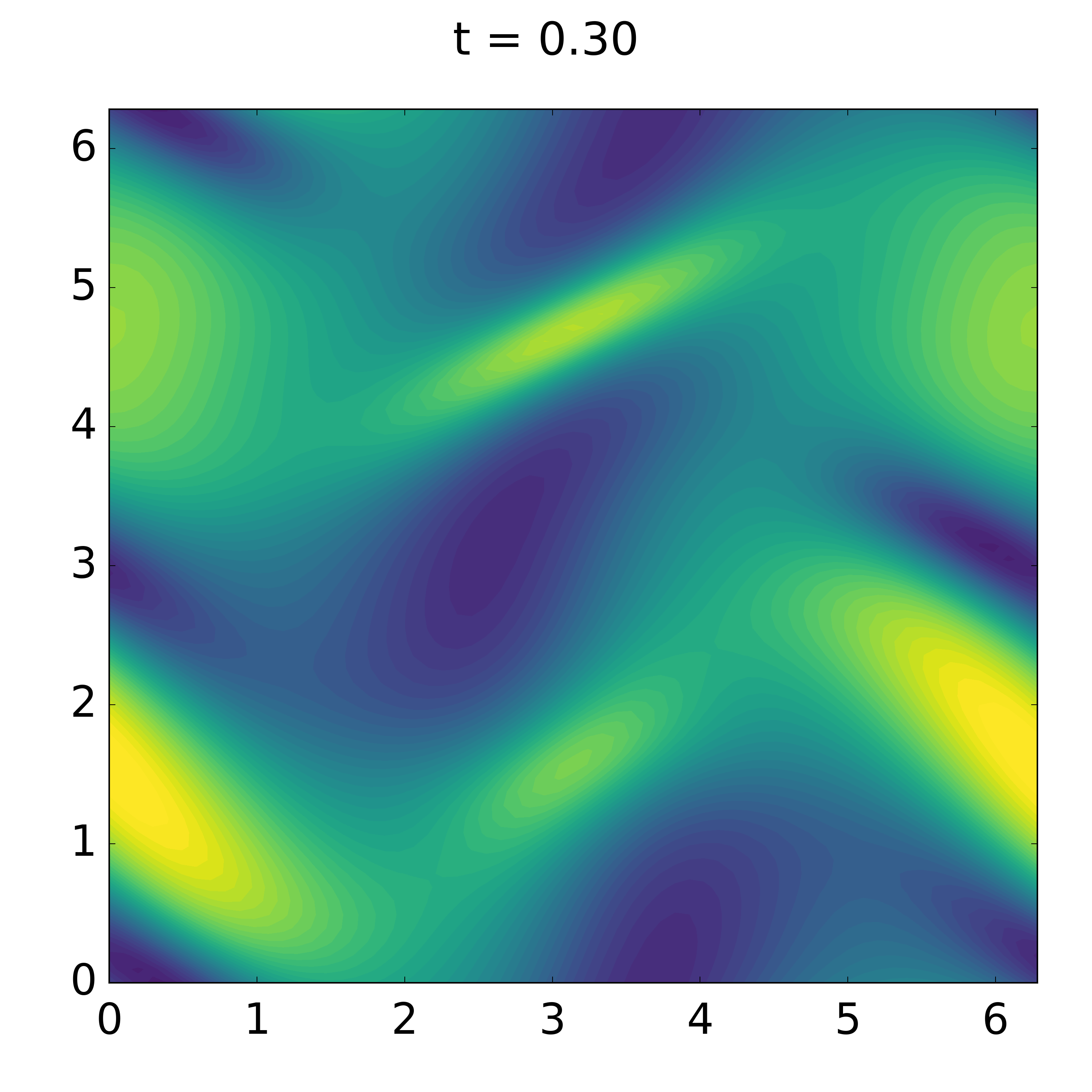}
}

\subfloat{
\includegraphics[width=.42\textwidth]{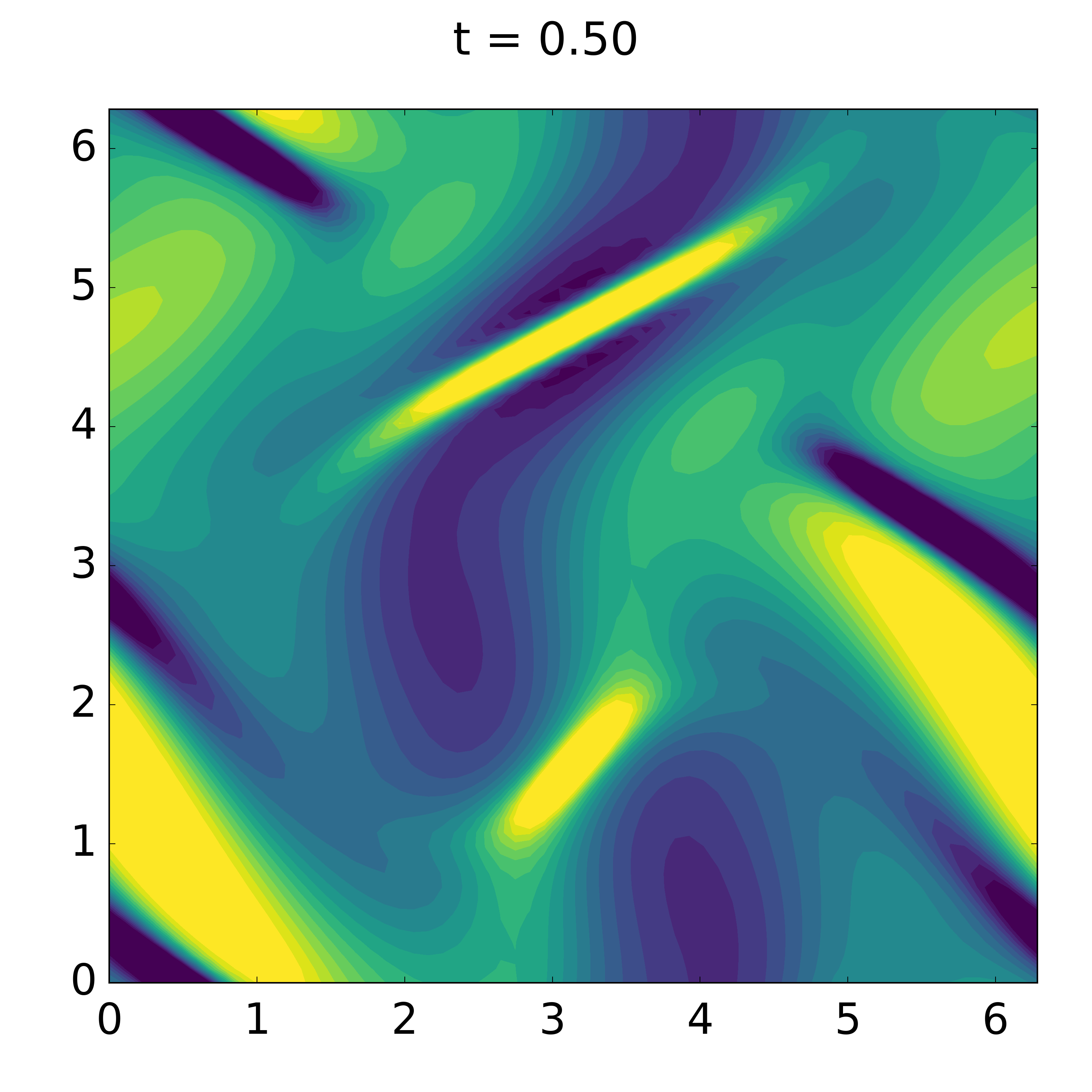}
}
\subfloat{
\includegraphics[width=.42\textwidth]{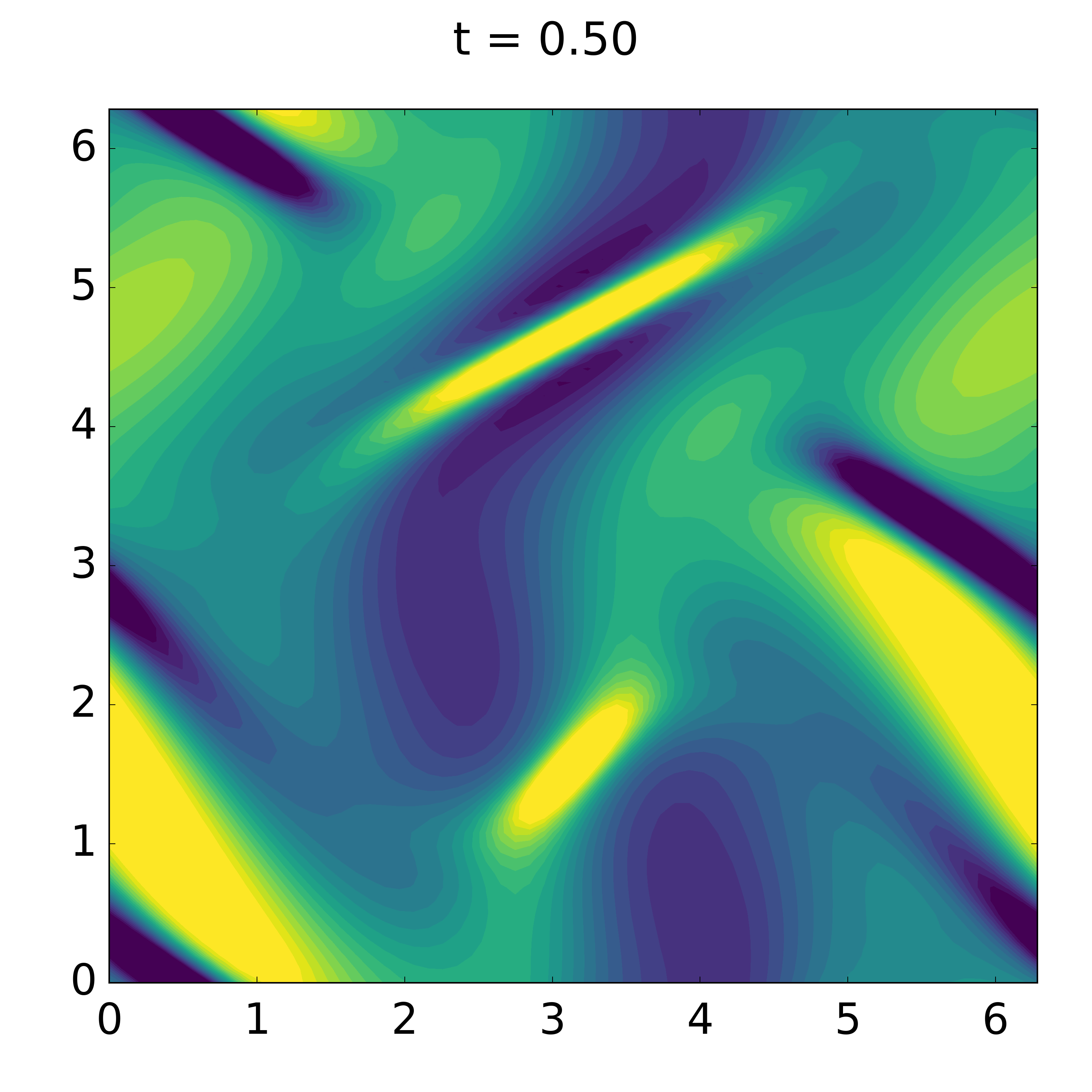}
}

\subfloat{
\includegraphics[width=.42\textwidth]{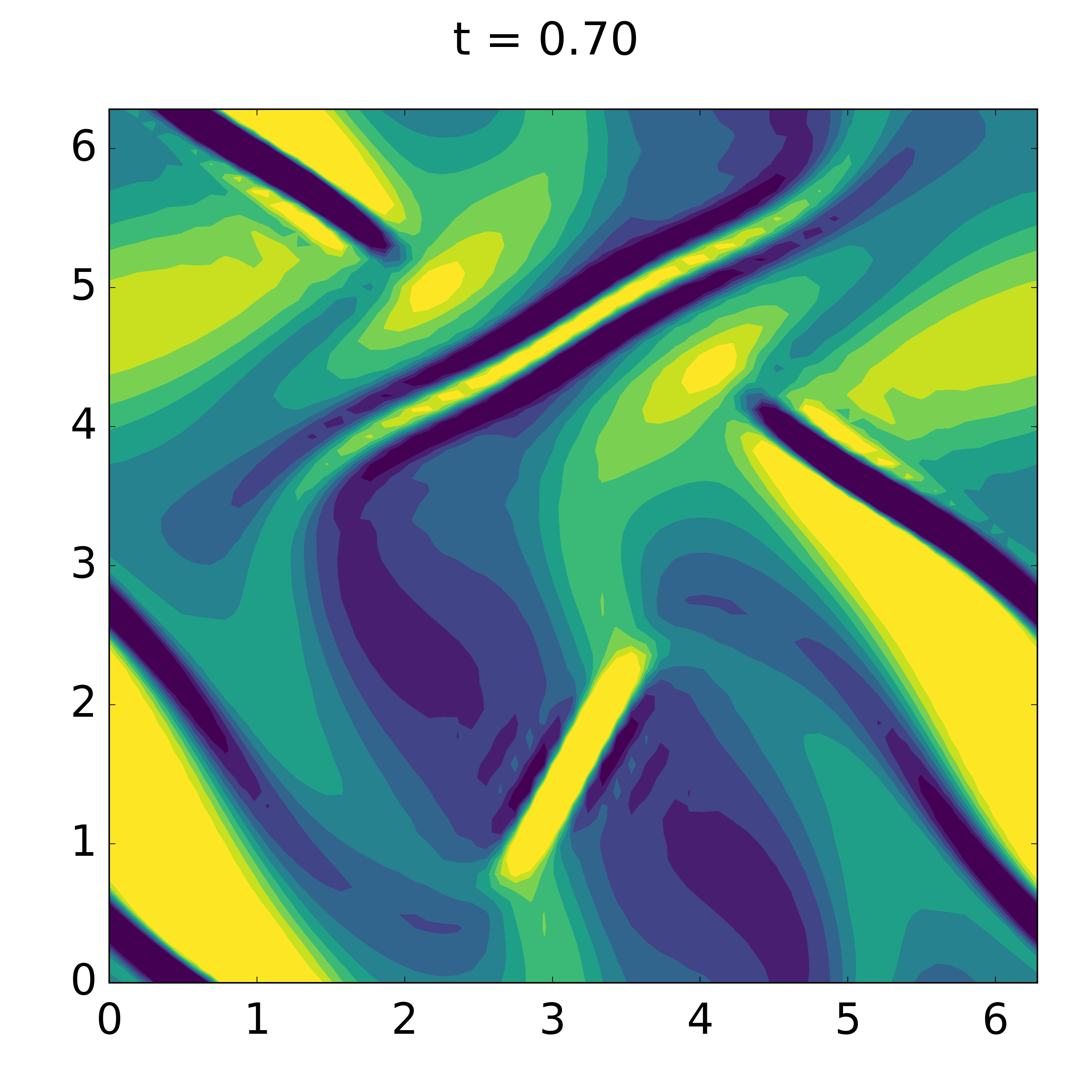}
}
\subfloat{
\includegraphics[width=.42\textwidth]{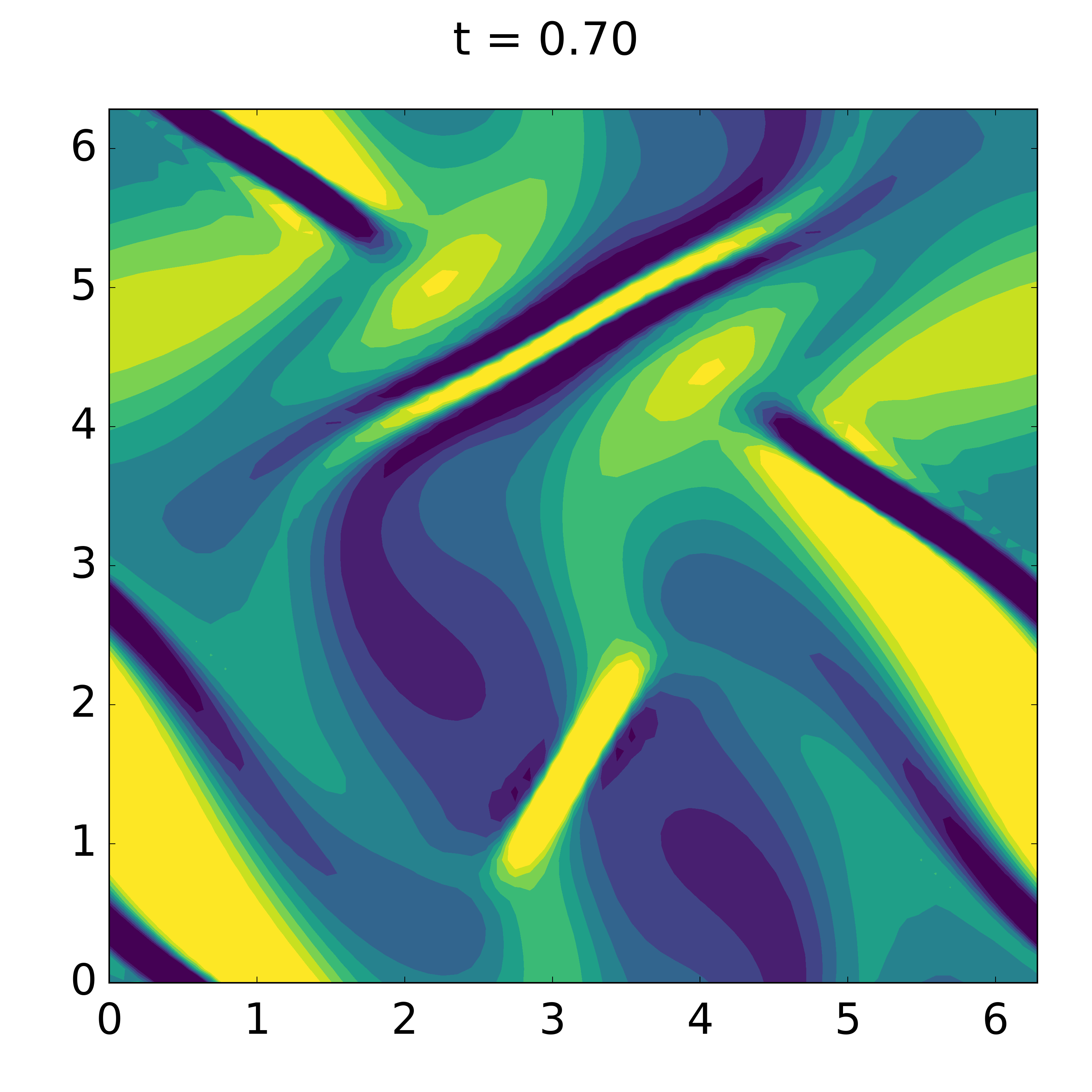}
}

\caption{Orszag-Tang Vortex with the pseudo-spectral integrator (left) and the variational integrator (right). Current density $j$. Fixed colour scale.}
\label{fig:orszag_tang_vortex_current_density}
\end{figure}

\begin{figure}[h]
\centering
\includegraphics[width=.45\textwidth]{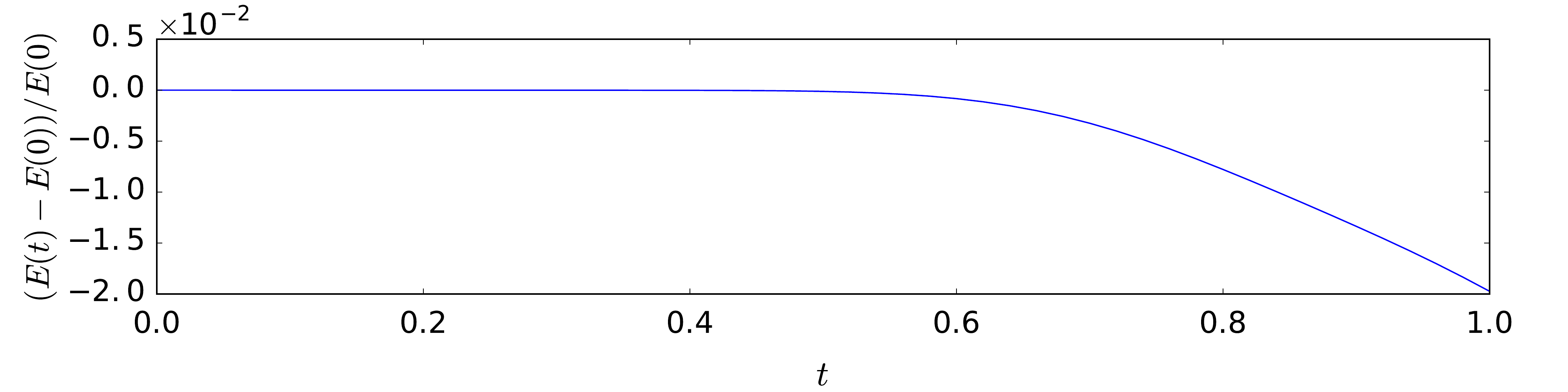}
\includegraphics[width=.45\textwidth]{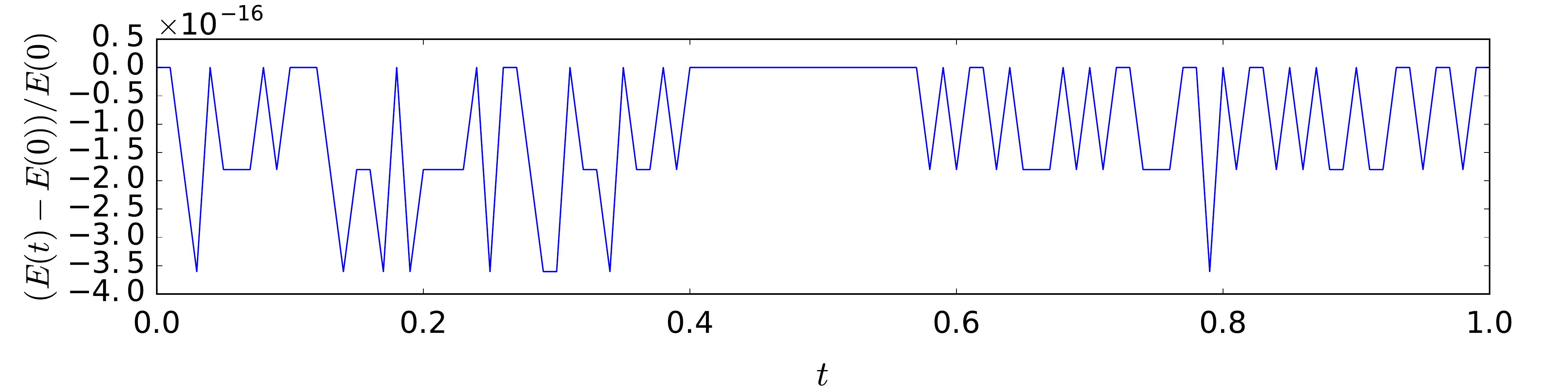}

\includegraphics[width=.45\textwidth]{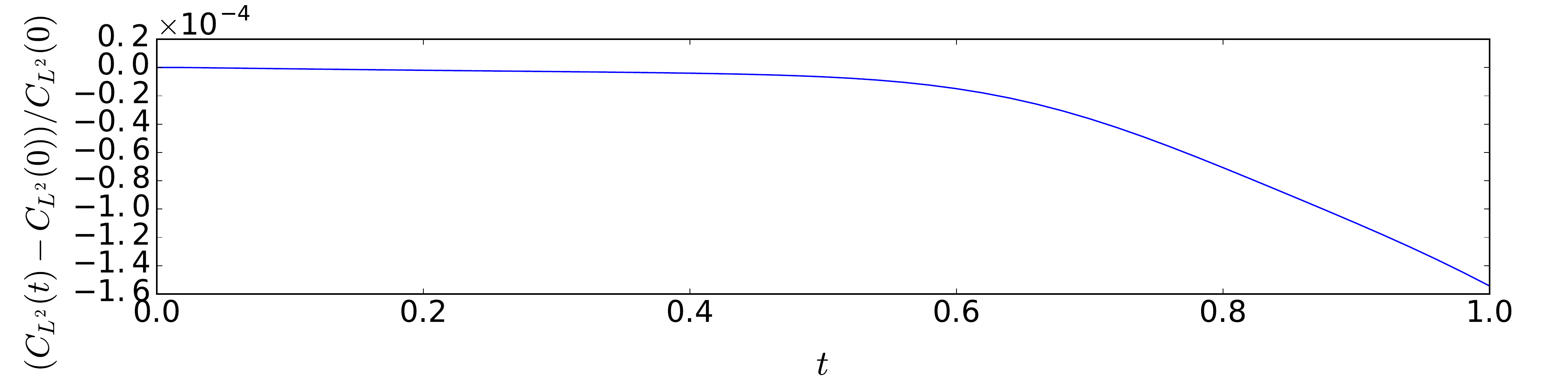}
\includegraphics[width=.45\textwidth]{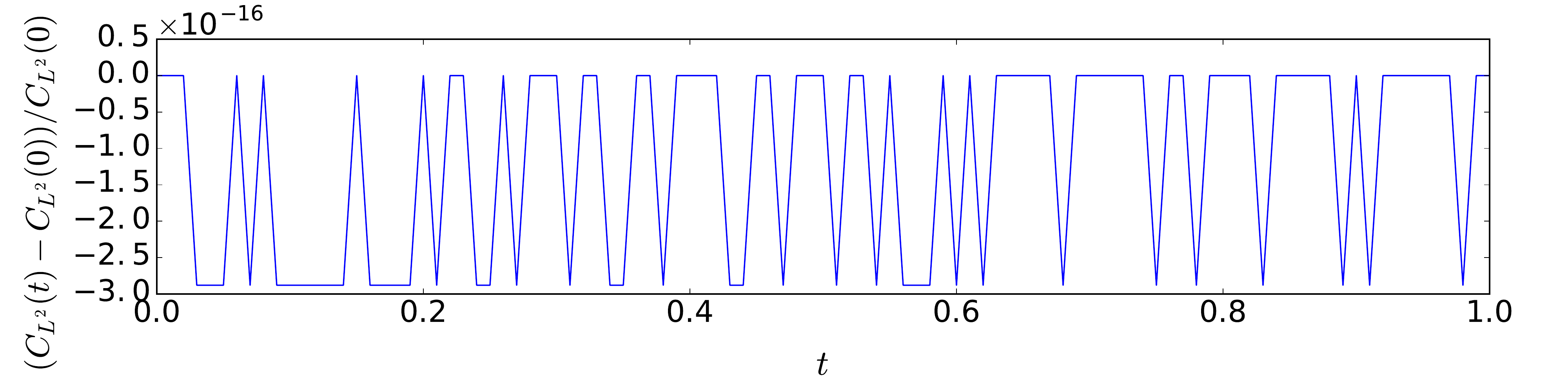}

\includegraphics[width=.45\textwidth]{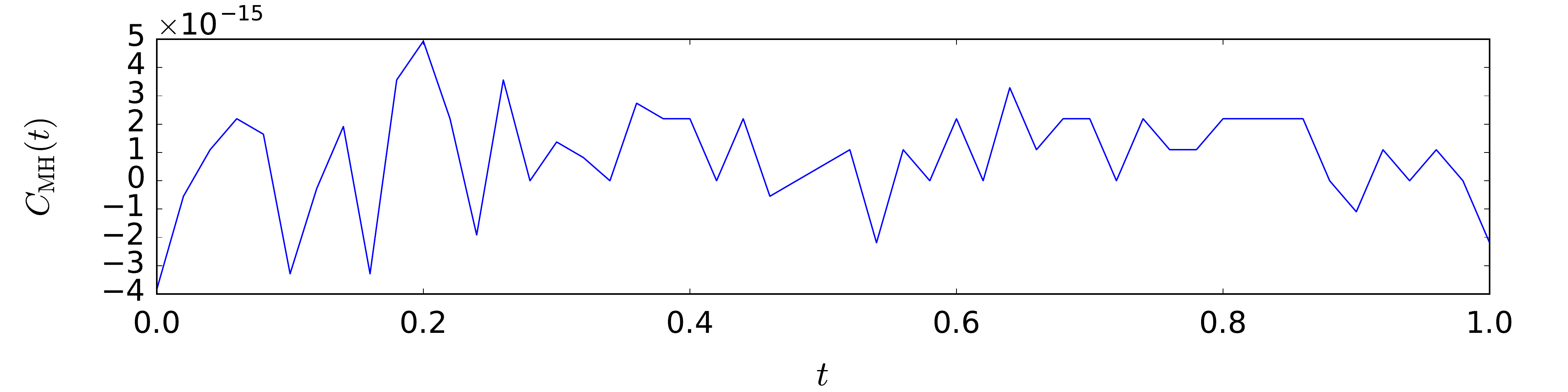}
\includegraphics[width=.45\textwidth]{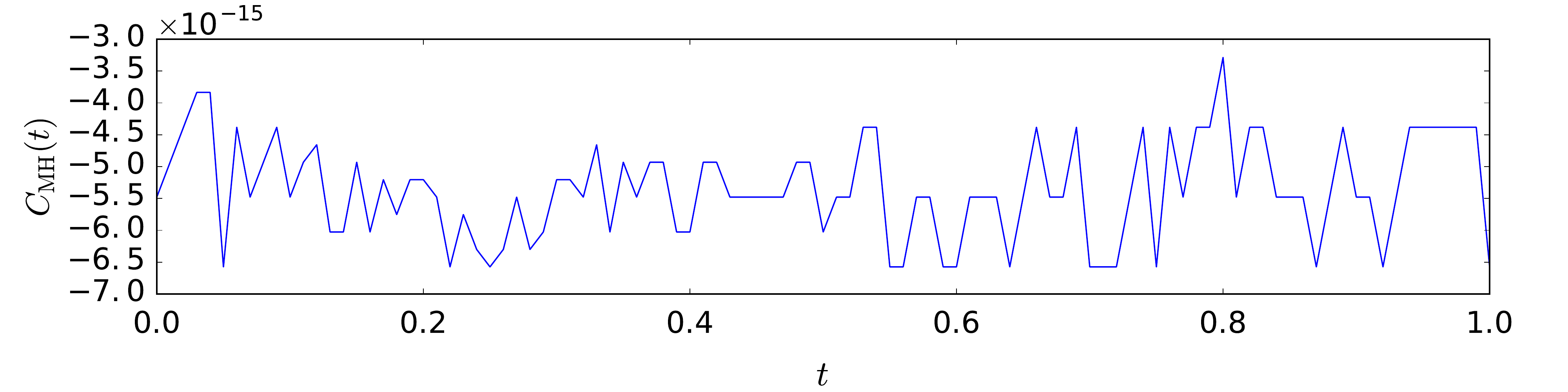}

\includegraphics[width=.45\textwidth]{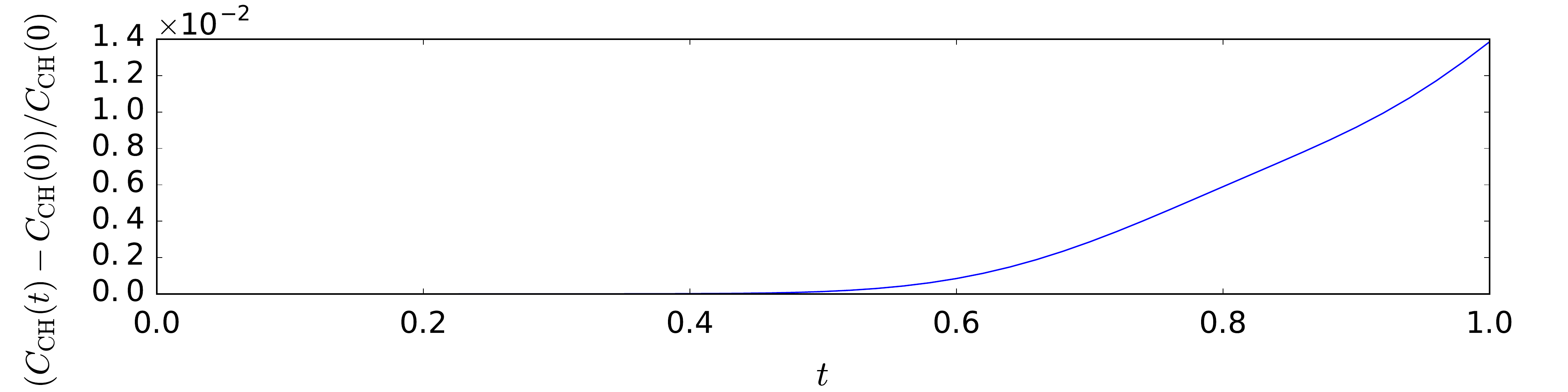}
\includegraphics[width=.45\textwidth]{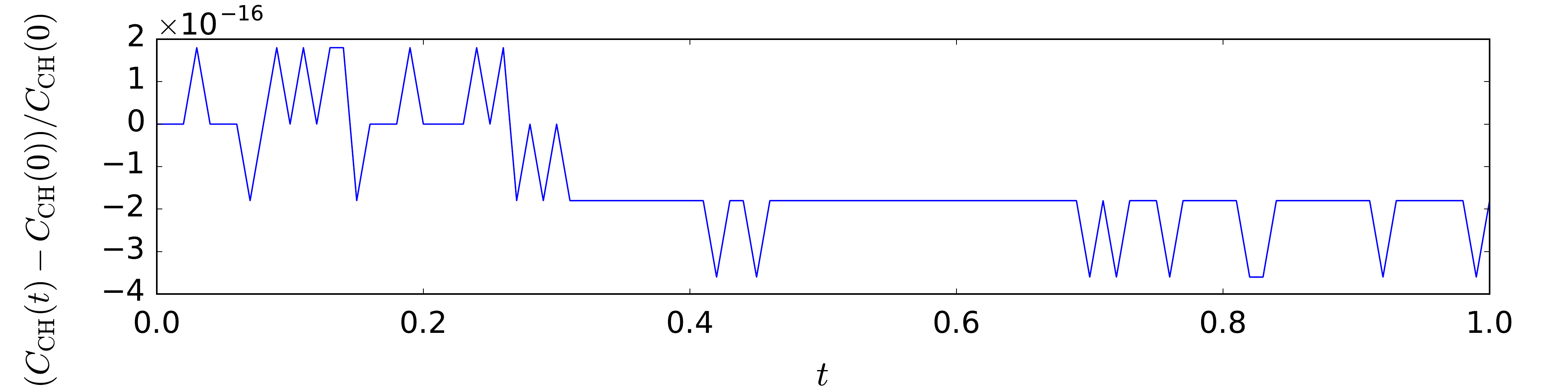}
\caption{Orszag-Tang vortex with the pseudo-spectral integrator (left) and the variational integrator (right). Error of the total energy $E$, the $L^{2}$ norm $C_{L^{2}}$ of $\psi$, magnetic helicity $C_{\mathrm{MH}}$ and cross helicity $C_{\mathrm{CH}}$. Note the common factor on the top of the axis.}
\label{fig:orszag_tang_vortex_timetraces}
\end{figure}

The Orszag-Tang vortex describes a turbulent setting in which current sheets develop. These are areas where the magnetic field changes sign such that the current density becomes very large.
Figure~\ref{fig:orszag_tang_vortex_current_density} compares the current density computed with the variational integrator~\eqref{eq:rmhd_integrator_simplified} and the pseudo-spectral integrator.
The current sheets are located in those parts of the plot where the colour changes from blue to yellow within a small region.
The results shown here are in good agreement with those of \citet{CordobaMarliani:2000} as well as those of another variational integrator for ideal MHD \cite{KrausMaj:2016} and a method based on discrete Euler-Poincaré reduction~\cite{Gawlik:2011}.
Moreover, the variational integrator and the pseudo-spectral integrator also show good quantitative agreement as can be seen in Figure~\ref{fig:orszag_tang_vortex_norms}, which shows a comparison of the $L_{2}$ norms of the current density $j$ and the vorticity $\omega$ as well as the kinetic and magnetic energy.
After $t = 0.7$, the simulation is under-resolved. Note that in the original work, \citet{CordobaMarliani:2000} use an adaptive mesh refinement approach with an initial resolution of $1024 \times 1024$ points and several refinement cycles.
The variational integrator, on the other hand, captures the correct behaviour and preserves energy, magnetic helicity and cross helicity to machine precision even at very low resolution (see Figure~\ref{fig:orszag_tang_vortex_timetraces}).

\clearpage

\subsection{Current Sheet}\label{ssec:current_sheet}

Also in this example we adopt the variational integrator \eqref{eq:rmhd_integrator_simplified} to solve the RMHD equations~\eqref{eq:rmhd_equations}. We consider the following initial condition
\begin{align}\label{eq:current_sheet_initcondcs}
\phi &= \phi_{0} \, \big( \cos (x+y) - \cos (x-y) \big) , &
\psi &= \dfrac{\psi_{0}}{\cosh^{2} (x)} ,
\end{align}
with $\psi_{0} = 1.29$ and $\phi_{0} = 10^{-3}$,
which leads to the formation of a current sheet centred at $x=0$. The same initial condition was adopted for collisionless reconnection studies~\cite{Grasso:2006, TassiGrasso:2010}.
The spatial domain is $(x,y) \in [-\pi , +\pi) \times [-\pi , +\pi)$ with periodic boundaries, resolved by $n_{x} \times n_{y} = 1024 \times 512$ grid points.
In order to satisfy the periodicity condition in the $x$ direction a Fourier series representation of the equilibrium flux function has been adopted. Namely, we expand the expression for $\psi$ in Equation~\eqref{eq:current_sheet_initcondcs} in a Fourier series, truncated up to $22$ modes. This truncation has already been shown~\cite{Grasso:2006} to provide a good representation of the equilibrium flux function. 

\begin{figure}[h]
\centering
\subfloat[t=0]{
\includegraphics[width=.32\textwidth]{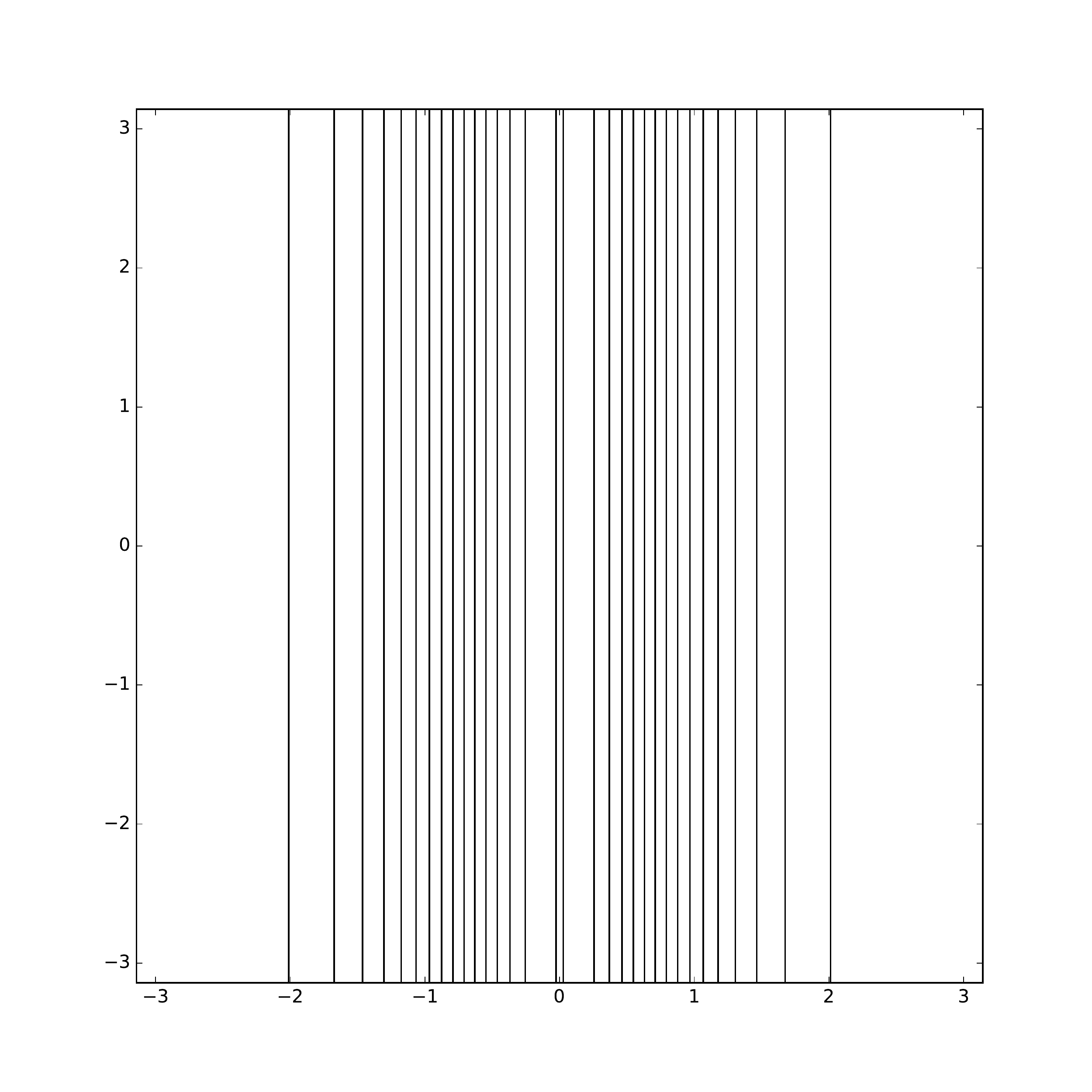}
}
\subfloat[t=50]{
\includegraphics[width=.32\textwidth]{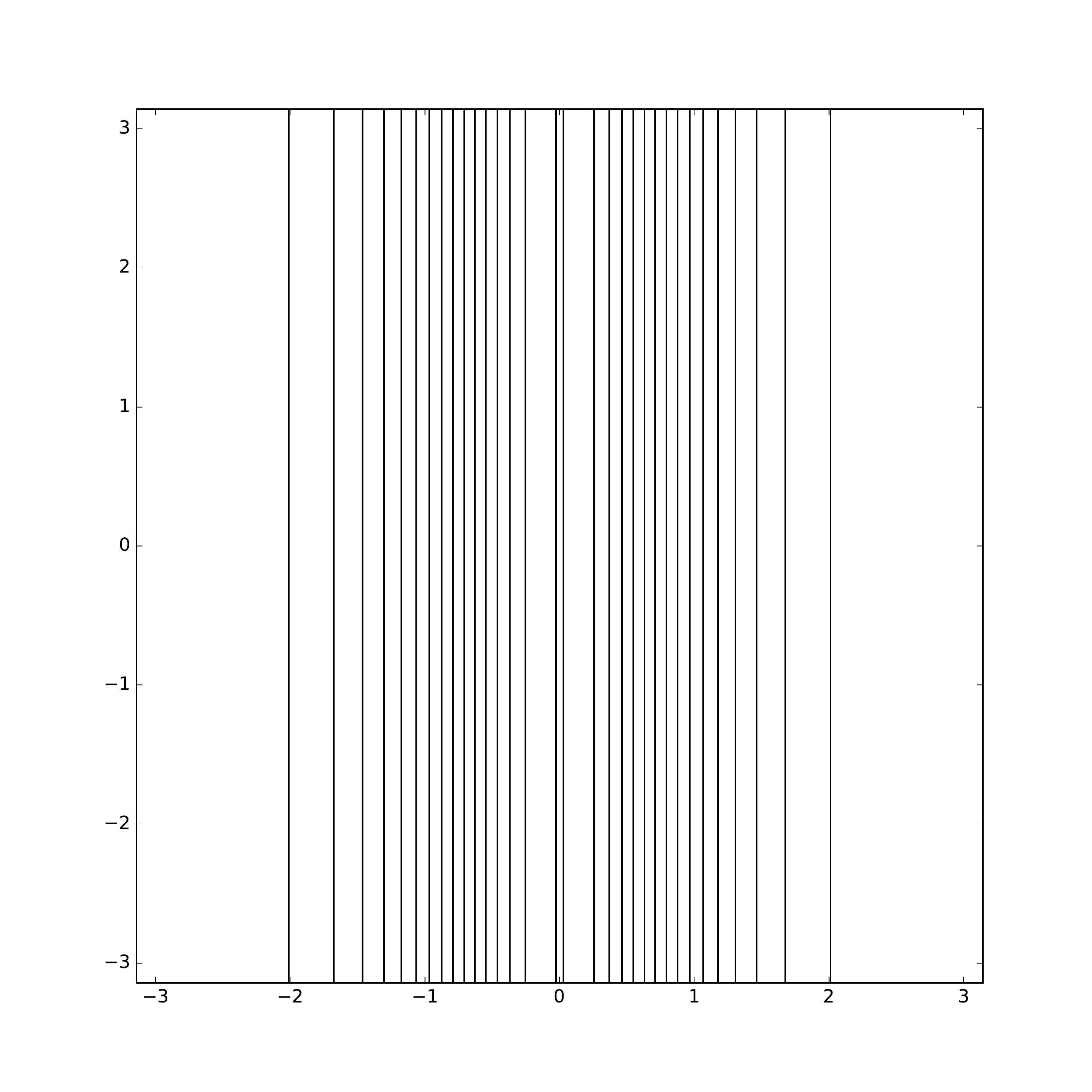}
}
\subfloat[t=100]{
\includegraphics[width=.32\textwidth]{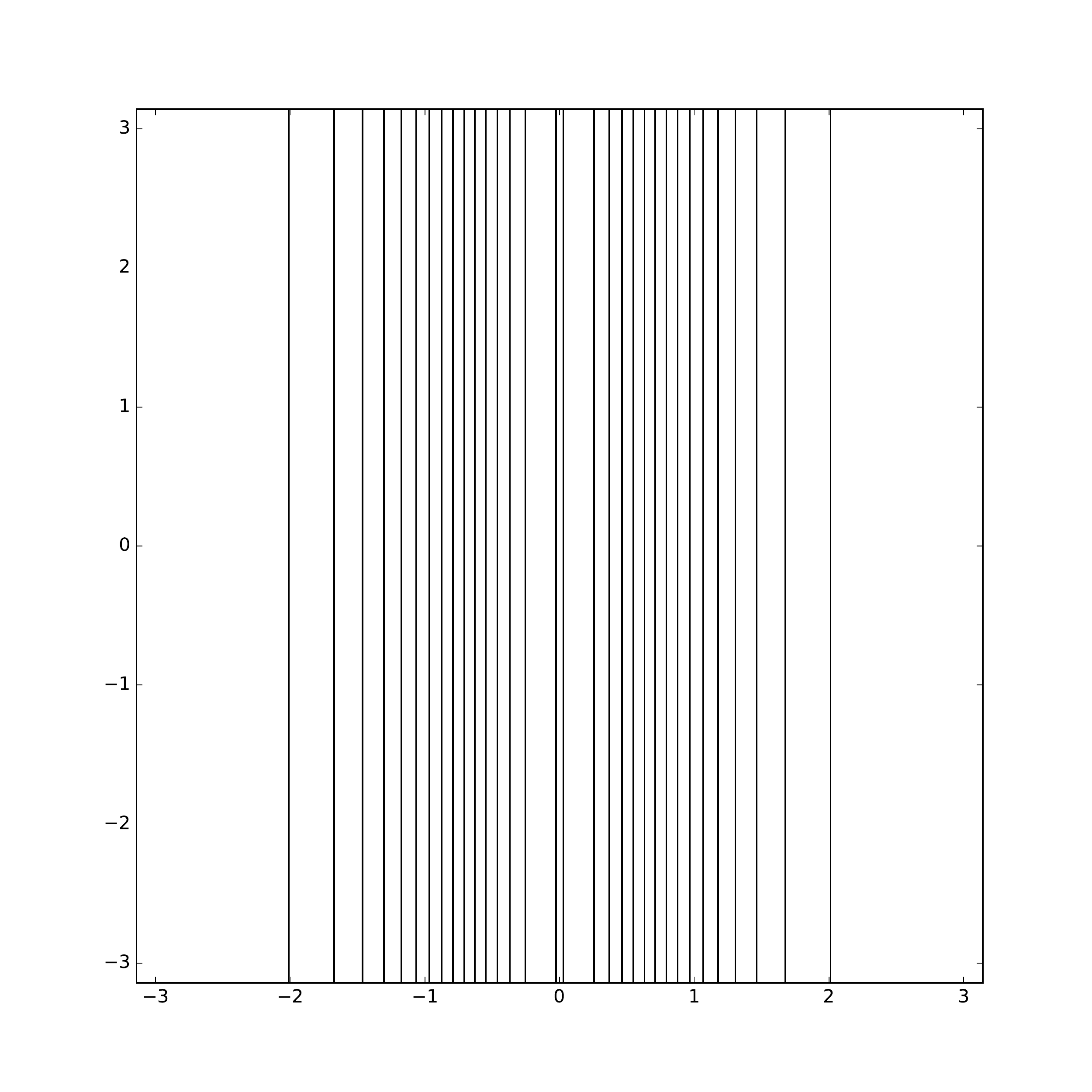}
}

\caption{Current sheet. Contour lines of the flux function $\psi$.}
\label{fig:mhd_current_sheet_psi}
\end{figure}

\begin{figure}[h]
\centering
\includegraphics[width=.45\textwidth]{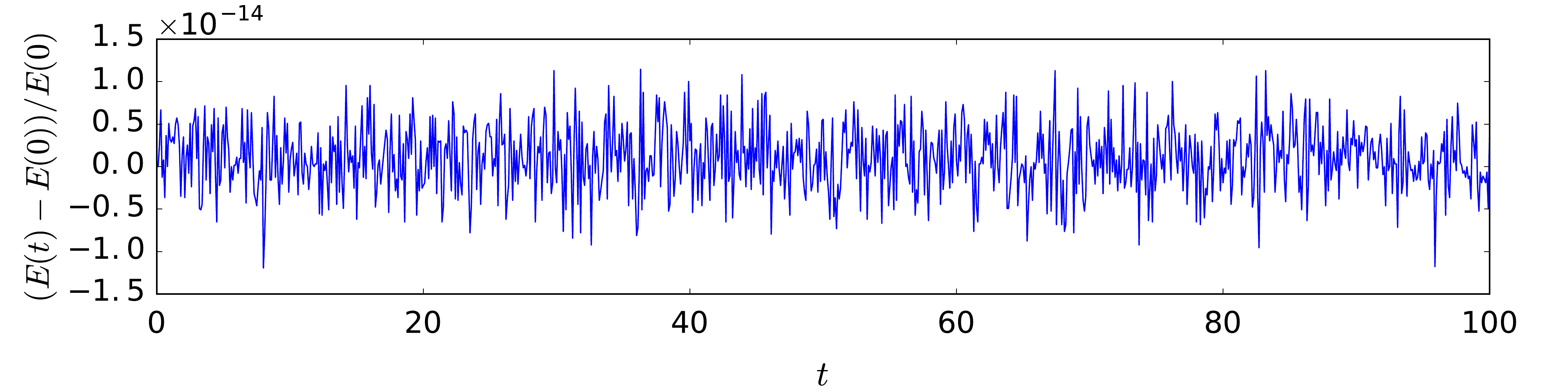}
\includegraphics[width=.45\textwidth]{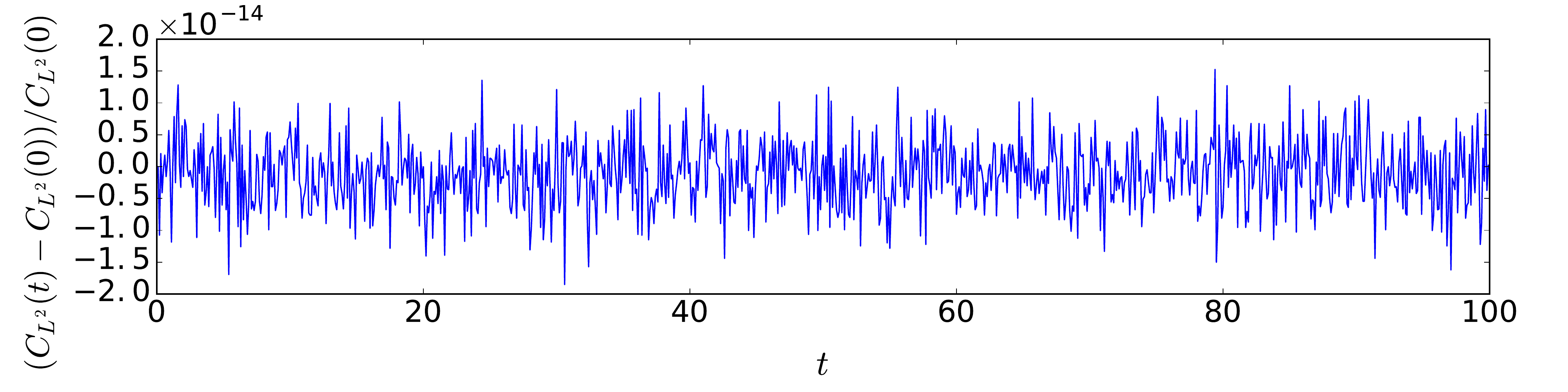}
\includegraphics[width=.45\textwidth]{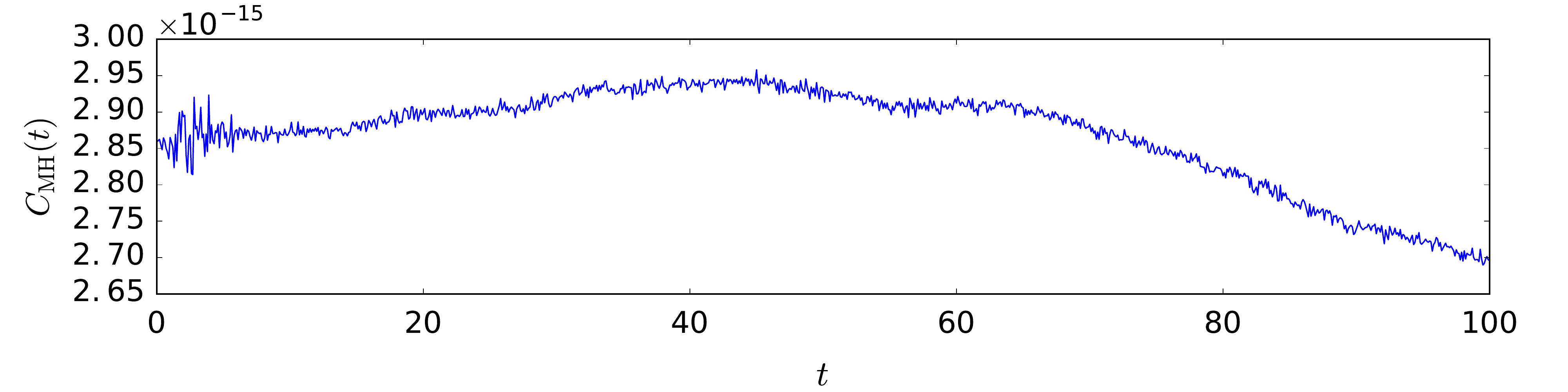}
\includegraphics[width=.45\textwidth]{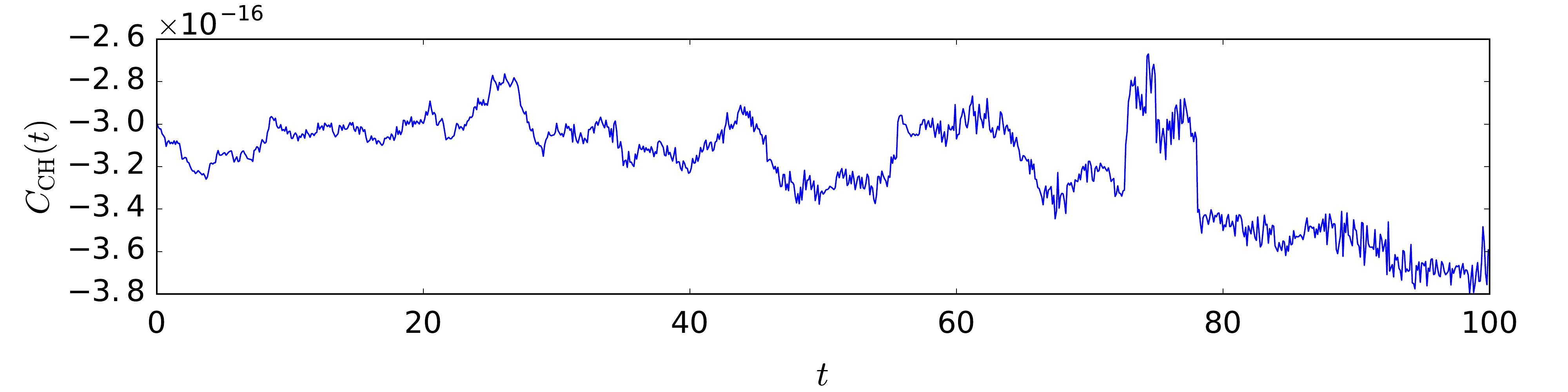}
\caption{Current sheet with the variational integrator. Error of the total energy $E$, the $L^{2}$ norm $C_{L^{2}}$ of $\psi$, magnetic helicity $C_{\mathrm{MH}}$ and cross helicity $C_{\mathrm{CH}}$.}
\label{fig:mhd_current_sheet_timetraces}
\end{figure}

In Figure~\ref{fig:mhd_current_sheet_psi} the magnetic potential $\psi$ is plotted at various points in time.
It is seen that the topology of the contour lines of the magnetic potential is preserved.
Artificial reconnection due to spurious effects of the numerics is absent.
This is also reflected in the good conservation properties regarding energy, magnetic helicity and cross helicity (see Figure~\ref{fig:mhd_current_sheet_timetraces}).

\subsection{Collisionless Reconnection}

\begin{figure}[t]
	\centering
	\includegraphics[width=.7\textwidth]{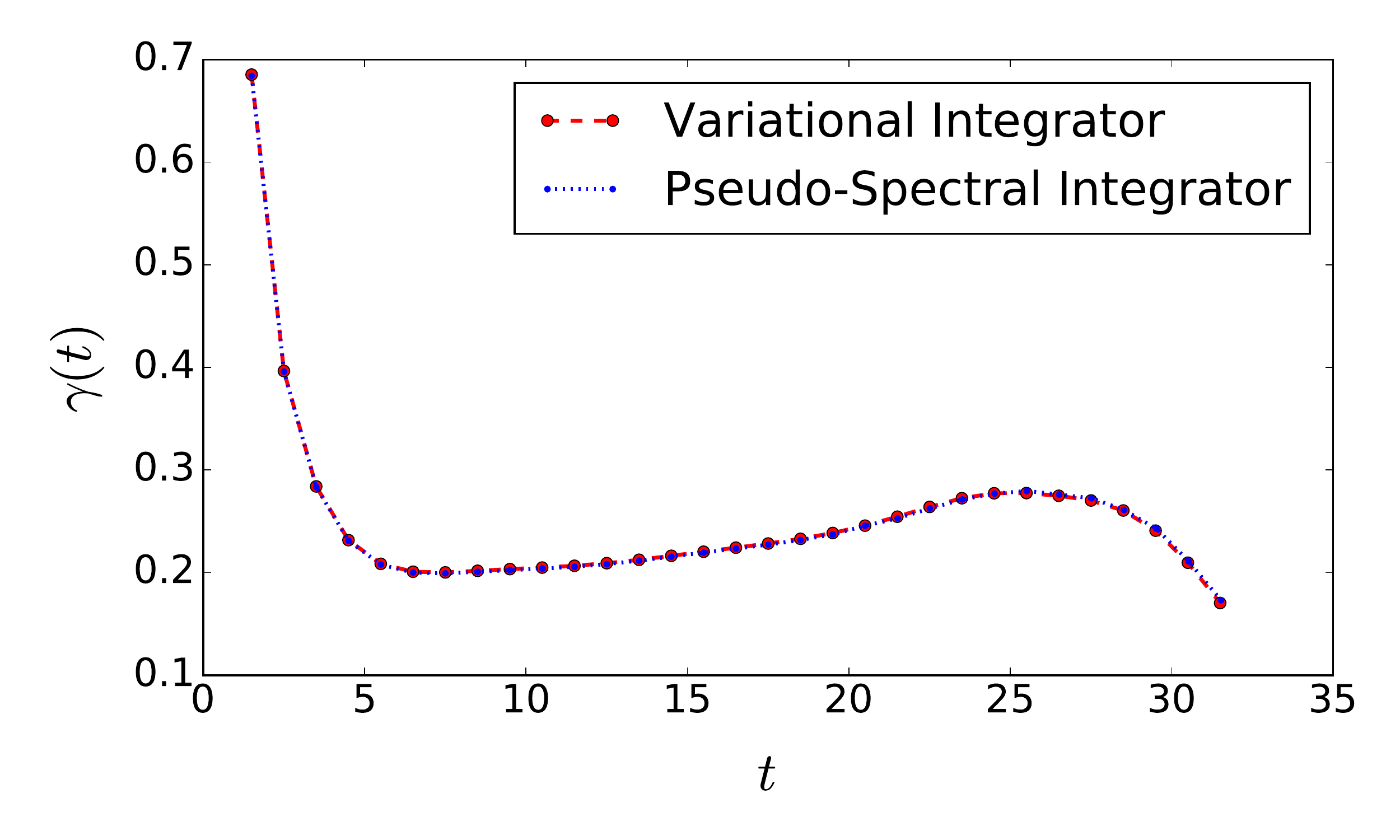}
	\caption{Growth rate $\gamma$ of the magnetic island computed with both integrators.}
	\label{fig:reconnection_growth_rate}
\end{figure}

In the previous example we verified that in the ideal case, the variational integrator~\eqref{eq:rmhd_integrator_simplified} is free of artificial reconnection due to numerical resistivity or other spurious effects.
Here, we use the same setup and the same initial condition as in the example of Section~\ref{ssec:current_sheet}, but we solve the RMHD model with electron inertia effects corresponding to equations~\eqref{eq:rmhd_equations_ei},  so that reconnection of magnetic field lines is expected to take place. We compare the results obtained from the variational integrator with those obtained from a pseudo-spectral code.
The simulation reported here has been performed on a spatial domain of $(x,y) \in [-\pi , +\pi) \times [-\pi , +\pi)$ with periodic boundaries using $n_{x} \times n_{y} = 1024 \times 512$ grid points.
The value of the electron skin depth $d_e$ has been set equal to $0.2$. 
For the variational integrator we use a time step of $h_{t} = 0.01$. For the pseudo-spectral code a minimum value of $h_{t} = 0.001$ in the nonlinear phase has been adopted.
A first important quantity to consider in reconnection studies is the linear growth rate of the initial perturbation, which is defined as
\begin{align}
\gamma(t) = \frac{d}{dt} \ln (\psi(t,0,\pi)-\psi(0,\pi,0)) .
\end{align}
From Figure~\ref{fig:reconnection_growth_rate} one sees that the growth rate follows the expected behaviour consisting of a transient phase up to $t=6$, approximately, followed by the linear phase, from $t=6$ to $t=12$, where $\gamma$ is nearly constant, before entering the nonlinear phase for $t>12$. We observe that growth rates determined with the variational integrator and with the pseudo-spectral code are almost identical.
The same level of agreement is observed for a case with weaker initial perturbation ($\phi_0 = 10^{-8}$), where the linear phase lasts longer than in the case shown here.

\begin{figure}[p]
\centering
\subfloat[t=28]{
\includegraphics[width=.42\textwidth]{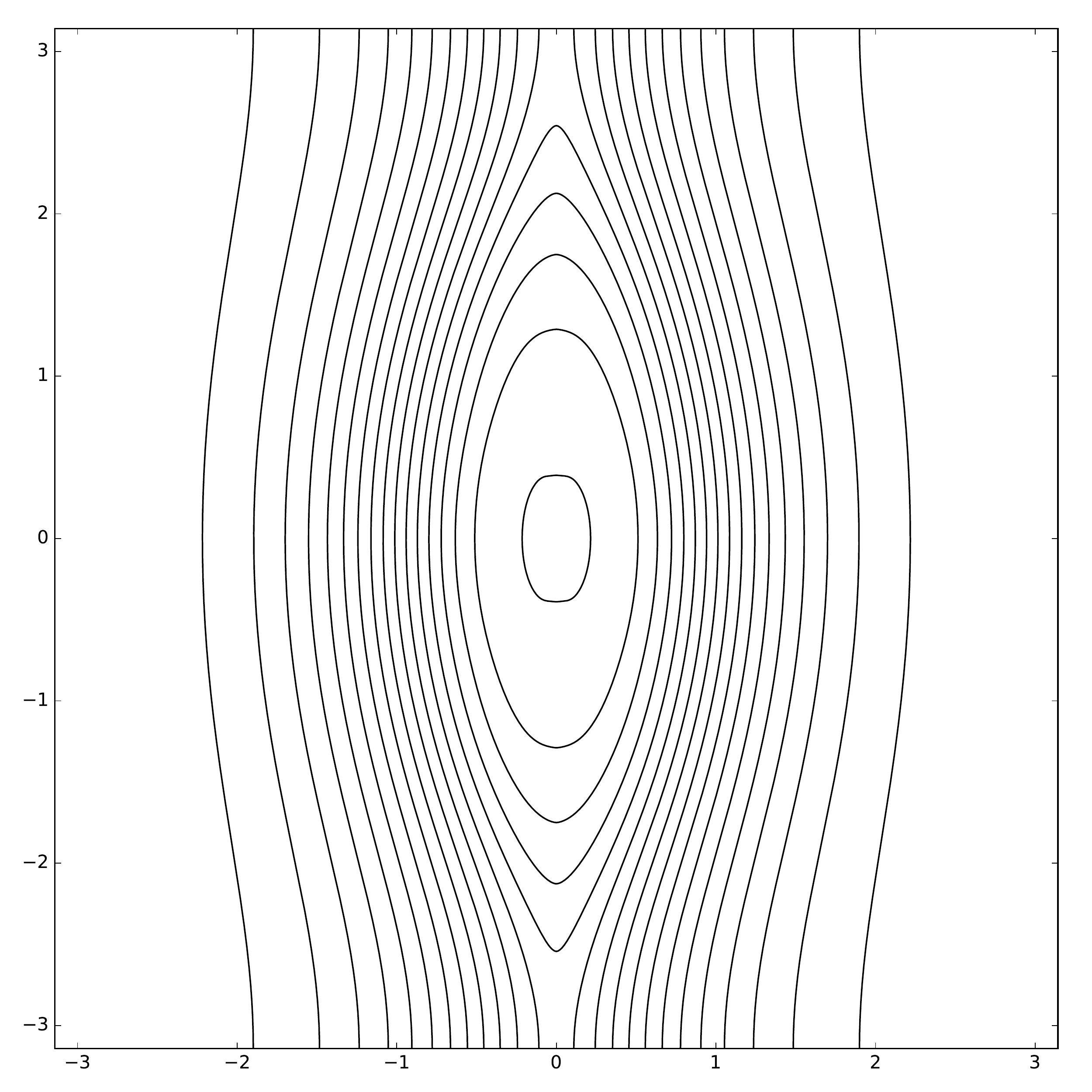}
}
\subfloat[t=28]{
\includegraphics[width=.42\textwidth]{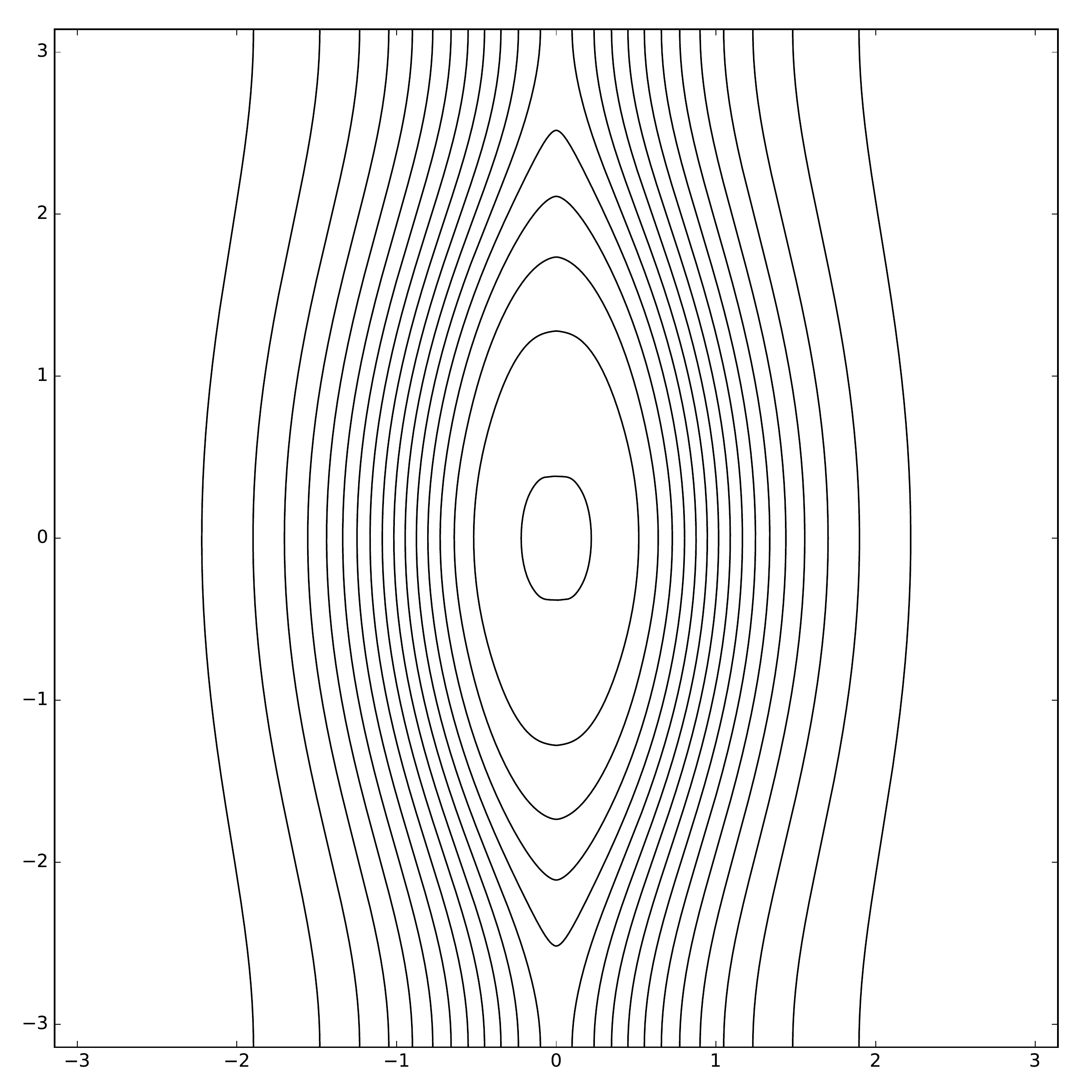}
}

\subfloat[t=30]{
\includegraphics[width=.42\textwidth]{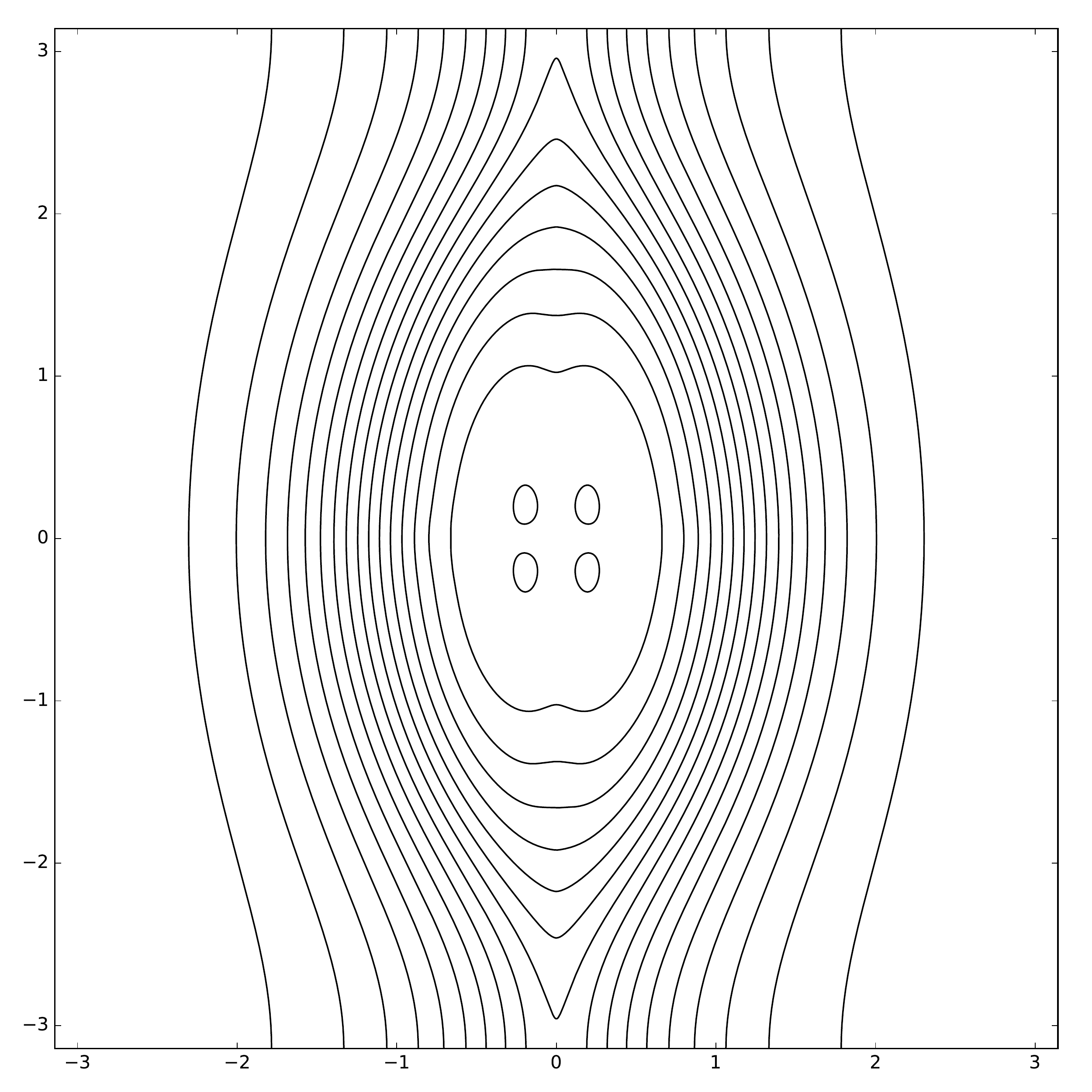}
}
\subfloat[t=30]{
\includegraphics[width=.42\textwidth]{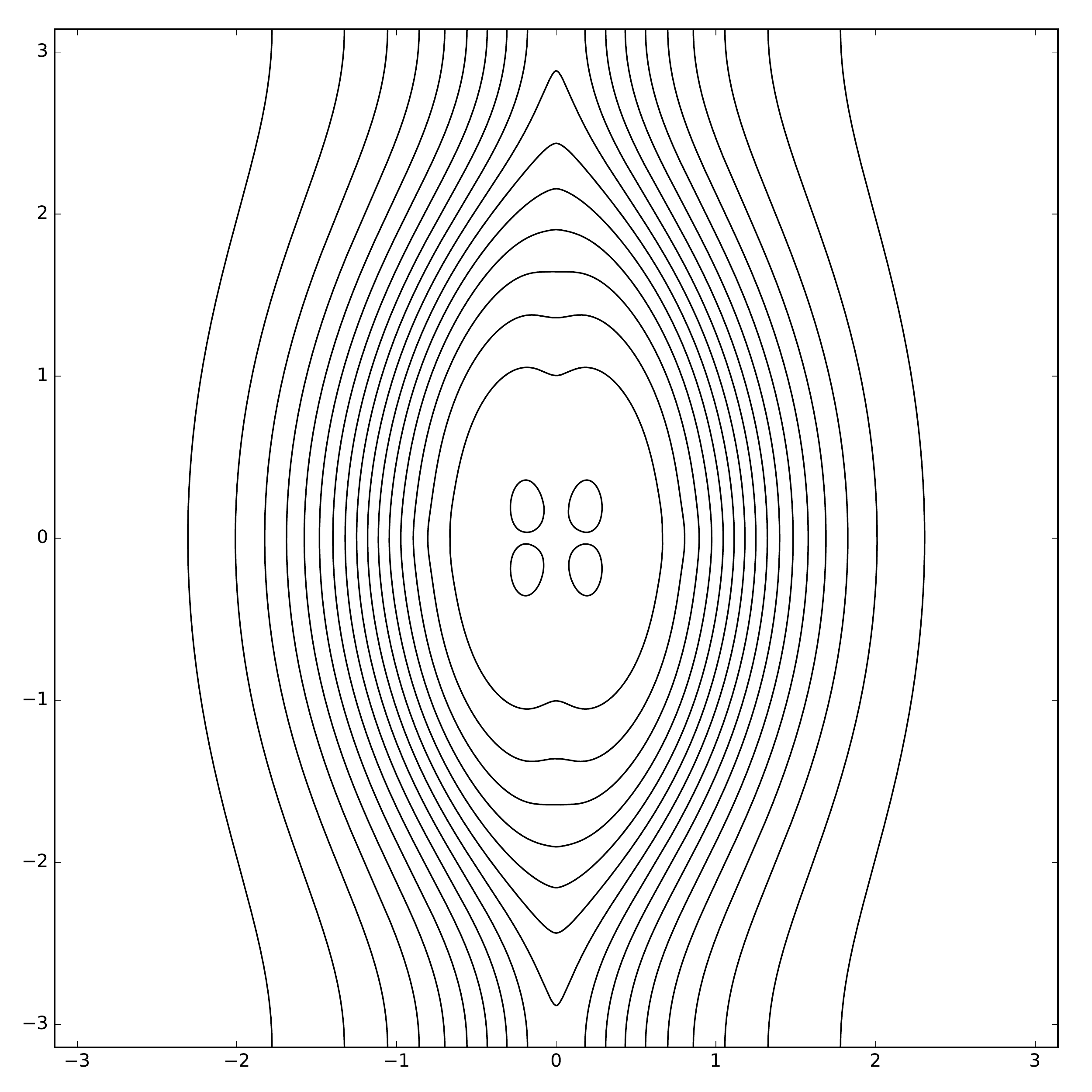}
}

\subfloat[t=32]{
\includegraphics[width=.42\textwidth]{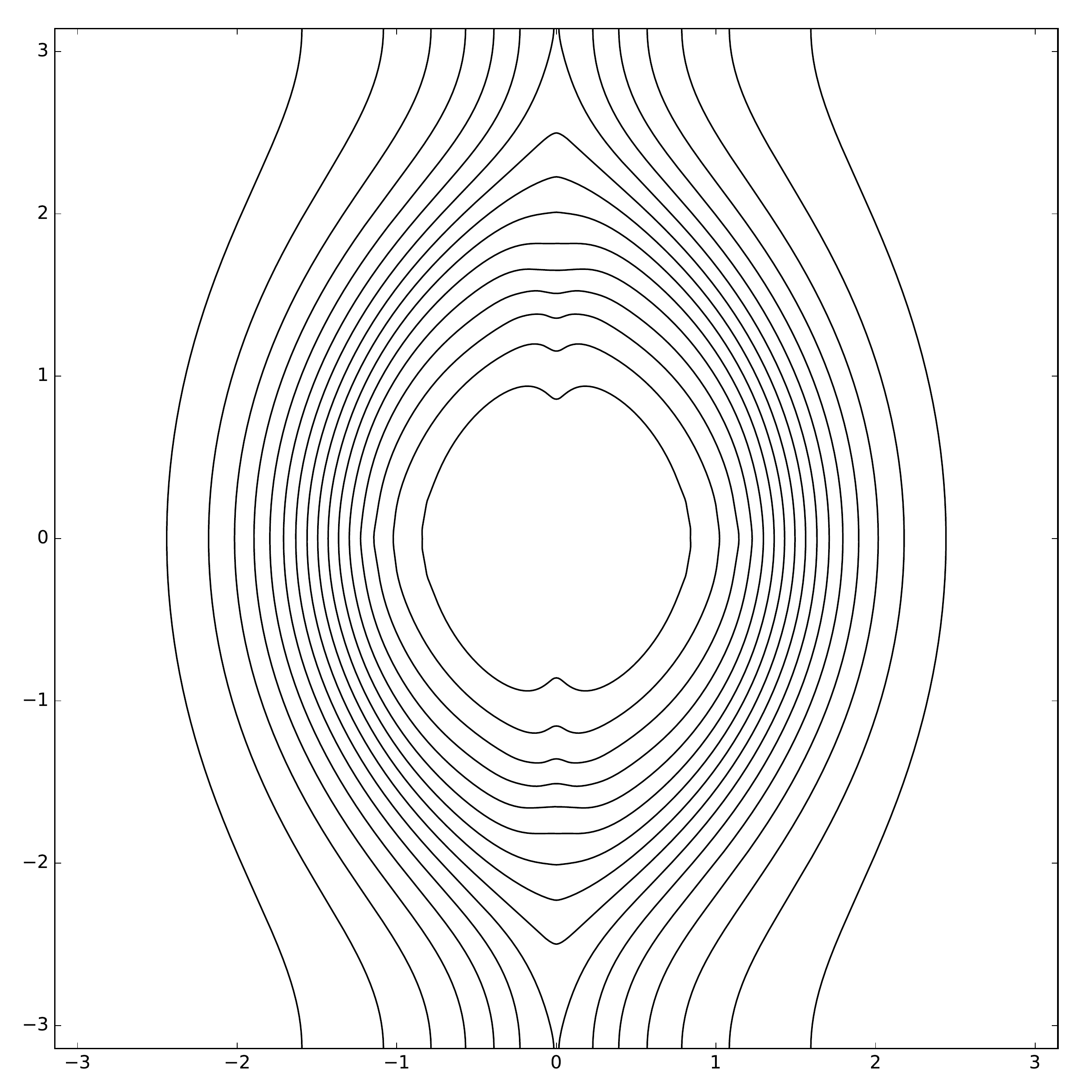}
}
\subfloat[t=32]{
\includegraphics[width=.42\textwidth]{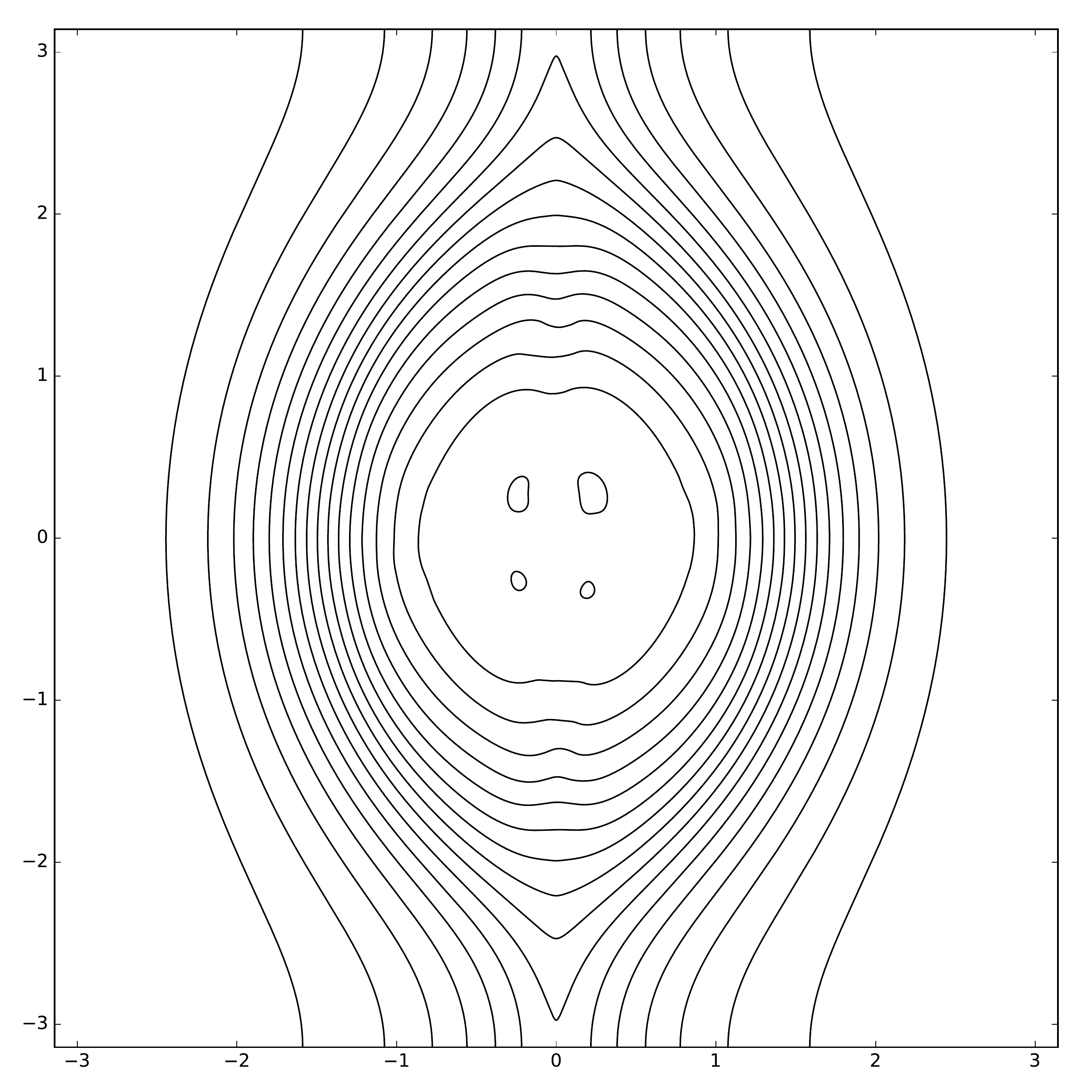}
}

\caption{Magnetic reconnection with the pseudo-spectral integrator (left) and the variational integrator (right). Vector potential $\psi$.}
\label{fig:mhd_reconnection_psi}
\end{figure}

The dynamics of the island growth is practically identically modelled by both, the variational and the pseudo-spectral integrator (see Figure~\ref{fig:mhd_reconnection_psi}). This does not appear too surprising, given the exponential convergence of the pseudo-spectral method and its good conservation properties until shortly before the end of the simulation. Only then we observe minor differences in the solutions at the inside of the island.
At this point, after $t=30$, power spectra, not shown here, indicate that the simulation is under-resolved for both integrators, with non-negligible amounts of energies residing in the smallest scales. This phase of the evolution corresponds to the onset of a secondary Kelvin-Helmholtz-type of instability, that characterises this collisionless reconnection regime \cite{Del03,Del05}. A turbulent regime follows, with energy continuously transferred to smaller scales but with the constraint of total energy conservation due to the Hamiltonian nature of the system.
In this situation, from Figure~\ref{fig:mhd_reconnection_psi} it emerges that the variational integrator does not preserve the intrinsic parity symmetry of the equations, $\psi(x,y) = \psi(-x,y)$ and $\psi(x,y) = \psi(x,-y)$, anymore. This, however, is expected in the turbulent regime as a consequence of chaotic dynamics. The same loss of parity, although not visible here, actually occurs also for the pseudo-spectral code, although at a slightly later time.

\begin{figure}[p]
\centering
\subfloat[t=28]{
\includegraphics[width=.42\textwidth]{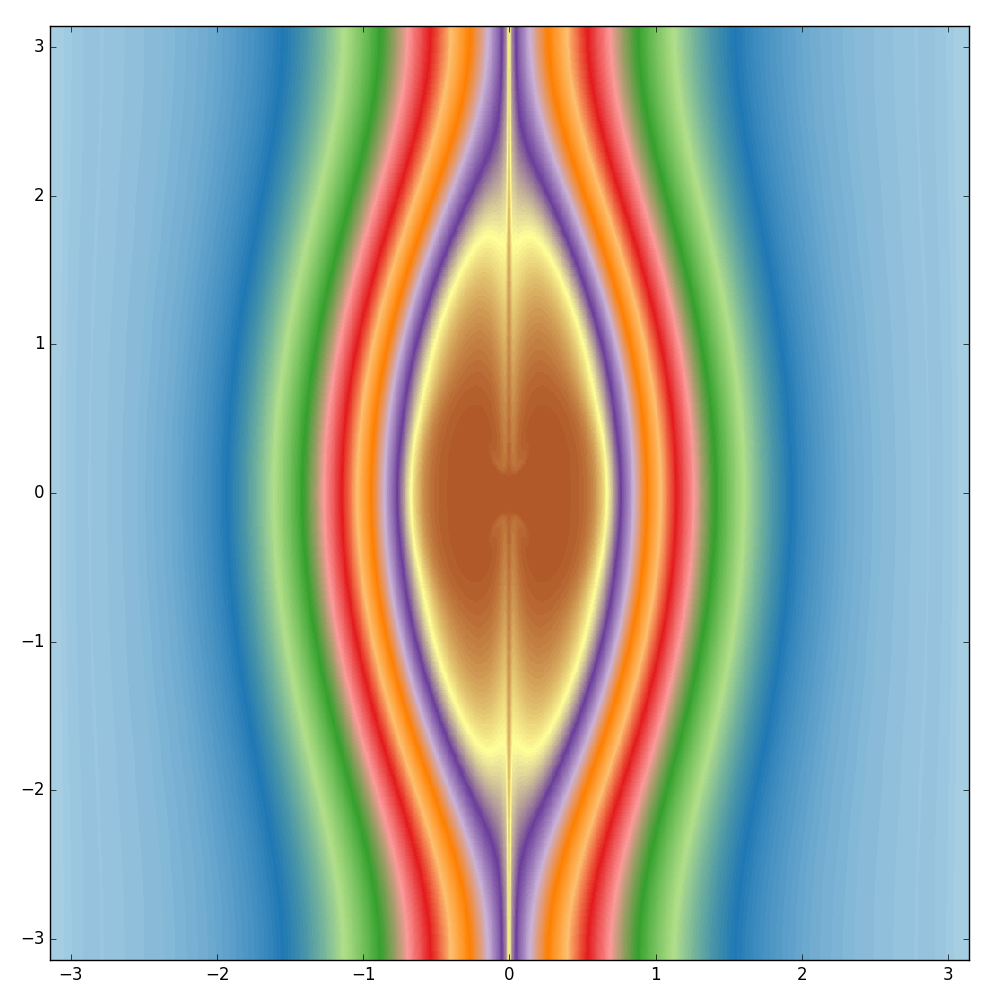}
}
\subfloat[t=28]{
\includegraphics[width=.42\textwidth]{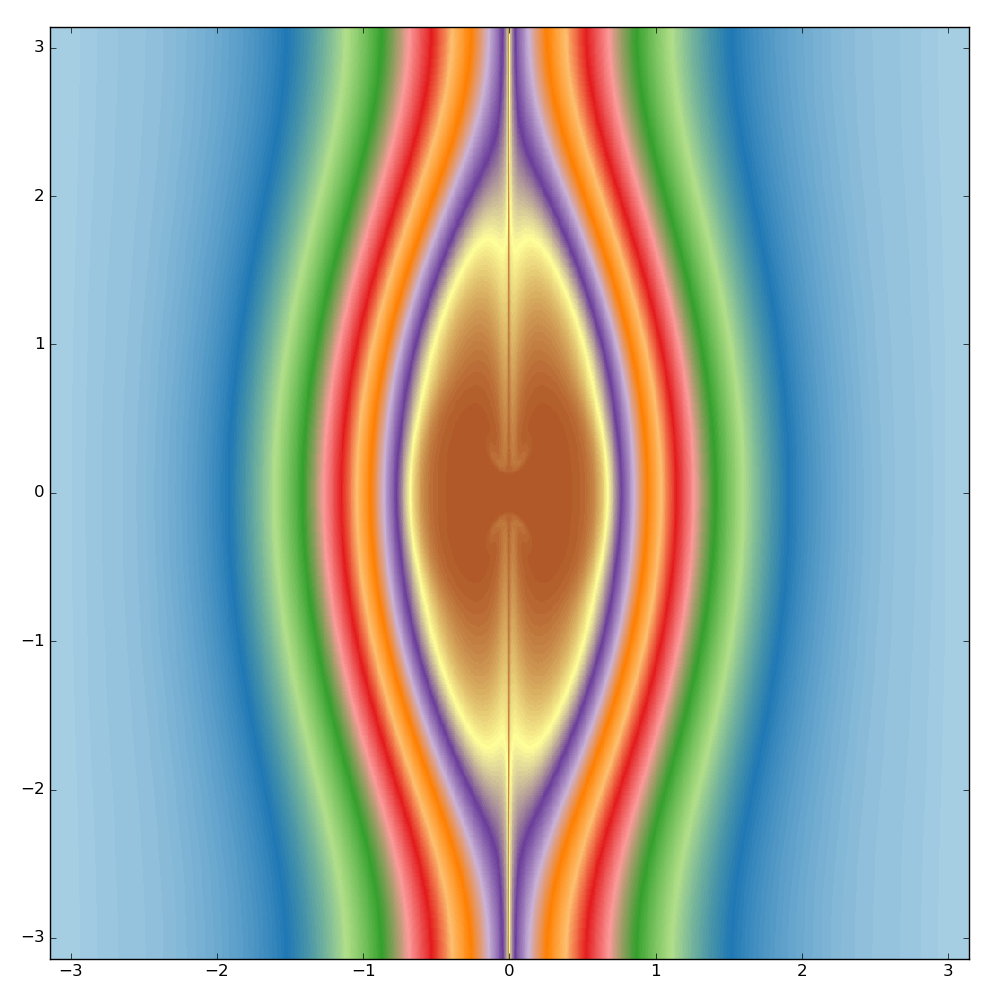}
}

\subfloat[t=30]{
\includegraphics[width=.42\textwidth]{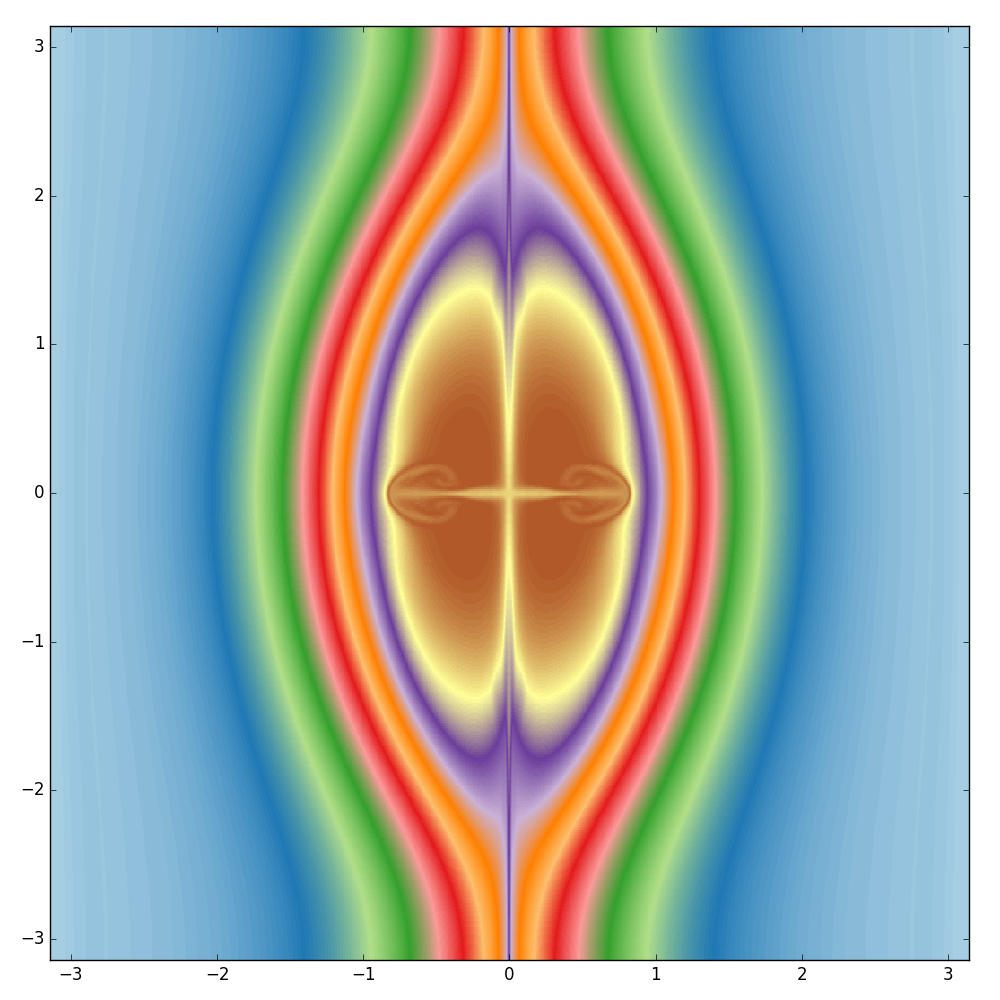}
}
\subfloat[t=30]{
\includegraphics[width=.42\textwidth]{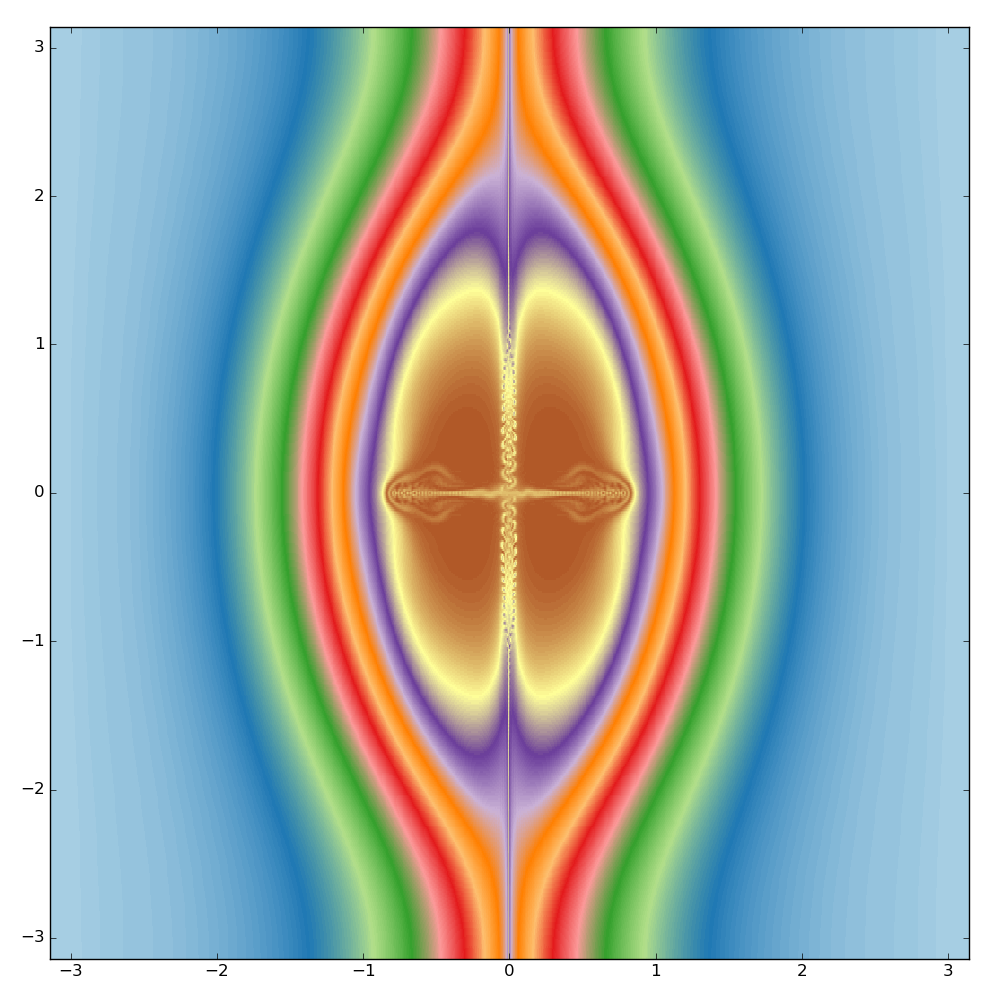}
}

\subfloat[t=32]{
\includegraphics[width=.42\textwidth]{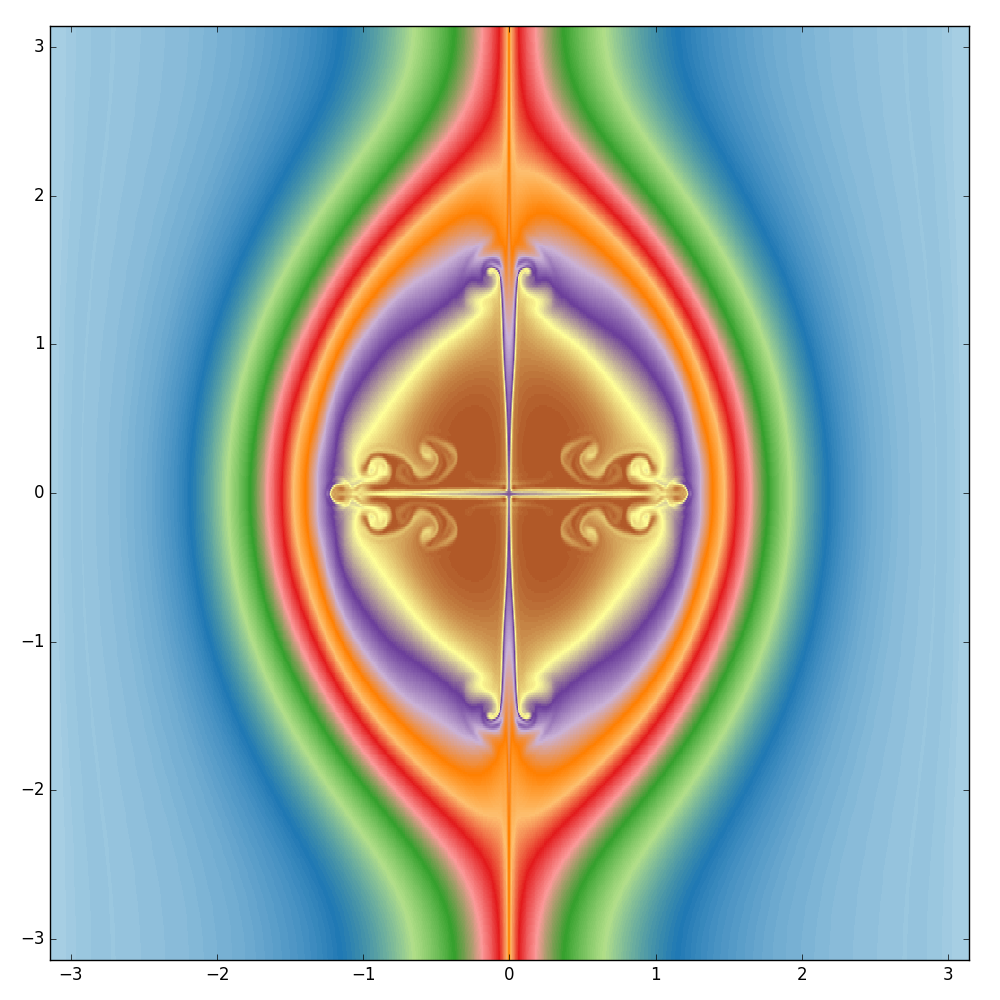}
}
\subfloat[t=32]{
\includegraphics[width=.42\textwidth]{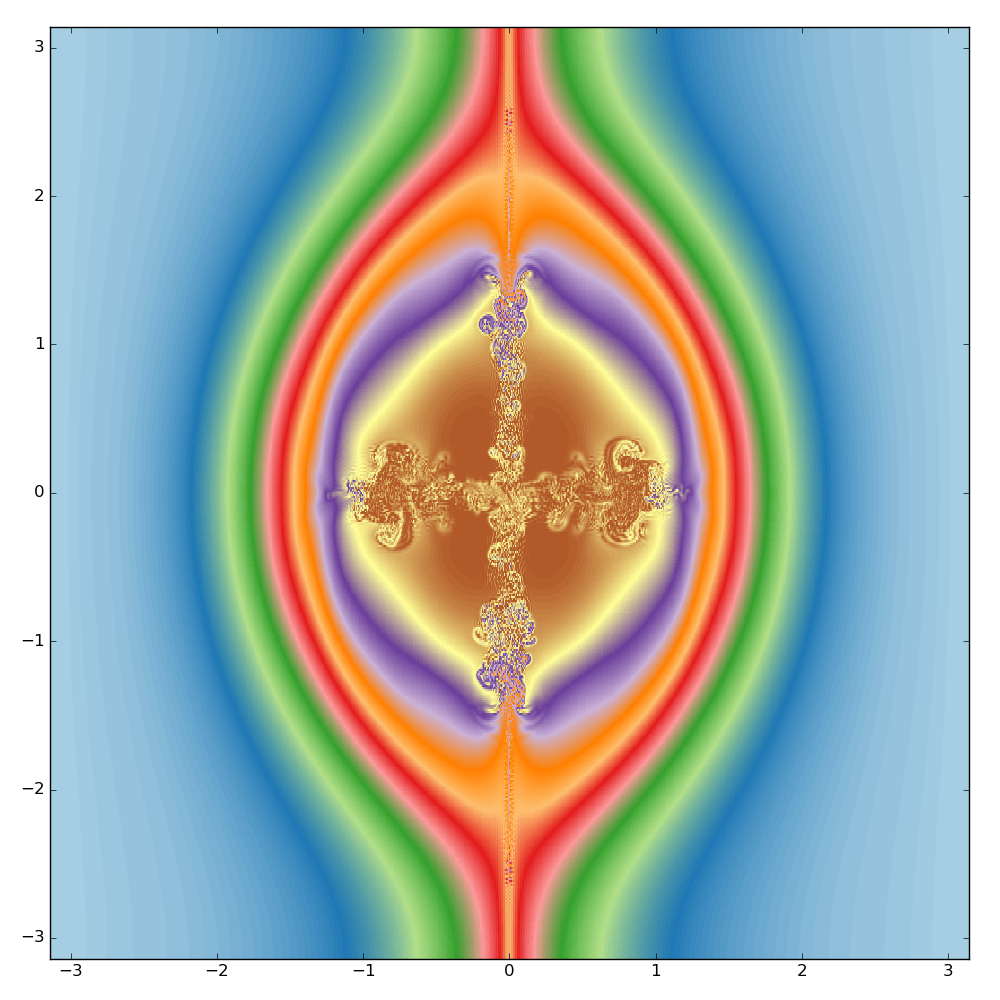}
}

\caption{Magnetic reconnection with the pseudo-spectral integrator (left) and the variational integrator (right). Generalised vector potential $\bar{\psi}$. Fixed colour scale.}
\label{fig:mhd_reconnection_psie}
\end{figure}

In general, the results of both integrators agree very well also in the generalised magnetic potential $\bar{\psi}$ (see Figure~\ref{fig:mhd_reconnection_psie}) and the vorticity $\omega$ (not shown here), except for the turbulent fine scale structures along the $x=0$ and more prominently the $y=0$ axes, which appear at about $t=30$. For a given resolution, these structures appear later in the solution of the pseudo-spectral method, which might also explain why the pseudo-spectral integrator retains parity symmetry longer than the variational integrator. This could be due to the dissipation introduced by the discretisation of the time derivative, which has a smoothing effect on the solution and causes the dissipation of energy and the $L^{2}$ norm of $\bar{\psi}$. We remark that, when increasing resolution, the appearance of the small scale turbulent structures occurs earlier. The growth rate of the related instability seems indeed to increase with decreasing scales, as is the case for the standard Kelvin-Helmholtz instability. 

\begin{figure}[h]
\centering
\includegraphics[width=.45\textwidth]{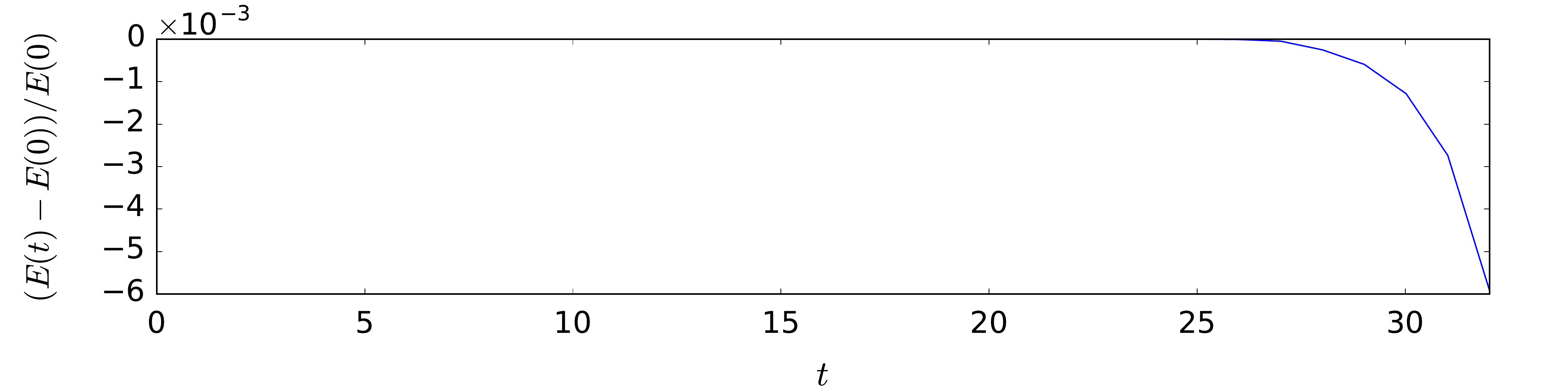}
\includegraphics[width=.45\textwidth]{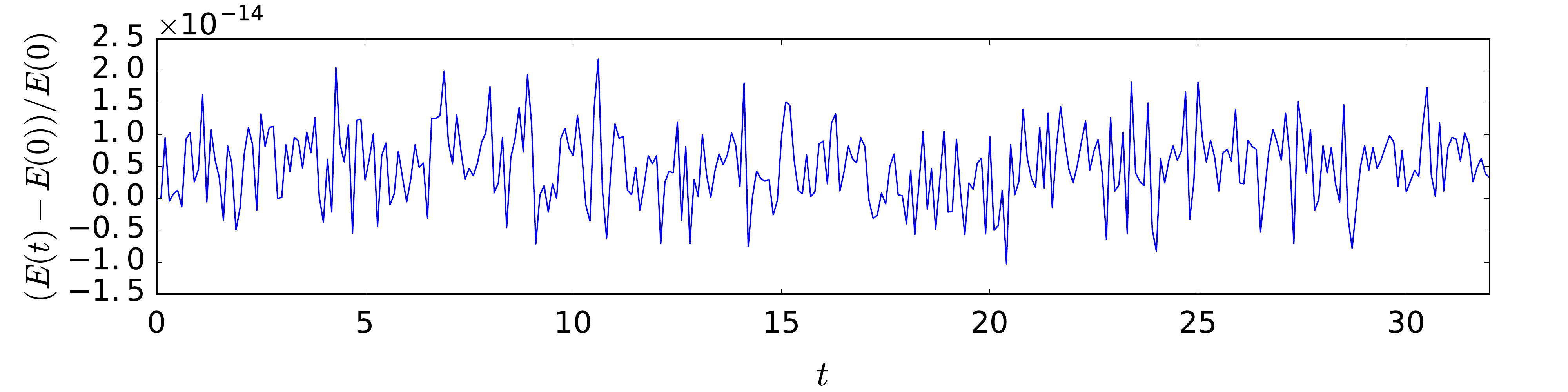}

\includegraphics[width=.45\textwidth]{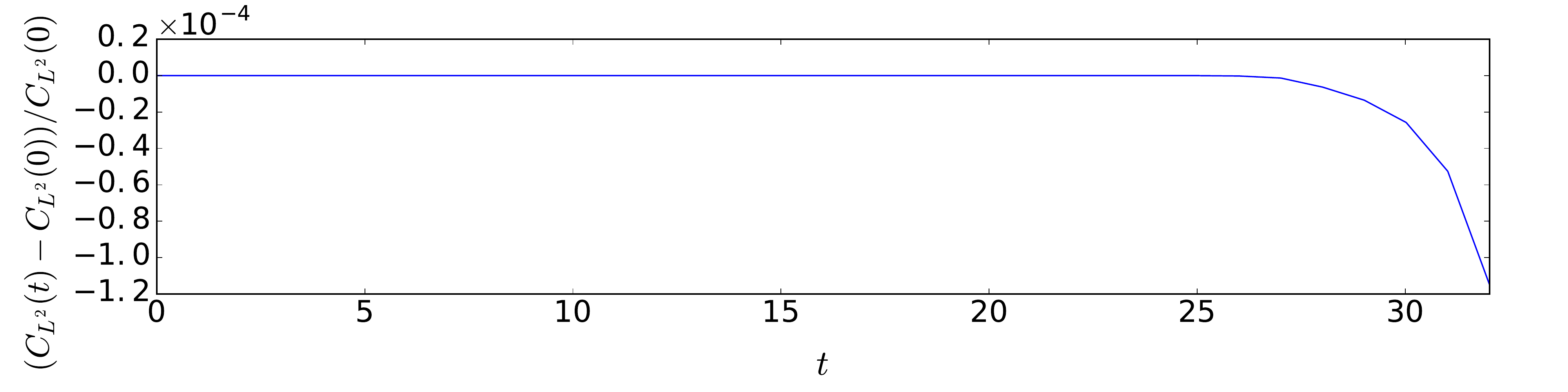}
\includegraphics[width=.45\textwidth]{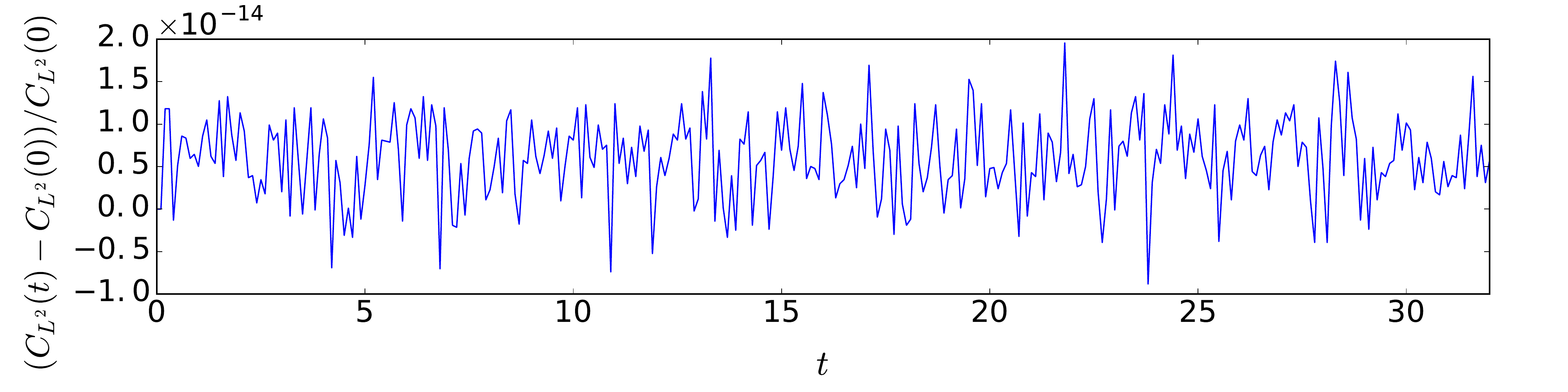}

\includegraphics[width=.45\textwidth]{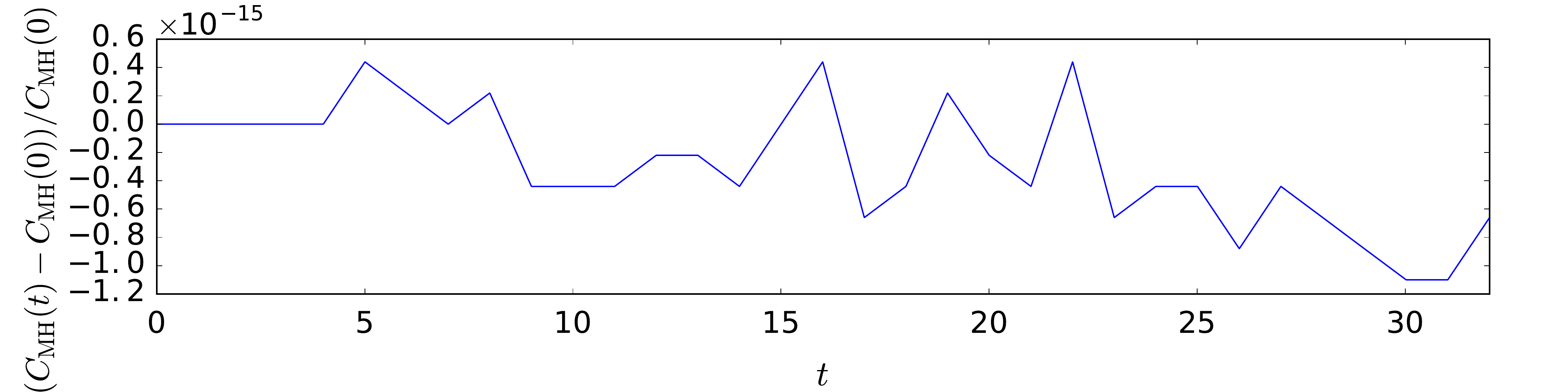}
\includegraphics[width=.45\textwidth]{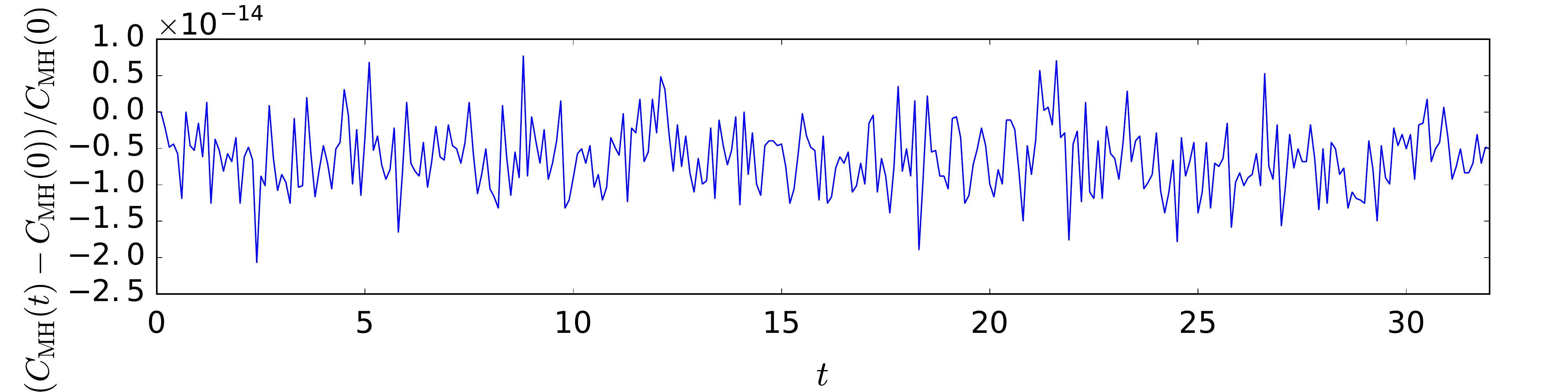}

\includegraphics[width=.45\textwidth]{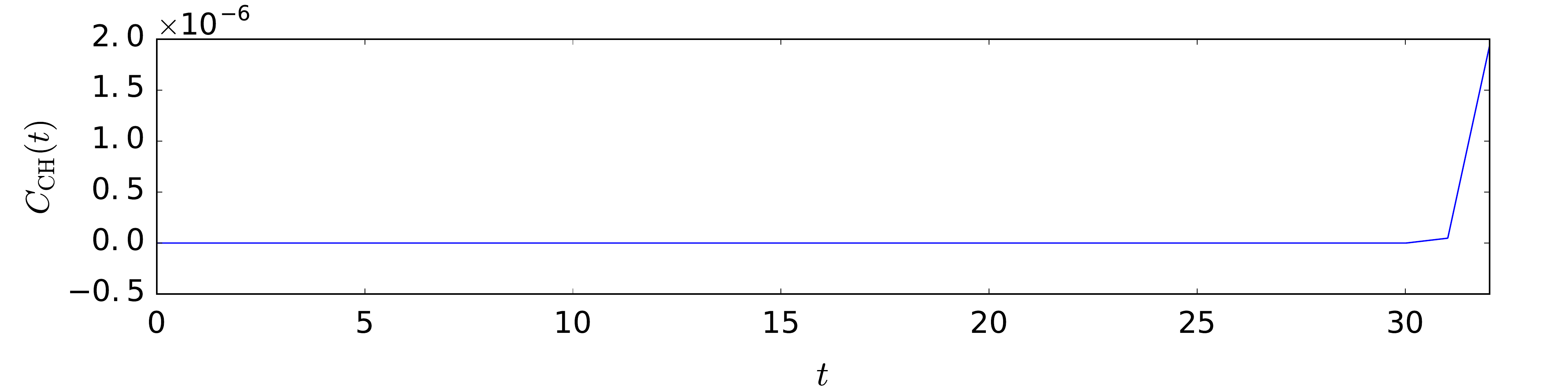}
\includegraphics[width=.45\textwidth]{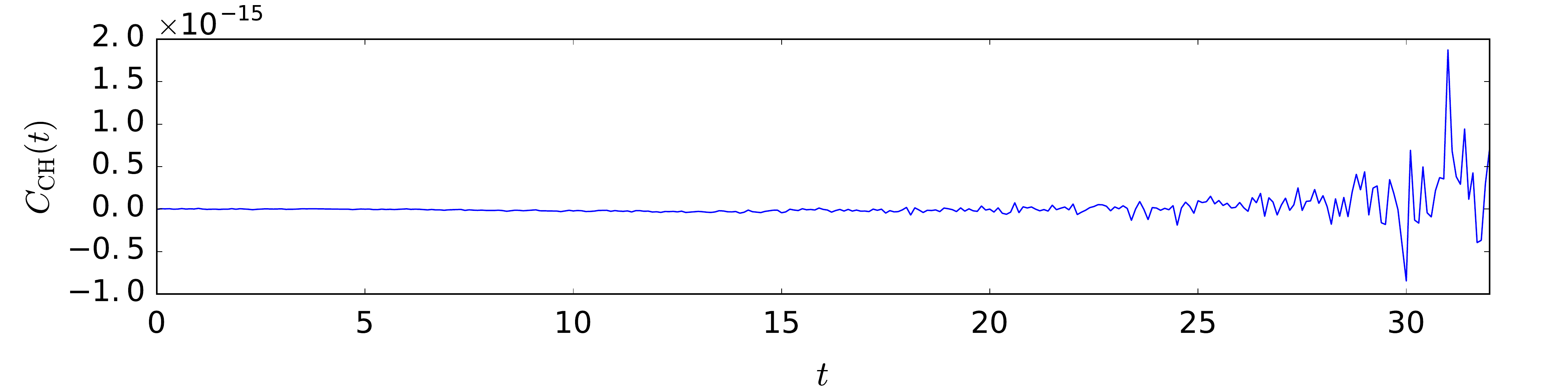}

\caption{Reconnection with the pseudo-spectral integrator (left) and the variational integrator (right). Error of the total energy $E$, the $L^{2}$ norm $C_{L^{2}}$ of $\bar{\psi}$, generalised magnetic helicity $C_{\mathrm{MH}}$ and cross helicity $C_{\mathrm{CH}}$.}
\label{fig:mhd_reconnection_timetraces}
\end{figure}

Throughout the simulation, the variational integrator shows excellent preservation of energy, the $L^{2}$ norm of $\bar{\psi}$, generalised magnetic helicity and cross helicity (see Figure~\ref{fig:mhd_reconnection_timetraces}). The pseudo-spectral integrator dissipates energy and the $L^{2}$ norm of $\bar{\psi}$, but preserves magnetic helicity and cross helicity (see Figure~\ref{fig:mhd_reconnection_timetraces}). Only towards the end of the simulation, conservation of cross helicity degrades rapidly. Without electron inertia and for the simulation time considered here, the pseudo-spectral integrator also preserves energy and the $L^{2}$ norm of $\psi$ exactly. With the variational integrator we also see an increase in cross helicity when the island develops, but the order of magnitude of the total cross helicity still remains close to machine accuracy.

\clearpage

\section{Summary and Outlook}

In this work we developed a variational discretisation of ideal and inertial RMHD. As we do not have a suitable physical Lagrangian available, we adopted a formal Lagrangian approach~\cite{KrausMaj:2015, Ibragimov:2006, Ibragimov:2007}. A particularly interesting feature of this formulation is that not only conservation laws related to variational symmetries but also the conservation of Casimirs can be analysed by the Noether theorem.

In the spatial discretisation of the action principle, we used a simple finite difference approach, which is a straight forward extension of our previous work on the vorticity equation~\cite{KrausMaj:2015}. Here, however, we used a different strategy for the discretisation in time, which directly leads to one-step time integrators of Runge-Kutta type. Even though we restricted ourselves to such a simple discretisation of the Lagrangian, leading to the Arakawa scheme combined with the Crank-Nicholson scheme, the generalisation to higher-order methods is straight forward. Due to the flexibility of the variational integrator framework, both the spatial and temporal discretisation can easily be extended towards more elaborate methods \cite{Leok:2005, Leok:2012, Hall:2015, Kraus:2016:Evolution, Kraus:2016:Splines, Chen:2008}.
In that sense the current work can be understood as a proof of principle, showing that even simple finite difference schemes obtained as variational integrators of a formal Lagrangian have excellent conservation properties when applied to nonlinear, coupled systems like RMHD.

The particular method obtained in this work preserves the total energy as well as linear and quadratic Casimir invariants of the system, namely magnetic helicity, cross helicity and the $L^{2}$ norm of the magnetic potential. In simulations of the Orszag-Tang vortex and of a current sheet model, we verified the favourable properties of the variational integrator, especially that it respects the aforementioned conservation laws of the RMHD system exactly (up to machine accuracy). A remarkable feature is that in the ideal case magnetic reconnection is absent. This is inherent to the physics but rarely respected by the numerics. Only when effects of finite electron inertia are added to the system, reconnection is observed as is expected from the theory.

\pagebreak

We compared the results of the variational integrator with a well established pseudo-spectral integrator \cite{Grasso:2009, GrassoTassi:2010, TassiGrasso:2010, TassiMorrison:2010} and found excellent qualitative and quantitative agreement.
Moreover, we found that at a given resolution the variational integrator, due to absence of dissipation effects, retains more structures at fine scales than the pseudo-spectral integrator. Therefore, in order to perform a simulation with a certain level of detail, less resolution is needed when using the variational integrator than when using the pseudo-spectral method.

The absence of dissipation can cause numerical instabilities and is problematic in the presence of turbulence, e.g., in the setting of a Kelvin-Helmholtz instability, where energy is transferred to ever finer scales, so that at some point of the simulation the resolution will always be insufficient.
For such problems, however, ideal reduced magnetohydrodynamics is not an appropriate model.
For realistic simulations in turbulent regimes, dissipation needs to be included in the model, e.g., viscosity and resistivity in the case of MHD, in order to remove energy and smoothen small structures at the correct scale.
Even though adding dissipation can lead to the loss of conserved quantities like energy, we believe that it is of great importance to preserve such conservation laws in the discretisation of the ideal part of the system. Only that way uncontrolled dissipation, as it is often introduced with other discretisation schemes, can be avoided, so that the amount of dissipation in the simulation is exactly the amount put in by physical mechanisms while spurious numerical dissipation is absent.

In the current work, we do not address questions regarding the performance of the proposed method but only provide a reference implementation in order to verify the favourable conservation properties. The work by~\citet{Chacon:2002} shows that efficient and scalable solvers for such methods can be constructed based on Jacobian-free Newton-Krylov methods with physics-based preconditioning. We expect the performance of the variational integrator derived in Section~\ref{sec:vi} to be similar to that of the scheme used by \citet{Chacon:2002} when applying the same solver strategies.
Let us note that the variational integrator, due to being a finite difference method which requires communication only between nearest neighbours, is much more amenable to parallelisation than the pseudo-spectral method, which relies on Fast Fourier Transforms and therefore global operations. Hence the variational integrator is better suited for the extension towards computationally more challenging problems, e.g., in three dimensions.

A natural extension of this work is the development of higher-order discretisations~\cite{Kraus:2016:Evolution, Kraus:2016:Splines} as well as the treatment of more physics-comprehensive models like the gyrofluid four-field model proposed by~\citet{Wae12}, models for electrostatic turbulence \cite{WaelMorHor} and models for magnetic reconnection evolving temperature and heat flux fluctuations \cite{Gra15}. As the structure of the respective equations is very similar to the ones treated here, the extension of the scheme and its implementation is expected to be straight forward.

\subsection*{Acknowledgements}

MK is grateful to Omar Maj and Akihiro Ishizawa for useful discussions on the presented matters and related topics. ET acknowledges the financial support from CNRS through the PICS project NEICMAR.

\appendix

\section{Nonlinear Solver and Preconditioner}
\label{sec:app_solver}

The system~\eqref{eq:rmhd_integrator_simplified} constitutes a nonlinear algebraic system of equations, which is solved using Newton's method.
Denoting $\phy = (\omega, \psi, \phi, j)$, we can write~\eqref{eq:rmhd_integrator_simplified} as $F(\phy^{n+1}) = 0$, with
\begin{subequations}\label{eq:app_residuals}
\begin{align}
\nonumber
F_{\omega}
&= \omega^{n+1} - \omega^{n} + \dfrac{h_{t}}{4} \big[ A (\phi^{n+1}, \omega^{n+1}) + A (\phi^{n}, \omega^{n+1}) + A (\phi^{n+1}, \omega^{n}) + A (\phi^{n}, \omega^{n}) \big]
\\
&\hphantom{ = \omega^{n+1} - \omega^{n} } \; + \dfrac{h_{t}}{4} \big[ A (j^{n+1}, \psi^{n+1}) + A (j^{n}, \psi^{n+1}) + A (j^{n+1}, \psi^{n}) + A (j^{n}, \psi^{n}) \big] ,
\\
F_{\psi}
&= \psi^{n+1} - \psi^{n} + \dfrac{h_{t}}{4} \big[ A (\phi^{n+1}, \psi^{n+1}) + A (\phi^{n}, \psi^{n+1}) + A (\phi^{n+1}, \psi^{n}) + A (\phi^{n}, \psi^{n}) \big] ,
\\
F_{\phi} &= \omega^{n+1} + \Delta \phi^{n+1} ,
\\
F_{j} &= j^{n+1} + \Delta \psi^{n+1} .
\end{align}
\end{subequations}
Newton's method consists of solving a series of systems of the form
\begin{align}\label{app:newton}
J (\phy^{n+1}_{m}) \, \delta \phy^{n+1}_{m} = - F (\phy^{n+1}_{m}) ,
\end{align}
where $J$ denotes the Jacobian matrix,
\begin{align}
J (\phy^{n+1}) = F'(\phy^{n+1}) ,
\end{align}
$\phy^{n+1}_{m}$ is the $m$th field vector, $\delta \phy^{n+1}_{m}$ is the $m$th increment, from which the $(m+1)$th field vector is computed as
\begin{align}
\phy^{n+1}_{m+1} = \phy^{n+1}_{m} + \delta \phy^{n+1}_{m} ,
\end{align}
$F (\phy^{n+1}_{m})$ is the vector of residuals as defined in~\eqref{eq:app_residuals} and $m$ is the nonlinear iteration level.
This iterative procedure is stopped when some convergence criterion is satisfied, specifically
\begin{align}
\norm{ F (\phy^{n+1}_{m}) }_{2} < n \varepsilon_{\text{abs}} + \varepsilon_{\text{rel}} \norm{ F (\phy^{n+1}_{0}) }_{2} ,
\end{align}
where $\norm{\cdot}_{2}$ denotes the $l^{2}$ norm, $n = n_{x} \times n_{y}$ is the total number of grid points $\varepsilon_{\text{abs}} = 5 \times 10^{-16}$ is the absolute tolerance and $\varepsilon_{\text{rel}}$ is the relative tolerance (usually set to $10^{-10}$ in this work).
The $4 \times 4$ block matrix $J$ can easily be computed as
\begin{align}
J = \begin{pmatrix}
\unity + \tfrac{h_{t}}{2} A ( \phi^{n+1/2}_{m}, \cdot ) & \hphantom{\unity + } \; \tfrac{h_{t}}{2} A ( j^{n+1/2}_{m}, \cdot ) & - \tfrac{h_{t}}{2} A ( \omega^{n+1/2}_{m}, \cdot ) & - \tfrac{h_{t}}{2} A ( \psi^{n+1/2}_{m}, \cdot ) \\
\mathbb{0} & \unity + \tfrac{h_{t}}{2} A ( \phi^{n+1/2}_{m}, \cdot ) & - \tfrac{h_{t}}{2} A ( \psi^{n+1/2}_{m}, \cdot ) & \mathbb{0} \\
\unity & \mathbb{0} & \Delta & \mathbb{0} \\
\mathbb{0} & \Delta & \mathbb{0} & \unity \\
\end{pmatrix} ,
\end{align}
where $\phy^{n+1/2}_{m} = ( \phy^{n} + \phy^{n+1}_{m} ) / 2$.
With that, the system~\eqref{app:newton} is solved iteratively using GMRES. In each iteration of the GMRES algorithm, we only have to compute a matrix-vector product of the form $J \upsilon$, which can be implemented in a matrix-free way so that the actual Jacobian matrix $J$ does not need to be constructed and stored.

In order to increase the efficiency of the solver, we introduce a preconditioning matrix $P$ into~\eqref{app:newton}, that is
\begin{align}\label{app:newton_pc}
\big( J (\phy^{n+1}_{m}) \, P^{-1} \big) \, \big( P \, \delta \phy^{n+1}_{m} \big) = - F (\phy^{n+1}_{m}) ,
\end{align}
so that GMRES will solve
\begin{align}
\big( J (\phy^{n+1}_{m}) \, P^{-1} \big) \, \upsilon = - F (\phy^{n+1}_{m}) .
\end{align}
In each GMRES iteration, we have to solve $P \hat{\upsilon} = \upsilon$ for $\hat{\upsilon}$ and compute the matrix-vector product $J \hat{\upsilon}$.
The Newton increment is found upon solving $P \, \delta \phy^{n+1}_{m} = \delta \hat{\phy}^{n+1}_{m}$, where $\delta \hat{\phy}^{n+1}_{m}$ is the result of the GMRES solver.
The matrix $P$ should be such that it approximates $J$ but is easier to solve for. In its construction, we use a physics-based approach, following the work of \citet{Chacon:2002} (see also~\citet{Chacon:2003}). The derivation is summarised in the following section.

\subsection*{Derivation of the Preconditioner}

We start the construction of the preconditioner with the linearised, semi-discrete (Crank-Nicolson in time, c.f.~\eqref{eq:rmhd_integrator_simplified}) version of~\eqref{eq:rmhd_equations}, that is 
\begin{subequations}\label{app:linearised_integrator}
\begin{align}
\label{app:linearised_integrator_omega}
\delta \omega^{n+1}
\nonumber
&+ \tfrac{h_{t}}{2} \{ \phi^{n+1/2} , \delta \omega^{n+1} \}
 + \tfrac{h_{t}}{2} \{ \delta \phi^{n+1} , \omega^{n+1/2} \} \\
&+ \tfrac{h_{t}}{2} \{ j^{n+1/2} , \delta \psi^{n+1} \} 
 + \tfrac{h_{t}}{2} \{ \delta j^{n+1} , \psi^{n+1/2} \}
= - F_{\omega} , \\
\label{app:linearised_integrator_psi}
\delta \psi^{n+1}
&+ \tfrac{h_{t}}{2} \{ \phi^{n+1/2} , \delta \psi^{n+1} \}
 + \tfrac{h_{t}}{2} \{ \delta \phi^{n+1} , \psi^{n+1/2} \}
= - F_{\psi} , \\
\label{app:linearised_integrator_phi}
\delta \omega^{n+1} &+ \Delta \delta \phi^{n+1}
= - F_{\phi} , \\
\label{app:linearised_integrator_j}
\delta j^{n+1} &+ \Delta \delta \psi^{n+1}
= - F_{j} .
\end{align}
\end{subequations}
Insert~\eqref{app:linearised_integrator_phi} and~\eqref{app:linearised_integrator_j} as well as the definitions of $\omega$ and $j$ into~\eqref{app:linearised_integrator_omega},
\begin{align}
\Delta \delta \phi^{n+1}
\nonumber
&- \tfrac{h_{t}}{2} \{ \Delta \phi^{n+1/2}, \delta \phi^{n+1} \}
 + \tfrac{h_{t}}{2} \{ \phi^{n+1/2}, \Delta \delta \phi^{n+1} \}
 + \tfrac{h_{t}}{2} \{ \phi^{n+1/2}, F_{\phi} \} \\
&+ \tfrac{h_{t}}{2} \{ \Delta \psi^{n+1/2}, \delta \psi^{n+1} \}
 - \tfrac{h_{t}}{2} \{ \psi^{n+1/2}, \Delta \delta \psi^{n+1} \}
 - \tfrac{h_{t}}{2} \{ \psi^{n+1/2}, F_{j} \}
= F_{\omega} - F_{\phi} .
\end{align}
Following Equation~(22) in Reference~\cite{Chacon:2002}, we introduce the approximations
\begin{subequations}
\begin{align}
  \{ \phi^{n+1/2}, \Delta \delta \phi^{n+1} \}
- \{ \Delta \phi^{n+1/2}, \delta \phi^{n+1} \} 
&\approx \Delta \{ \phi^{n+1/2}, \delta \phi^{n+1} \} ,
\\
  \{ \psi^{n+1/2}, \Delta \delta \psi^{n+1} \}
- \{ \Delta \psi^{n+1/2}, \delta \psi^{n+1} \} 
&\approx \Delta \{ \psi^{n+1/2}, \delta \psi^{n+1} \} ,
\end{align}
\end{subequations}
so that
\begin{multline}
\Delta \delta \phi^{n+1}
 + \tfrac{h_{t}}{2} \Delta \{ \phi^{n+1/2}, \delta \phi^{n+1} \}
 - \tfrac{h_{t}}{2} \Delta \{ \psi^{n+1/2}, \delta \psi^{n+1} \}
= \\
= F_{\omega} - F_{\phi}
- \tfrac{h_{t}}{2} \{ \phi^{n+1/2}, F_{\phi} \} 
+ \tfrac{h_{t}}{2} \{ \psi^{n+1/2}, F_{j} \} .
\end{multline}
Upon inversion of the Laplace operator, we obtain
\begin{align}
\delta \phi^{n+1}
= \tfrac{h_{t}}{2} \{ \psi^{n+1/2}, \delta \psi^{n+1} \}
- \tfrac{h_{t}}{2} \{ \phi^{n+1/2}, \delta \phi^{n+1} \}
+ L ,
\end{align}
with
\begin{align}\label{eq:app_L}
L = \Delta^{-1} \big( F_{\omega} - F_{\phi}
- \tfrac{h_{t}}{2} \{ \phi^{n+1/2}, F_{\phi} \} 
+ \tfrac{h_{t}}{2} \{ \psi^{n+1/2}, F_{j} \} \big) .
\end{align}
In order to solve this equation together with~\eqref{app:linearised_integrator_psi}, we use the following Jacobi iteration, 
\begin{subequations}\label{app:iterative_solver}
\begin{align}
\label{app:iterative_solver_phi}
\delta \phi^{n+1}_{l+1}
&= \tfrac{h_{t}}{2} \{ \phi^{n+1/2}, \delta \phi^{n+1}_{l  } \}
+ \tfrac{h_{t}}{2} \{ \psi^{n+1/2}, \delta \psi^{n+1}_{l+1} \}
+ L , \\
\label{app:iterative_solver_psi}
\delta \psi^{n+1}_{l+1}
& - \tfrac{h_{t}}{2} \{ \psi^{n+1/2} , \delta \phi^{n+1}_{l+1} \}
= - \tfrac{h_{t}}{2} \{ \phi^{n+1/2} , \delta \psi^{n+1}_{l} \} - F_{\psi} ,
\end{align}
\end{subequations}
where $l$ denotes the iteration count.
Note, that here we evaluate the advective term with $\psi^{n+1}_{l}$ and not with $\psi^{n+1}_{l+1}$ as in Reference~\cite{Chacon:2002}.
Insert~\eqref{app:iterative_solver_phi} into~\eqref{app:iterative_solver_psi}, so that
\begin{multline}\label{eq:app_delta_psi}
\delta \psi^{n+1}_{l+1}
- \tfrac{h_{t}^{2}}{4} \{ \psi^{n+1/2} , \{ \psi^{n+1/2}, \delta \psi^{n+1}_{l+1} \} \}
= \\
= \tfrac{h_{t}^{2}}{4} \{ \psi^{n+1/2} , \{ \phi^{n+1/2}, \delta \phi^{n+1}_{l  } \} \}
+ \tfrac{h_{t}}{2} \{ \psi^{n+1/2} , L \}
- \tfrac{h_{t}}{2} \{ \phi^{n+1/2} , \delta \psi^{n+1}_{l} \}
- F_{\psi} .
\end{multline}
In order to solve this equation, we have to invert a parabolic operator $Q$, given by
\begin{align}\label{eq:app_parabolic_operator}
Q = \unity - \tfrac{h_{t}^{2}}{4} \{ \psi^{n+1/2} , \{ \psi^{n+1/2}, \cdot \} \} .
\end{align}
This is done by a matrix-free conjugate gradient solver\footnote{Let us remark that the optimal solution strategy for this problem is a multi-grid approach as outlined in Reference~\cite{Chacon:2002}. However, as we are concerned with a reference implementation and not with obtaining the best possible performance we adopted a simplified strategy that is easier to implement.} with a low number of iterations (depending on the problem, we use $3-5$).
Once the Jacobi iteration is terminated (after $1-3$ iterations in this work), the vorticity and current density are computed by~\eqref{app:linearised_integrator_phi} and~\eqref{app:linearised_integrator_j}.
We use the same discretisations for the Poisson brackets $\{ \cdot , \cdot \}$ and the Laplacian $\Delta$ as in~\eqref{eq:arakawa_bracket_1} and~\eqref{eq:rmhd_integrator_poisson_definition}.
The iterative Jacobi solver provides an approximate solution of~\eqref{app:newton} by $\delta \phy^{n+1}_{m} \approx P^{-1} \big( - F (\phy^{n+1}_{m}) \big)$. In the preconditioner, we have to solve more general systems of the form $P \hat{\upsilon} = \upsilon$. This is accomplished by replacing $-F$ in~\eqref{eq:app_L} and~\eqref{eq:app_delta_psi} with $\upsilon$.
For each Jacobi iteration we have one inversion of $Q$. In addition, for each GMRES iteration we have one inversion of the Laplacian $\Delta$ in order to compute the right-hand sides. This is done with conjugate gradients, preconditioned with the HYPRE/BoomerAMG multi-grid solver~\cite{hypre-web-page}.

For the case with electron inertia (c.f., Section~\ref{sec:rmhd_electron_inertia}), Equation~\eqref{eq:app_delta_psi} is replaced by
\begin{multline}\label{eq:app_delta_psie}
\delta \psi^{n+1}_{l+1}
- d_{e}^{2} \, \Delta \delta \psi^{n+1}_{l+1}
- \tfrac{h_{t}^{2}}{4} \{ \psi^{n+1/2} , \{ \psi^{n+1/2} + d_{e}^{2} \, j^{n+1/2}, \delta \psi^{n+1}_{l+1} \} \}
= \\
= \tfrac{h_{t}^{2}}{4} \{ \psi^{n+1/2} + d_{e}^{2} \, j^{n+1/2} , \{ \phi^{n+1/2}, \delta \phi^{n+1}_{l  } \} \}
+ \tfrac{h_{t}}{2} \{ \psi^{n+1/2} + d_{e}^{2} \, j^{n+1/2} , L \}
\\
- \tfrac{h_{t}}{2} \{ \phi^{n+1/2} , \delta \psi^{n+1}_{l} \}
- d_{e}^{2} \tfrac{h_{t}}{2} \{ \phi^{n+1/2} , \delta j^{n+1}_{l} \}
- F_{\psi} ,
\end{multline}
so that the parabolic operator that has to be inverted becomes
\begin{align}
\bar{Q} = \unity - d_{e}^{2} \, \Delta - \tfrac{h_{t}^{2}}{4} \{ \psi^{n+1/2} , \{ \psi^{n+1/2}, \cdot \} \} .
\end{align}
Here, the same solution strategy as above is used. However, convergence of the preconditioner can be further accelerated by preconditioning the conjugate gradient solver with HYPRE/BoomerAMG applied to $\unity - d_{e}^{2} \, \Delta$, as this expression becomes the dominating part of $\bar{Q}$, and the corresponding matrix can easily be implemented.

\pagebreak

\bibliographystyle{plainnat}
\bibliography{vi_reduced_mhd2d}

\end{document}